\documentclass{ieeeaccess}
\usepackage[utf8]{inputenc}
\usepackage[T1]{fontenc}
\usepackage{cite}
\usepackage{amsfonts}
\usepackage{algorithmic}
\usepackage{textcomp}
\usepackage{gensymb}
\usepackage{amssymb}
\usepackage{soul,framed} 
\usepackage{multirow}
\newcommand{\cmmnt}[1]{}
\usepackage[pdftex]{graphicx}
\usepackage[cmex10]{amsmath}
\usepackage{array}
\usepackage{mdwmath}
\usepackage{mdwtab}
\usepackage{physics}
\usepackage{eqparbox}
\usepackage{url}
\usepackage{comment}
\usepackage[acronym]{glossaries}
\usepackage{hyperref}
\usepackage{cleveref}
\usepackage{makerobust}
\usepackage{tikz}
\NewSpotColorSpace{PANTONE}
\AddSpotColor{PANTONE} {PANTONE3015C} {PANTONE\SpotSpace 3015\SpotSpace C} {1 0.3 0 0.2}
\SetPageColorSpace{PANTONE}%
\usetikzlibrary{arrows,shapes,positioning,shadows,trees}
\tikzset{
  basic/.style  = {draw, text width=4cm, drop shadow, rectangle},
  root/.style   = {basic, rounded corners=2pt, thin, align=center, fill=white},
  level 2/.style = {basic, rounded corners=6pt, thin,align=center, fill=white, text width=8em},
  level 3/.style = {basic, thin, align=left, fill=white, text width=8em}
}
\def\BibTeX{{\rm B\kern-.05em{\sc i\kern-.025em b}\kern-.08em
    T\kern-.1667em\lower.7ex\hbox{E}\kern-.125emX}}

\makeglossaries

\newacronym{ADC}{ADC}{analog-to-digital converter}
\newacronym{AP}{AP}{Access Point}
\newacronym{ARNS}{ARNS}{Augmented Relative Navigation System}
\newacronym{ASP}{ASP}{Adaptive Synchronous Parallel}
\newacronym{BioRARSA}{BioRARSA}{Robust Adaptive Random Search Algorithm}
\newacronym{BS}{BS}{base station}
\newacronym{BSP}{BSP}{Bulk Synchronous Parallel}
\newacronym{CDSS}{CDSS}{Cohesive Distributed Satellite System}
\newacronym{CF}{CF}{cell-free}
\newacronym{CFO}{CFO}{carrier frequency offset}
\newacronym{CNN}{CNN}{Convolutional Neural Network}
\newacronym{CoMP}{CoMP}{Coordinated Multipoint}
\newacronym{COTS}{COTS}{Commercial Off-The-Shelf}
\newacronym{CPU}{CPU}{central processing unit}
\newacronym{CRLB}{CRLB}{Cramer-Rao lower bound}
\newacronym{CSI}{CSI}{Channel State Information}
\newacronym{DBF}{DBF}{Distributed Beamforming}
\newacronym{DCA}{DCA}{Distributed Consensus Algorithm}
\newacronym{DOWR}{DOWR}{Dual One-Way Ranging}
\newacronym{DOWT}{DOWT}{Dual One-Way Time}
\newacronym{DRT}{DRT}{Dehop-Rehop transponder}
\newacronym{DSP}{DSP}{Digital Signal Processor}
\newacronym{DSS}{DSS}{Distributed Satellite System}
\newacronym{DSSS}{DSSS}{Direct Sequence Spread Spectrum Signal }
\newacronym{DTB}{DTB}{Distributed Transmit Beamforming}
\newacronym{D1BF}{D1BF}{Deterministic One-bit Feedback}
\newacronym{EKF}{EKF}{extended Kalman Filter}
\newacronym{E1BF}{E1BF}{Enhanced One-Bit Feedback}
\newacronym{FDD}{FDD}{Frequency Division Duplex}
\newacronym{FFT}{FH-FDMA}{fast Fourier transform}
\newacronym{FH-FDMA}{FH-FDMA}{Frequency Hopping-Frequency Division Multiple Access}
\newacronym{FPGA}{FPGA}{Field Programmable Gate Arrays}
\newacronym{F-RT}{F-RT}{frequency-slotted round-trip}
\newacronym{FSS}{FSS}{Federated Satellite System}
\newacronym{FTSP}{FTSP}{Flooding Time Synchronization Protocol}
\newacronym{GEO}{GEO}{Geostationary Orbit}
\newacronym{GNSS}{GNSS}{Global Navigation Satellite System}
\newacronym{GPS}{GPS}{Global Positioning System}
\newacronym{GRACE}{GRACE}{Gravity Recovery and Climate Experiment}
\newacronym{GRAIL}{GRAIL}{Gravity Recovery and Interior Laboratory}
\newacronym{IoSat}{IoSat}{Internet of Satellites}
\newacronym{INS}{INS}{Inertial navigation system}
\newacronym{InSAR}{InSAR}{Interferometric Synthetic Aperture Radar}
\newacronym{ISL}{ISL}{inter-satellite link}
\newacronym{KBR}{KBR}{K-Band Microwave Ranging}
\newacronym{KF}{KF}{Kalman filter}
\newacronym{LAMBDA}{LAMBDA}{Least-squares Ambiguity Decorrelation Adjustment}
\newacronym{LEO}{LISA}{Low-Earth Orbit}
\newacronym{LISA}{LISA}{Laser Interferometer Space Antenna}
\newacronym{LO}{LO}{local oscillator}
\newacronym{LOS}{LOS}{line of sight}
\newacronym{LSTM}{LSTM}{Long short-term memory}
\newacronym{MANET}{MANET}{MobileAd-hoc Network}
\newacronym{MEO}{MEO}{Medium-Earth Orbit}
\newacronym{MIMO}{MIMO}{multiple-input and multiple-output}
\newacronym{ML}{ML}{Machine Learning}
\newacronym{MMSE}{MMSE}{minimum mean square error}
\newacronym{MSE}{MSE}{mean square error}
\newacronym{NTP}{NTP}{Network Time Protocol}
\newacronym{OCB}{OCB}{Opportunistic Collaborative Beamforming}
\newacronym{OFDM}{OFDM}{Orthogonal Frequency Division Multiplexing}
\newacronym{OLFAR}{OLFAR}{Orbiting Low Frequency Antennas for Radio Astronomy}
\newacronym{OPLL}{OPLL}{Optical \acrshort{PLL}}
\newacronym{OTA}{OTA}{over-the-air}
\newacronym{PA}{PA}{Pairwise Algorithm}
\newacronym{PBS}{PBS}{Pairwise Broadcast Synchronization}
\newacronym{PLL}{PLL}{Phase-Locked Loop}
\newacronym{PoC}{PoC}{Proof of Concept}
\newacronym{POD}{POD}{precise orbit determination}
\newacronym{PRISMA}{PRISMA}{Prototype Research Instruments and Space Mission technology Advancement}
\newacronym{PRN}{PRN}{pseudo-random noise}
\newacronym{PTP}{PTP}{Precision Time Protocol}
\newacronym{RBS}{RBS}{reference broadcast synchronization}
\newacronym{RF}{RF}{radio frequency}
\newacronym{RMS}{RMS}{root mean square}
\newacronym{R-RT}{R-RT}{robust round-trip}
\newacronym{RSS}{RSS}{received signal strength}
\newacronym{SAR}{SAR}{synthetic aperture radar}
\newacronym{SDDB}{SDDB}{Successive Deterministic Distributed Beamforming}
\newacronym{SDR}{SDR}{Software-Defined Radio}
\newacronym{SIN}{SIN}{Space Information Network}
\newacronym{SNR}{SNR}{Signal-to-Noise Ratio}
\newacronym{SSP}{SSP}{Stale Synchronous Parallel}
\newacronym{TDD}{TDD}{Time Division Duplex}
\newacronym{TOA}{TOA}{time of arrival}
\newacronym{TOF}{TOF}{time of flight}
\newacronym{TPSN}{TPSN}{Timing Synchronization Protocol for Sensor Networks}
\newacronym{T-RT}{T-RT}{time-slotted round trip}
\newacronym{TWTT}{TWTT}{two-way time transfer}
\newacronym{TWR}{TWR}{Two-Way Ranging}
\newacronym{UE}{UE}{User Equipment}
\newacronym{USRP}{USRP}{Universal Software Radio Peripherals}
\newacronym{UAV}{UAV}{Unmanned Aerial Vehicles}
\newacronym{UWB}{UWB}{ultrawideband}
\newacronym{WSN}{WSN}{Wireless Sensor Network}
\newacronym{0F}{0F}{zero feedback}
\newacronym{1BF}{1BF}{One-bit Feedback}
\newacronym{2BF}{2BF}{Two-bits Feedback}
\newacronym{2WS}{2WS}{Two-way Synchronization}
\newacronym{3D}{3D}{3-Dimensional}

\MakeRobustCommand\acrshort
\MakeRobustCommand\cite

\begin{document}
\history{Date of publication xxxx 00, 0000, date of current version xxxx 00, 0000.}
\doi{10.1109/ACCESS.2017.DOI}

\title{Architectures and Synchronization Techniques for Distributed Satellite Systems: A Survey}
	\author{\uppercase{Liz Martinez Marrero}\authorrefmark{1}, \IEEEmembership{Student Member, IEEE},
	\uppercase{Juan C. Merlano Duncan}\authorrefmark{2}, \IEEEmembership{Senior Member, IEEE},
	\uppercase{Jorge~Querol}\authorrefmark{3}, \IEEEmembership{Member, IEEE},
	\uppercase{Sumit Kumar}\authorrefmark{4}, \IEEEmembership{Member, IEEE},
	\uppercase{Jevgenij~ Krivochiza}\authorrefmark{5}, \IEEEmembership{Student Member, IEEE},
	\uppercase{Shree~Krishna~Sharma}\authorrefmark{6}, \IEEEmembership{Senior Member, IEEE}, 
	\uppercase{Symeon~Chatzinotas}\authorrefmark{7}, \IEEEmembership{Senior Member, IEEE},
	\uppercase{Adriano Camps}\authorrefmark{8,9}, \IEEEmembership{Fellow Member, IEEE},
	 and \uppercase{Bj\"orn Ottersten}\authorrefmark{10}, \IEEEmembership{Fellow Member, IEEE}}
\address[1-7,10]{Interdisciplinary Centre for Security, Reliability and Trust (SnT), University of Luxembourg, 1855, Luxembourg City, Luxembourg}
\address[8]{Universitat Polit\`ecnica de Catalunya, Barcelona, Spain}
\address[9]{Institut d'Estudis Espacials de Catalunya, Barcelona, Spain}

\tfootnote{This research was funded in whole by the Luxembourg National Research Fund (FNR), under the CORE project COHESAT: Cognitive Cohesive Networks of Distributed Units for Active and Passive Space Applications, grant reference [FNR11689919]. For the purpose of open access, the author has applied a Creative Commons Attribution 4.0 International (CCBY4.0) license to any Author Accepted Manuscript version arising from this submission.}

\markboth
{Marrero \headeretal: Architectures and Synchronization Techniques for Distributed Satellite Systems: A Survey}
{Marrero \headeretal: Architectures and Synchronization Techniques for Distributed Satellite Systems: A Survey}

\corresp{Corresponding author: Liz Martinez Marrero (e-mail: liz.martinez-marrero@ uni.lu).}

\begin{abstract}
\acrshort{CDSS} is a key enabling technology for the future of remote sensing and communication missions. However, they have to meet strict synchronization requirements before their use is generalized. When clock or local oscillator signals are generated locally at each of the distributed nodes, achieving exact synchronization in absolute phase, frequency, and time is a complex problem. In addition, satellite systems have significant resource constraints, especially for small satellites, which are envisioned to be part of the future \acrshort{CDSS}. Thus, the development of precise, robust, and resource-efficient synchronization techniques is essential for the advancement of future \acrshort{CDSS}. In this context, this survey aims to summarize and categorize the most relevant results on synchronization techniques for \acrshort{DSS}. First, some important architecture and system concepts are defined. Then, the synchronization methods reported in the literature are reviewed and categorized. This article also provides an extensive list of applications and examples of synchronization techniques for \acrshort{DSS} in addition to the most significant advances in other operations closely related to synchronization, such as inter-satellite ranging and relative position. The survey also provides a discussion on emerging data-driven synchronization techniques based on \acrshort{ML}. Finally, a compilation of current research activities and potential research topics is proposed, identifying problems and open challenges that can be useful for researchers in the field.
\end{abstract}


\titlepgskip=-15pt

\maketitle

\section{Introduction}
\label{section::Intro}

\PARstart{D}{istributed} Satellite Systems is a very promising architecture for future remote sensing and communication missions \cite{Graziano2013}\cite{Radhakrishnan2016}. The term \acrshort{DSS} refers to satellite systems with two or more spacecraft communicating to accomplish the mission goal. Nowadays, most missions are \acrshort{DSS}, in part influenced by the space industry's paradigm shift toward smaller and cheaper satellites \cite{Radhakrishnan2016}. Its potential increases with the use of signal coherent processing methods from different platforms, especially coherent transmission and reception, which implies that the system behaves as a single unit with collocated transceivers, also known as \acrshort{DBF}. This entails the improvement of several \acrshort{DSS} applications such as Earth observation, geolocation, navigation, imaging, communications, among others. However, it requires strict time, frequency, and phase synchronization among the distributed nodes.

For example, deep space communications could be accomplished by \acrshort{DSS} applying distributed \acrshort{MIMO} techniques \cite{Savazzi} if the terminals at each end of the link are synchronized at symbol level. This implies that the stations need to maintain synchronized clocks with sub-nanosecond accuracy to guarantee a bandwidth of a few hundred MHz \cite{Barton2014}. Another example can be Earth observation applications based on distributed \acrshort{SAR} interferometry. In this case, the final resolution depends on the accuracy of the phase synchronization achieved between the satellites \cite{Chen2020}. For some applications, the synchronization requirement is the critical limiting factor in making practical implementations possible \cite{Merlano-Duncan2021}. In those cases, the level of effort, power consumption, and complexity required for the synchronization algorithm may be higher than those needed for the mission itself.

Achieving exact phase, frequency, and time synchronization is a complex problem when the clock reference signals are generated locally at each distributed node. It is significantly more challenging when the distance between the distributed nodes is much larger than the signal wavelength, and especially when this distance varies over time due to the relative movement of the nodes, as it typically happens in \acrshort{DSS} \cite{Gun2009}. Furthermore, the electrical distance, determined by the inter-satellite channels and the \acrshort{RF} chains at each spacecraft, is not constant either. In these cases, it is not possible to rely on network backhaul links or to use the exact location of the collaborating nodes as can be done in some terrestrial networks \cite{Sundaramoorthy2013}\cite{Sundaramoorthy2016}.

The synchronization of distributed wireless systems is highly challenging and has received plenty of attention in the literature. For example, \cite{Nasir2016} summarizes some of the timing and carrier synchronization techniques proposed for wireless communication systems. Another survey on the synchronization in wireless systems is given in \cite{SundararamanClockSurvey}, where the authors summarize the clock synchronization protocols in \acrshort{WSN}. However, both articles are limited to terrestrial systems, and they do not address synchronization for \acrshort{DSS}. 

On the other hand, the opportunities and challenges of \acrshort{DSS} have been explored in many publications. For example, \cite{Newman2008} summarizes the use of \acrshort{WSN} for planetary exploration. In this article, the authors mentioned synchronization as one of the problems encountered in distributed systems design. Besides, \cite{Selva2017} provided a comprehensive assessment of modern concepts and technologies of \acrshort{DSS} and analyzed the technical barriers to \acrshort{DSS} implementation. In addition, \cite{Kodheli2020} stated the revolutionary strength of \acrshort{DSS} such as satellite swarms and the limitations in its implementation due to synchronization requirements.

However, the synchronization of \acrshort{DSS} is still an open and challenging research question that has attracted the attention of the scientific community in the last few years. This has resulted in a large body of work appearing in conferences and journals from different fields. To the best of authors' knowledge, a comprehensive survey document to cover topics on synchronization applied to \acrshort{DSS} is still missing. This article intends to fill this gap. With this idea in mind, this paper summarizes and categorizes the most relevant publications on synchronization techniques for \acrshort{DSS}. The result is a comprehensive survey that can be used as a guide for researchers and developers working in this field. 

\subsection{Contributions of This Paper}

This survey describes the most relevant results on synchronization for \acrshort{DSS} and the strict requirements and new technologies related to distributed satellite missions. For the sake of clarity, the main contributions of this paper can be summarized as:

\begin{itemize}
	\item A brief survey of the \acrshort{DSS} architectures is provided, classifying them into five general groups: Constellations, Clusters, Swarms, Fractionated and Federated spacecraft, which main features are identified.
	\item The distributed time, phase, and frequency wireless synchronization methods reported in the literature are summarized and compared, analyzing their feasibility for \acrshort{DSS}.
	\item Other operations, closely related to synchronization in \acrshort{DSS} such as inter-satellite ranging and relative positioning are also analyzed.
	\item It is offered an extensive compilation of the missions and \acrshort{PoC} implementations reported up to the present. 
	\item Some of the most relevant current research activities and potential research topics are presented, identifying problems and open challenges.
\end{itemize}

\subsection{Paper Organization}

The paper is organized as follows: After the Introduction, some important concepts on the architectures and system models for \acrshort{DSS} are defined in Section \ref{section::Architectures}. The synchronization methods reported in the literature are related and categorized in Section \ref{section::Synchronization}. Section \ref{section::Ranging} deals with other operations closely related to the synchronization in \acrshort{DSS} such as: inter-satellite ranging and relative position. Sections \ref{section::App} and \ref{section::Examples} comment examples of synchronization methods in \acrshort{DSS}, whereas Section \ref{section::App} lists some ideas and new methods that have not been implemented yet. Section \ref{section::Examples} provides to the reader examples of \acrshort{PoC} and launched missions of \acrshort{DSS} with special emphasis on the synchronization methods implemented. Section \ref{section::ML} presents the use of \acrshort{ML} for synchronization purposes. The most relevant open questions and future research directions are highlighted in Section \ref{section::Challenges}. Finally, the reader can find the general conclusions in Section \ref{section::Concl}. The list of acronyms is provided in Section \ref{section::acronyms} to make the reading more accessible.

\subsection{Acronyms}
\label{section::acronyms}

\printglossary[type=\acronymtype]

\section{Architectures and System Model for Distributed Satellite Systems}
\label{section::Architectures}

\subsection{Overview of Existing Architectures}
\label{section::arch_class}
\acrshort{DSS} can be classified into five general groups: Constellations, Clusters, Swarms, Fractionated, and Federated spacecraft. Table \ref{tab::summary_architectures} summarizes the main characteristics of these groups.

\begin{itemize}
    \item \textbf{Constellations} refer to a traditional approach where tens of medium to large satellites (over 500 kg each) are distributed around Earth to guarantee global or regional coverage of a service. Some of the most famous satellite constellations are: the \acrshort{GNSS} constellations, such as \acrshort{GPS} and Galileo and; the satellite communication constellations, such as Globalstar \cite{wwwGlobalstar}, Iridium \cite{Maine1995}, and OneWeb \cite{wwwOneWeb}. Inter-satellite communication in these networks is scarce or null, except in the Iridium constellation where each satellite can have four Ka-band \acrshort{ISL} \cite{Maine1995}.
    
    \item A \textbf{cluster} is a group of at least two mini or micro satellites (from 10 to 500 kg each) deliberately positioned closely together to enhance or create new system capabilities. These \acrshort{DSS} cover a smaller portion of the Earth and mainly require inter-satellite communications to keep a close flying formation. Some satellite clusters are \acrshort{GRACE} \cite{Tapley2004}, \acrshort{LISA} \cite{Racca2010}, \acrshort{PRISMA} \cite{wwwPrisma}, PROBA-3 \cite{wwwProba3}, and TanDEM-X \cite{Maurer2016}. 
    
    \item Satellite \textbf{swarms} are similar to clusters, except they contain a significantly higher number of satellites, often smaller and less expensive (less than 10 kg each). They are envisioned to contain hundreds and even thousands of nanosatellites operating together in a loose flying formation. They will require inter-satellite communications, as each member determines and controls relative positions to the others. Unlike previous \acrshort{DSS} that have several examples, satellite swarms are a new concept yet to be demonstrated. Examples of satellite swarms projects are QB-50 \cite{wwwQB50} and \acrshort{OLFAR} \cite{Engelen2010}. 
   
    \item \textbf{Fractionated spacecraft} is a revolutionary satellite architecture that distributes the functions of a single large satellite into numerous modules that communicate by \acrshort{ISL} in a highly dynamic topology \cite{Brown2006a}. Among the multiple advantages of this concept can be mentioned the flexibility,  robustness, and the significant decrease in the required time to launch and deploy a satellite mission. Fractionated satellite systems is a very recent concept that have not yet been implemented.
    
    \item Finally, the \textbf{Federated Satellite Systems (\acrshort{FSS})} paradigm visualizes opportunistic collaboration among fully independent and heterogeneous spacecraft \cite{Ruiz-De-Azua2020}. This is one of the most recent \acrshort{DSS} concepts and is inspired by the current peer-to-peer networks and cloud computing. The main idea is to benefit from the potential of under-utilized space commodities by trading and sharing previously unused resources available in space assets at any given time. It is worth noting that much of \acrshort{FSS}'s potential relies on the spacecraft capabilities to establish communications through \acrshort{ISL}. The recently launched FSSCat mission \cite{Camps2021} is an example of \acrshort{FSS}.
\end{itemize}

\begin{table*}[]
\centering
\caption{Main characteristics of the \acrshort{DSS} groups}
\resizebox{\textwidth}{!}{%
\begin{tabular}{|c|c|c|c|c|}
\hline
\textbf{Classification} & \textbf{Number of Satellites} & \textbf{Satellites Weight} & \textbf{Inter-satellite Comm} & \textbf{Examples} \\
\hline
Constellation & tens & $> 500$ kg   & scarce or null    & \acrshort{GNSS}, Globalstar, Iridium \\ \hline
Cluster & at least 2 satellites & 10 to 500 kg   & \begin{tabular}[c]{@{}c@{}}required to keep a \\ close flying formation\end{tabular} & \begin{tabular}[c]{@{}c@{}}\acrshort{GRACE}, \acrshort{LISA}, \acrshort{PRISMA}, \\ PROBA-3, TanDEM-X\end{tabular} \\ \hline
Swarm & \begin{tabular}[c]{@{}c@{}}hundreds in a loose\\flying formation\end{tabular}  & $< 10$ kg   & \begin{tabular}[c]{@{}c@{}}required to determine and \\ control relative position\end{tabular} & QB-50, OLFAR \\ \hline
\begin{tabular}[c]{@{}c@{}}Fractionated \\ spacecraft\end{tabular} & at least 2 satellites & $< 500$ kg   & \begin{tabular}[c]{@{}c@{}}required to achieve\\the mission objectives\end{tabular} & not implemented yet \\ \hline
\acrshort{FSS} & dynamic & any & \begin{tabular}[c]{@{}c@{}}required to share\\unused resources\end{tabular} & FSSCat \\ \hline
\end{tabular}%
}
\label{tab::summary_architectures}
\end{table*}

In addition to this classification, \acrshort{CDSS} can be categorized according to their synchronization scheme as centralized or distributed. The former refers to distributed systems where all the nodes adjust their carriers to follow one controller node, which has the most stable oscillator in the system. This synchronization scheme has a relatively simple implementation, but it may have robustness drawbacks. On the other hand, distributed synchronization satellite systems do not rely on a single node but try to find a common carrier considering all the oscillators in the system. This characteristic overcomes the robustness drawback of the former group but makes the synchronization algorithms more complex. The synchronization methods suitable for \acrshort{CDSS} will be addressed in detail in the following sections.

Moreover, considering the communication links between the nodes of the \acrshort{DSS}, the space segment can be classified as:  
\begin{itemize}
    \item \textbf{Ring}, in which each node connects to two other nodes, forming a single continuous path through all the elements of the \acrshort{DSS}. This topology is not suitable for centralized synchronization algorithms.  
    
    \item \textbf{Star}, in which each node connects to a central node that performs specific tasks ranging from communications with the ground segment to leading the synchronization. This topology generally uses centralized synchronization methods. 
    
    \item \textbf{Mesh}, in which each node is connected to every other node in the distributed system. This topology can be implemented fully or partially depending on the complexity of the \acrshort{DSS}, and it accepts the implementation of both distributed and centralized synchronization algorithms.  
    \item \textbf{Hybrid} topologies combine two or more of the previous ones.
\end{itemize}

\begin{figure}[!t]
\centering
\includegraphics[width = \columnwidth]{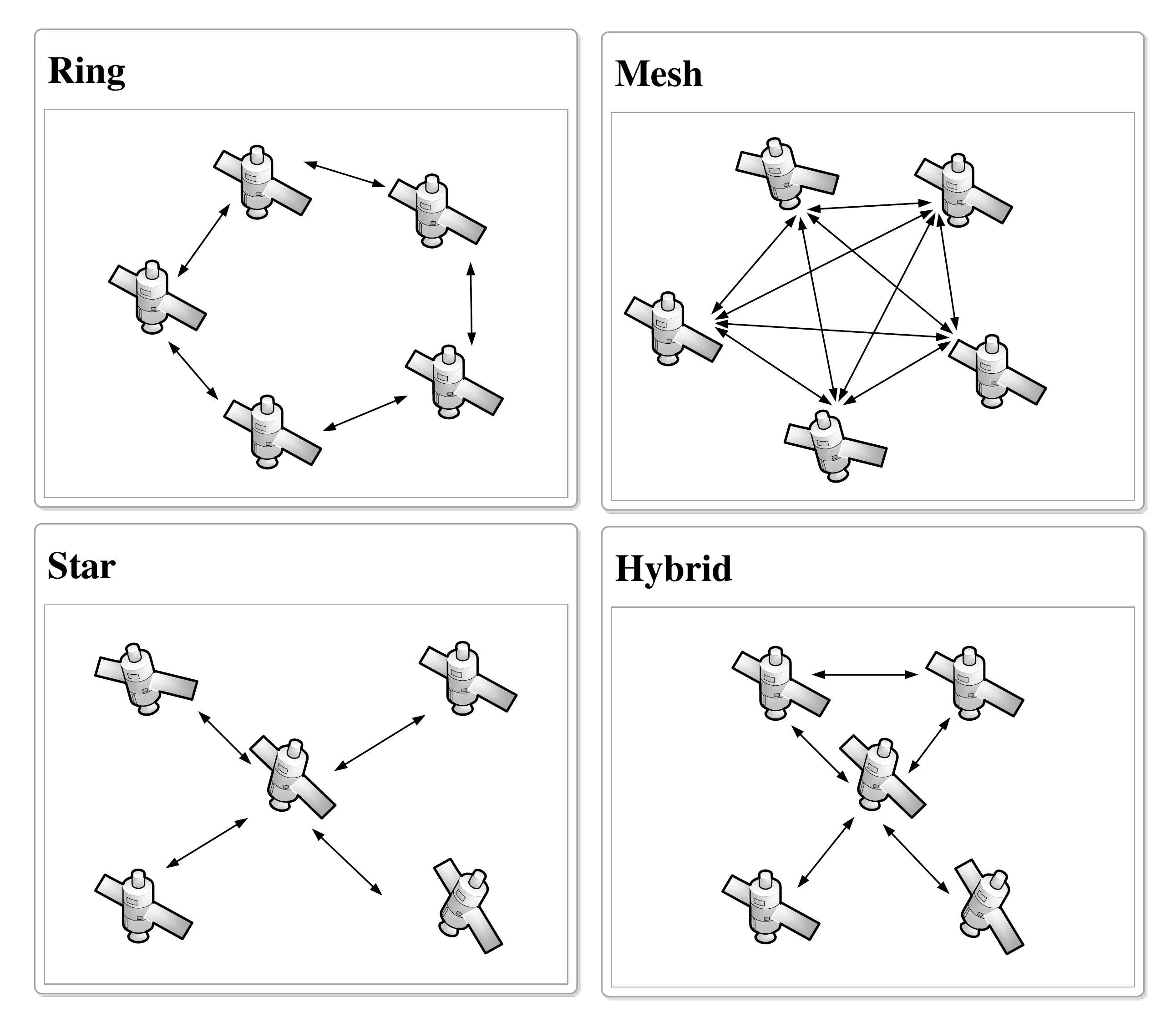}
\caption{Classification of \acrshort{DSS} considering the \acrshort{ISL}}
\label{fig::arch_class}
\end{figure}

These topologies are shown in Fig. \ref{fig::arch_class}, where the arrows represent the \acrshort{ISL}.

\subsection{Generalized System Model}
\label{section::Generalized system model}

The general \acrshort{DSS} considered in this article is a distributed array of autonomous nodes which collaborate to perform distributed beamforming towards an intended target node outside the array. From a general perspective, the nodes of the distributed arrays considered in this survey can be moving while the whole system tries to stay at a fixed position, or the entire array can be following a trajectory or orbit.

For both cases, the nodes require transmission and reception capabilities to synchronize the distributed system. For the general \acrshort{DSS}, no specific geometric distribution of the nodes is assumed, none of the previously mentioned classes neither. However, each of them can be described by a state variable $\overrightarrow{x_n}  = (\overrightarrow{p_n},\overrightarrow{v_n})$ where $\overrightarrow{p_n}$ and $\overrightarrow{v_n} \triangleq \dv{\overrightarrow{p_n}}{t}$ represent position and speed of the node $n$, respectively. 

\begin{figure}[!t]
\centering
\includegraphics[width = \columnwidth]{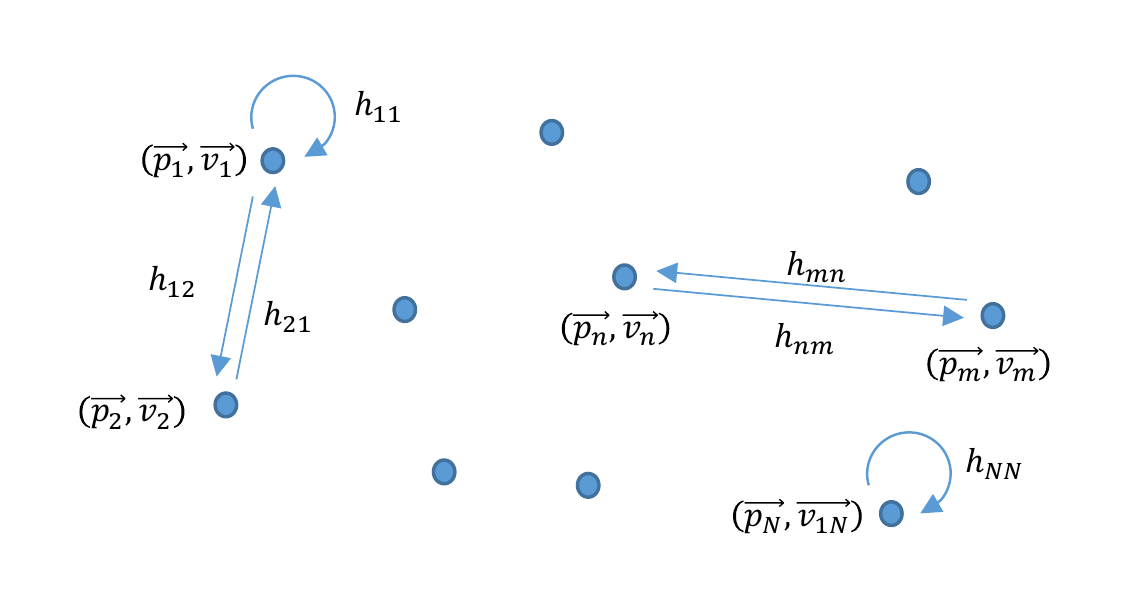}
\caption{General scheme of the \acrshort{DSS} performing the required synchronization for a \acrshort{DBF} transmission}
\label{fig::System model}
\end{figure}

Besides, it is assumed the existence of \acrshort{RF}-\acrshort{ISL} between all the elements of the \acrshort{DSS}, which implies a Mesh configuration, even if all the links are not activated at the same time. The complex transfer function of these \acrshort{ISL}s is formally represented as complex coefficients $h_{nm}$, where $n$ and $m$ are the node subscripts. The matrix $\mathbf{H}$, containing all the $h_{nm}$, can be used to describe the \acrshort{DSS}. It determines the most suitable synchronization procedure. Fig. \ref{fig::System model} shows the general \acrshort{DSS} considered in the next sections to analyze the synchronization techniques.

The quality of the \acrshort{ISL} is considered variable as a function of the position and orientation of the nodes, given that omnidirectional antennas are not assumed. Besides, the frequency responses of the \acrshort{RF} chains are considered different for each node. 

For synchronization purposes, each of the distributed nodes generates its own initial reference signal, which stability depends on the available hardware at each satellite. The objective of the synchronization algorithm is to make all of those initial reference signals converge to a common reference with the best possible accuracy.

\subsection{Summary and Lessons Learnt}
\begin{itemize}
    \item \acrshort{DSS} can be classified as Constellations, Clusters, Satellite swarms, Fractionated spacecraft, and, \acrshort{FSS}.
    \item Distributed synchronization algorithms are more robust than centralized synchronization algorithms, but more complex.
    \item According to the \acrshort{ISL}, there are four types of \acrshort{DSS}: Ring, Star, Mesh, or Hybrid topology.
    \item The general \acrshort{DSS} model considered in this article is a distributed array of autonomous spacecraft which collaborate to perform distributed beamforming towards an intended target node outside the array.
    \item The general model assumes a Mesh configuration with the reference signals locally generated at each distributed satellite.
\end{itemize}

\section{Overview of Synchronization Methods}
\label{section::Synchronization}

A critical aspect of the synchronization of distributed radio systems, in general, is the clock or time synchronization in addition to the phase synchronization. This section summarizes the most significant synchronization methods reported in the state of art. These algorithms can be classified as Closed-loop or Open-loop methods based on the use of feedback from a node external to the \acrshort{DSS}. The external node can be another satellite, an anchor point, or the intended communication target. The Closed-loop methods require a communication channel to transmit the feedback information between the external node and the \acrshort{DSS}. Whereas in Open-loop, the synchronization is achieved without the participation of any node other than the distributed satellites. 

\begin{figure}[!t]
\centering
\includegraphics[width = \columnwidth]{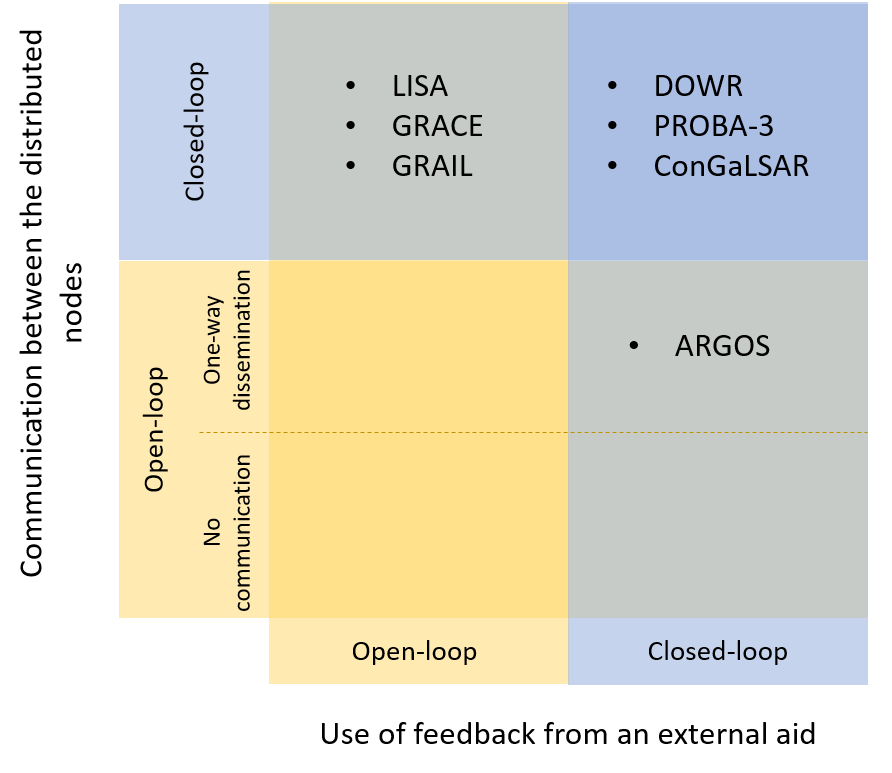}
\caption{Classification of the synchronization methods}
\label{fig::synchronization_classification}
\end{figure}

Another way to classify the synchronization methods considers the communication between the elements of the \acrshort{DSS}. In order to achieve synchronization, some algorithms require the exchange of information among the distributed satellites. This can be done as a two-way message exchange or closed loops, which requires a duplex channel between the nodes of the \acrshort{DSS}, or as a broadcast or one-way communication. Another option is to synchronize without any communication among the elements of the \acrshort{DSS}. In this case, it is possible to achieve coherence using the feedback from a node out of the \acrshort{DSS}. Both classifications can be superimposed, as it is represented in Fig. \ref{fig::synchronization_classification}. In this figure, some of the missions analyzed in the following sections were included as examples. 

\subsection{Time Synchronization}
\label{section::time_synch}

Time synchronization is critical in the application of many \acrshort{DSS}. The coherent transmission in communication applications, as an instance, requires aligned signals at the symbol level to achieve the full potential of the beamforming gains. The tight clock and timing synchronization is achieved with different levels of accuracy in other areas, such as wired and optical networks and wireless sensor networks. For example, some distributed applications such as computer networking, distributed signal processing, instrumentation, and earth observation applications require accurate timing or clock synchronization. The purpose of this section is to explain those timing synchronization methods used in other areas and give advice on how to translate them into DSS, in particular for communications applications.

Previous studies have identified the time or clock synchronization challenge for clock frequencies in the order of tens to thousands of MHz and accuracies ranging between orders of one fractional digit relative to the clock frequency. Recently, it has been studied for multi-static and \acrshort{MIMO} radar and distributed beamforming applications where the required precision is orders of magnitude more stringent \cite{Prager2020}. Additionally, these systems require what is known as absolute time synchronization, which is different from relative time synchronization on which the timing of an impingent signal is tracked. The problem of absolute time synchronization was first formally defined by Poincar\'e and Einstein in 1898 \cite{Poincare1898}, and 1905 \cite{Einsten1905} respectively. The formal definition of this problem and the \acrshort{TWTT} concept was provided in the framework of relativistic-event-simultaneity.
 
\begin{figure}[!t]
\centering
\includegraphics[width = \columnwidth]{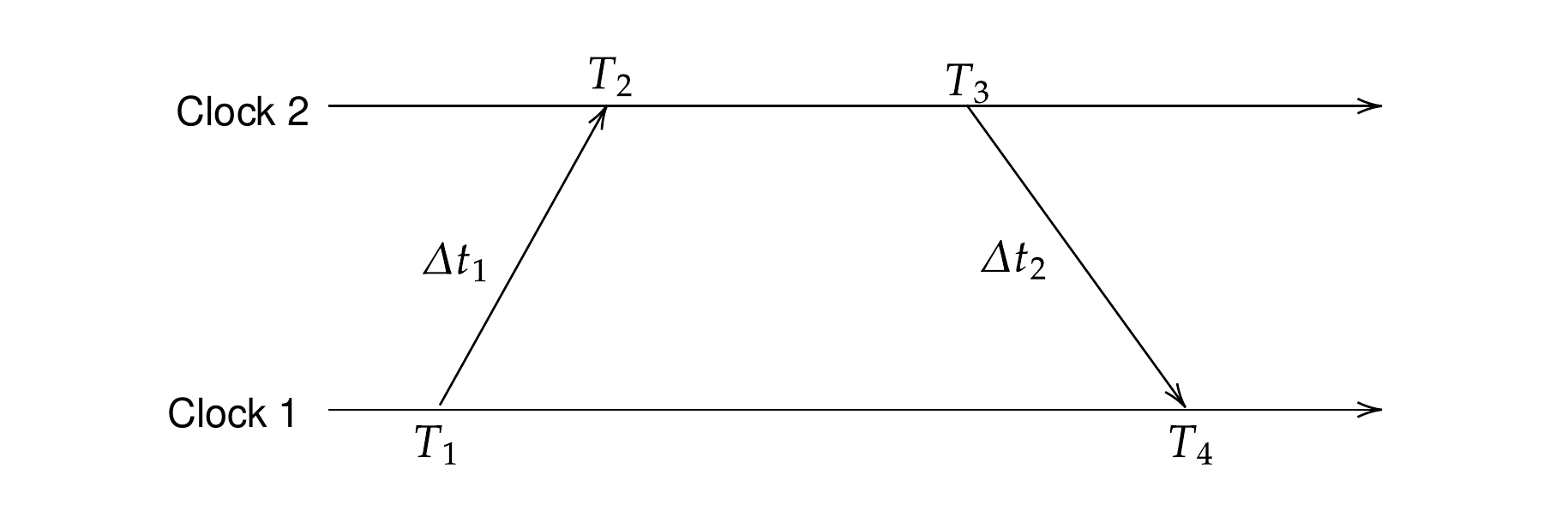}
\caption{Operating principle of the \acrshort{TWTT} clock synchronization \cite{Prager2020}}
\label{fig::Time synchronization diagram}
\end{figure}

Fig. \ref{fig::Time synchronization diagram} depicts the basic idea of the \acrshort{TWTT} concept. Here, an initiating or source node sends a signal (or packet) at time $T_1$. A slave (or follower node) receives the signal at time $T_2$ after a delay of $\Delta t_1 = T_2 - T_1$ and responds (or reflects) after a known delay at time $T_3$. The source node receives the response signal at time $T_4$. The time offset of the clocks is then $((T_2-T_1)-(T_4-T_3))/2 = (\Delta t_1- \Delta t_2)/2$, and the propagation delay is $((T_2-T_1)+(T_4-T_3))/2 = (\Delta t_1+\Delta t_2)/2$.  Therefore, proper knowledge of this propagation time offset will be used to achieve absolute synchronization. 

The \acrshort{TWTT} concept created the foundations for clock synchronization of twenty-century networks and complex systems, such as satellites and the Internet \cite{Kirchner1991}. This general concept can be applied to diverse kinds of systems and networks, including \acrshort{DSS}. This section will emphasize the case of \acrshort{DSS}. The following subsections describe timing synchronization methods that have been developed in different communications areas and could be extrapolated to \acrshort{DSS}. 
An example of the use of \acrshort{TWTT} in \acrshort{DSS} is presented in \cite{Shengkang2013}, where the authors compared the performance of three clocks offset prediction algorithms based on this method for a master-slave architecture. In addition, the authors in \cite{Gun2009} used \acrshort{TWTT} to synchronize four spacecraft in a distributed satellite formation flying, achieving time synchronization simulation errors smaller than $\pm$~10~ns. The work in \cite{Xiaochun2010} analyzed the effect of the motion of the satellite on the two-way time synchronization accuracy. Another example based on \acrshort{TWTT} calculates the time difference between two satellite clocks by measuring the phase difference of a pseudo-random noise code in a master-slave architecture \cite{Jiuling2018}. The method is proposed to achieve on-orbit synchronization in a micro-satellite cluster \cite{Jiuling2018}. Other authors combined \acrshort{RF} carrier ranging methods with \acrshort{TWTT} to obtain inter-satellite range and time synchronization simultaneously \cite{Pan2008}\cite{Ma2013}\cite{Yaowei2021}.

\subsubsection{Time synchronization in wired networks}  

The \acrshort{TWTT} concept is the basis of most synchronizations protocols in the literature. The \acrfull{NTP} and \acrfull{PTP} are extensively used in large-scale modern computer networks, and both operate with a \acrshort{TWTT} approach. The network nodes exchange timestamps employing (UDP) packets to measure the round-trip propagation latency. \acrshort{NTP} generates software timestamps with non-deterministic time offsets and achieves clock accuracies in the order of $\sim 10~\mu$s \cite{Mills2006}. The \acrshort{PTP} (IEEE-1588) is an evolution of the \acrshort{NTP}, which generates hardware timestamps using the waveforms of the associated clock. This protocol achieves clock alignment by exchanging synchronization packets among the involved nodes. The IEEE-1588 standard limits the maximum rate for the timestamp counters to 125~MHz, providing accuracies of $\sim 10$~ns \cite{IEEE_Standard}\cite{Shi2015}. The performance of these systems, such as \acrshort{NTP} and \acrshort{PTP}, rely directly on the clock rates used, and the achieved accuracy is given by the granularity associated with an integer counter.

Further refinements of the time synchronization mechanism use the clock phase information. A salient example is the White Rabbit project of the European Organization for Nuclear Research (CERN). White Rabbit operates with Ethernet frames to detect the phase difference between the local clock and a clock extracted from the received Ethernet signal. Across large wired networks, White Rabbit operates under specially designed network switches to measure and compensate fractional clock phase differences achieving $\sim 10$~ps accuracy \cite{Dierikx2016}. 

Even though these time synchronization mechanisms can not be used directly on \acrshort{DSS}, the main concepts can be extrapolated to \acrshort{DSS} scenarios. In the \acrshort{DSS}, the coarse synchronization methods such as \acrshort{NTP} and \acrshort{PTP} can be used to eliminate the ambiguity in the total channel delay. Then, the time synchronization can be refined with differential time mechanisms at the waveform level, using White Rabbit or similar approaches.

\subsubsection{Time synchronization in wireless sensor networks}

In wireless communications, absolute time synchronization is also frequently desired, and the wireless channel is typically used to exchange the synchronization messages. The algorithms for absolute time synchronization in wireless networks can be classified as sender-to-receiver and receiver-to-receiver synchronization methods. The former is based on the \acrshort{TWTT} between couples of nodes, whereas receiver-to-receiver methods use time readings of a standard signal broadcasted to a set of nodes from a common sender \cite{Noh2007}. 

Some receiver-to-receiver solutions include \acrshort{RBS} \cite{10.1145/844128.844143} and \acrshort{PBS} \cite{Noh2008}. The \acrshort{RBS} protocol implements a \acrshort{TOA} exchange between the distributed nodes disregarding the signal \acrshort{TOF} over the physical medium \cite{10.1145/844128.844143}. \acrshort{PBS} is a well-known timing synchronization scheme for \acrshort{WSN}, which is based on sets of node-pairs for network-wide synchronization. \acrshort{PBS} operates under the assumption that all the participating nodes will receive and detect the pairwise synchronization frames exchanged between two master/reference nodes. This approach assumes a hierarchically distributed structure \cite{Noh2008} and assumes that distances between nodes and their associated delays (\acrshort{TOF}) are identified in advance. 

The most well-known sender-to-receiver synchronization methods are \acrshort{TPSN} \cite{Rucksana2015} and \acrshort{FTSP} \cite{Maroti2004}. \acrshort{TPSN} implements \acrshort{TWTT} between pairs of nodes preceded by a discovery phase from where each node obtains a level. In \acrshort{FTSP} and its variations \cite{Shi2020}, the distributed nodes synchronize to a signal broadcasted from the root node or a previously synchronized node. 

For these protocols to work, synchronization must be performed several times. Additionally, the nature of the \acrshort{WSN}, in which the network observes a physical phenomenon (temperature, pressure, etc.), determines synchronization requirements in the order of microseconds. However, this level of accuracy is inadequate to perform distributed coherent (beamforming) radio applications.    

For example, the required time synchronization to achieve beamforming maintaining the performance at acceptable levels is around $\pm$~7.5 percent of a symbol duration. For single-carrier communication baud rates of a hundred MHz, this represents a required accuracy of $\pm$~0.75~ns. This accuracy could be achieved with a refinement of the methods mentioned above, such as the work in \cite{Roehr2007}. This article proposes a step forward into the timing accuracy increase by using frequency-modulated continuous-wave (FMCW) signals with relatively high bandwidth of 150-MHz. The method performs the synchronization between two stations using a \acrshort{TWTT} approach (similar to the one proposed in \cite{Fiedler2008TheOverview}). It uses the aforementioned radar-like waveform to provide a joint carrier-phase and timing synchronization with an accuracy of 66~ps.  

\subsubsection{Ultrawideband Pulse Synchronization} 

Wireless synchronization approaches using \acrshort{UWB} pulses instead of exchanges of a network packet have recently caught researchers' interest. To estimate the \acrshort{TOA} at sub-nanosecond levels, \acrshort{UWB} approaches take advantage of high-speed hardware, generally at sampling rates higher than 1~GHz. 

Several applications have exploited \acrshort{UWB} signaling using high speeds clocks and \acrshort{ADC}. Some examples are the sets of multiple active receivers locked, and synchronous to a single transmitter \cite{Zachariah2017}, distributed consensus techniques \cite{Segura2015}, and distributed sensor positioning \cite{Denis2006}. The works in \cite{Dongare2017} propose a propagation-aware \acrshort{TOF} protocol and provide validation for the system using an atomic clock integrated on a chip and a 64~GHz hardware clock timestamp counter \cite{Knappe2004}. As a result, these experiments achieved a distributed timing accuracy of 5~ns between two sensors. It is essential to mention that the transmission of \acrshort{UWB} pulses is not feasible for small satellites such as CubeSats due to power constraints. However, its advantages can be considered for \acrshort{DSS} with less strict power consumption requirements.   

Nevertheless, it is worth pointing out that increasing clock frequencies is not the only alternative to increase the timing accuracy in synchronization mechanisms. A common misconception in the literature regarding \acrshort{UWB} synchronization systems is that the \acrshort{ADC} frequency bounds the time resolution. As specified by the \acrshort{CRLB}  \cite{Exel2012}, accurate \acrshort{TOA} measurements within minuscule fractions of a sampling period is attainable, specially in \acrshort{LOS} scenarios.  The measurement of these time offsets, fractional to the sampling time, can be achieved by time offset mechanisms, also known as timing-error-detectors, such as the Gardner method \cite{1986Gardner}, and the Early-Late-Gate method \cite{ELG_Schmith} among others.

\subsection{Frequency and Phase Synchronization}
\label{section::phase_freq_synch}

As previously stated, considering the use of feedback from an external node, frequency, and phase synchronization methods can be classified as: 
\begin{itemize}
    \item Closed-loop methods, where the feedback from the target nodes could be either a single bit or a few bits or could have the form of rich feedback with limited or full \acrshort{CSI} \cite{Jayaprakasam2017}.
    \item Open-loop methods that require either intra-node communication or blind beamforming, also known as \acrshort{0F}  \cite{Jayaprakasam2017}.
\end{itemize}
In the following sections, some of the most notable examples from each category are discussed. 

\subsubsection{Closed-Loop Synchronization Methods} 

\begin{figure}
\centering
\begin{tikzpicture} [edge from parent fork down, sibling distance=40mm, level distance=15mm]
% root of the the initial tree, level 1
\node[root] {Closed-loop Methods}
% The first level, as children of the initial tree
  child {node[level 2] (c1) {Iterative Bit Feedback}}
  child {node[level 2] (c2) {Rich Feedback}};

% The second level, relatively positioned nodes
\begin{scope}[every node/.style={level 3}]
\node [below = 0.25cm of c1, xshift=15pt] (c11) {One/Two-bit feedback \cite{Mudumbai2006,Thibault2010}};
\node [below = 0.25cm of c11] (c12) {Cluster based \cite{Pun2009,Lee2018}};
\node [below = 0.25cm of c12] (c13) {Adjustable perturbation size \cite{Song2012,Tseng2014,Xie2018}};
\node [below = 0.25cm of c13] (c14) {Deterministic perturbation \cite{Thibault2010,Thibault2013}};

\node [below = 0.25cm of c2, xshift=15pt] (c21) {Explicit Channel Feedback \cite{Jeevan2008,Tushar2012,Kumar2014}};
\node [below = 0.25cm of c21] (c22) {Aggregate Rich Feedback \cite{Tu2002,Brown2012,Brown2015a,Goguri2016a,Amor2019}};
\node [below = 0.25cm of c22] (c23) {Reciprocity-based \cite{Brown,Brown2008,Manosas-Caballu2011}};

\end{scope}

% lines from each level 1 node to every one of its "children"
\foreach \value in {1,2,3,4}
  \draw[->] (c1.180) |- (c1\value.west);
\foreach \value in {1,2,3}
  \draw[->] (c2.180) |- (c2\value.west);
\end{tikzpicture}
\caption{Classification of the Closed-loop synchronization methods}
\label{fig::Closed-loop_classification}
\end{figure}
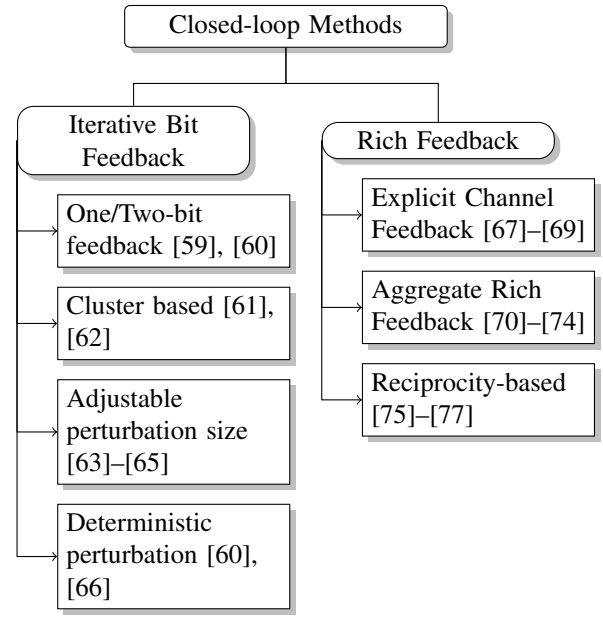

Fig. \ref{fig::Closed-loop_classification} lists the closed-loop synchronization methods found in the literature. They are classified into different groups according to the feedback type for a better understanding, although more detailed information is provided in this section.

The group Iterative Bit Feedback includes the algorithms where the distributed nodes modify their signals according to one or more decision bits received from the target node. Among them, the most well-known algorithm is the classical \acrshort{1BF}, proposed in \cite{Mudumbai2006}. This method considers the beamforming nodes synchronized in time and frequency and achieves phase synchronization by applying independent random phase rotations in each beamforming node. At each time slot, transmitter nodes add a random perturbation to their signals. The target node measures the \acrshort{RSS} and sends one bit indicating if the \acrshort{RSS} is better than the previous value. Depending on this bit, the transmitters keep or update the phase rotation. The process is repeated during the next time slot until each node's phase has been adjusted to its optimal value. The primary constraint of this algorithm is its convergence time, which was improved in \cite{Thibault2010} using \acrshort{2BF}.  

Other approaches improve convergence time by reducing the number of collaborative nodes. For example, \cite{Lee2018} proposed to separate the distributed nodes in clusters and perform the phase synchronization in two stages. The outcome of this algorithm with respect to \acrshort{1BF} is more evident as the number of nodes increases, when, for $N \approx 100$ nodes, it can decrease to half the required iterations. However, comparing both techniques only by the number of iterations is not fair since the epoch in \cite{Lee2018} is more complex and therefore takes a longer convergence time. Another iterative bit feedback method for phase synchronization named \acrshort{OCB} was proposed in \cite{Pun2009}. In this case, the destination node selects a subset of the distributed nodes whose transmitted signals produce the higher coherence gain at the destination.

Further enhancement of the \acrshort{1BF} is proposed in \cite{Tseng2014} by emulating the bacterial foraging techniques. In this algorithm which is called the \acrshort{BioRARSA}, the “swim” mechanism, along with the step size adjustment, enables the beamforming nodes to decrease convergence time as well as improve robustness against variation of initial conditions. A comprehensive explanation of this method and its performance can b found in \cite{Song2012}. Another synchronization method that increases the convergence speed by adjusting the perturbation size was presented in \cite{Xie2018}, where the authors proposed to exploit the cumulative positive feedback information additionally. 

The methods discussed above are based on random disturbances of the transmitted signal phase. However, \acrshort{D1BF} \cite{Thibault2010}, and its improved version, \acrshort{SDDB} \cite{Thibault2013} proposed to limit the possible phase perturbations to a set of discrete values from where choose the one that allows achieving the maximum \acrshort{RSS} possible. The number of elements in the perturbation set is proportional to the convergence time and determines the performance in terms of maximum achieved \acrshort{RSS}. Even though \acrshort{SDDB} shows steeper growth rates than \acrshort{D1BF}, both deterministic algorithms have limited performance due to the digitization of the perturbation set. Some combinations of deterministic and random methods were considered in \cite{Thibault2010}, where the hybrid methods obtained allow prioritizing the convergence time or the beamforming performance depending on the specific requirements of the network.

Iterative Bit Feedback algorithms are unsuited for \acrshort{DSS} due to their slow convergence characteristic. Generally, the long distance between the \acrshort{DSS} and the receiver implies a delay in the communication that, combined with the slow convergence of these methods, makes it not suitable to synchronize the system by Iterative Feedback algorithms.

The rich feedback methods use more information instead of just a few feedback bits to achieve synchronization. They can be classified in three categories according to the way the distributed nodes obtain the channel estimation:
\begin{itemize}
    \item Explicit Channel Feedback method, where each distributed node transmits a known sequence of training symbols to estimate the channel response. 
    \item Aggregate Rich Feedback methods, where the transmitters simultaneously send uncorrelated training sequences used to estimate each channel gain.
    \item Reciprocity-based methods, where the transmitters observe the uplink feedback signals sent by the target nodes and use reciprocity to estimate their downlink channel gains automatically.
\end{itemize}

These algorithms have proven to be more robust than the Iterative Bit Feedback methods at the price of considerable feedback overhead. For instance, the \acrshort{E1BF} \cite{Tushar2012}, which is an Explicit Channel Feedback algorithm, compensates for the effect of the average time-varying channel by combining explicit phase information with the \acrshort{1BF} method to provide better convergence and scalability as compared to \acrshort{1BF}. A comprehensive description of this algorithm can be found in \cite{Tushar2012}. Another Explicit Channel Feedback algorithm that outperforms \acrshort{1BF} for convergence time is the \acrshort{PA} for \acrshort{DTB} presented in \cite{Jeevan2008}, which is also more energetically efficient. 

Generally, nullforming requires far tighter synchronization than beamforming. For that reason, it mainly uses rich feedback approaches. For example, in \cite{Kumar2014}, a distributed gradient-descent algorithm is used to modify the signals' power and phase in consecutive time slots achieving beamforming and nullforming simultaneously. The method works in a time-scheduled way to estimate the \acrshort{CSI} and perform the collaborative transmission \cite{Kumar2014}.

In general, Aggregate Rich Feedback methods are more scalable than Explicit ones. For example, the algorithm proposed in \cite{Goguri2016a} is an Aggregate Rich Feedback method based in \acrshort{1BF} that allows each collaborating node to estimate its channel response to the receiver. Similarly, in the scheme proposed in \cite{Brown2012}, all the distributed nodes simultaneously transmit, and the receiver sends a phase compensation vector to achieve distributed transmit nullforming. However, in this method, the transmitter nodes use Code Division Multiple Access (CDMA) to facilitate signal separation at the receiver. The receiver uses an \acrshort{EKF} to generate state estimations for each transmitter, which implies some scalability limitations for this scheme. An improved version of the previous algorithm was proposed in \cite{Brown2015a}. 

The Reciprocity-based methods were first introduced as \acrshort{F-RT} synchronization method in \cite{Brown}. In this work, a distributed network of two nodes, each of them equipped with two \acrshort{PLL}, achieves frequency and phase synchronization by continuously transmitting three unmodulated beacons in a round-trip way. This strategy is effective in highly dynamic networks. However, in typical multipath channels, the frequency division duplexing intrinsic of \acrshort{F-RT} generates non-reciprocal phase shifts, which reduce the performance. To overcome the problem, in \cite{Brown2008} it is proposed a \acrshort{T-RT} algorithm, which is equivalent to the \acrshort{F-RT} but using the same frequency for all beacons. Frequency interference is avoided by time division duplexing. Even when this method gives the advantage of simultaneous frequency and phase synchronization, \acrshort{F-RT} and \acrshort{T-RT} schemes have as a drawback the extreme power consumption due to the extensive signaling. 

All the works listed in this section aim that all carriers arrive with the same initial phase at the destination based on perfect time alignment. Nevertheless, Mañosa-Caballu and Seco-Granados proposed a \acrshort{R-RT} synchronization protocol to achieve frequency, phase, and timing synchronization \cite{Manosas-Caballu2011}. This algorithm is based on the \acrshort{T-RT} method mentioned above. The robustness of the protocol allows it to work in dynamic environments in the sense that nodes can disappear without severely affecting the system's performance. Other approaches \cite{Nasir2016} addressed joint time delay and \acrshort{CFO} synchronization in closed-loop systems. For example, \cite{Amor2019} covers maximum likelihood synchronization in multi-node decode-and-forward cooperative relaying networks considering time-varying channels. In  \cite{Alvarez2018}, the same goal was approached through a weighted consensus algorithm to reach synchronization in a dense wireless network. 

\subsubsection{Open-loop Synchronization Methods} 

whereas rich feedback may give faster and better convergence than bit feedback, the signaling overhead is substantially more significant. The overhead implies latency problems in real-time applications, which can represent a considerable challenge for its implementation in cases such as satellite communications systems. In such scenarios where quick and reliable feedback from the target nodes is not possible, Open-loop synchronization methods are the recommended schemes.

\begin{figure}
\centering
\begin{tikzpicture} [edge from parent fork down, sibling distance=40mm, level distance=15mm]
% root of the the initial tree, level 1
\node[root] {Open-loop Methods}
% The first level, as children of the initial tree
  child {node[level 2] (c1) {Intra-node communication}}
  child {node[level 2] (c2) {Blind}};

% The second level, relatively positioned nodes
\begin{scope}[every node/.style={level 3}]
\node [below = 0.25cm of c1, xshift=15pt] (c11) {Master-slaves architectures \cite{Merlano-Duncan2021,Ochiai2005,Barriac2004}};
\node [below = 0.25cm of c11] (c12) {\acrshort{DCA} \cite{Rahman2012b}};
\node [below = 0.25cm of c12] (c13) {\acrshort{2WS} \cite{Preuss2011,Bidigare2015,NingXie2013a}};

\node [below = 0.25cm of c2, xshift=15pt] (c21) {\acrshort{0F} \cite{Bletsas2010,Bletsas2011}};

\end{scope}

% lines from each level 1 node to every one of its "children"
\foreach \value in {1,2,3}
  \draw[->] (c1.180) |- (c1\value.west);
\foreach \value in {1}
  \draw[->] (c2.180) |- (c2\value.west);
\end{tikzpicture}
\caption{Classification of the Open-loop synchronization methods}
\label{fig::Open-loop_classification}
\end{figure}
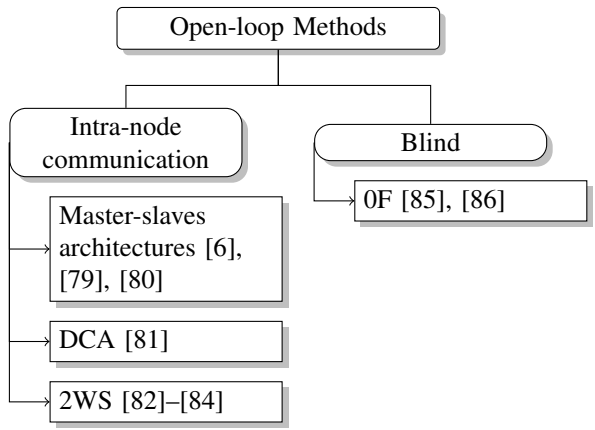

One of the simplest ways to do open-loop synchronization is through Master-slaves architectures, where one primary node broadcasts a beacon, and the secondary nodes lock their oscillators to this reference \cite{Merlano-Duncan2021}. This algorithm can be considered as a closed-loop method if the primary node is not part of the \acrshort{DSS} \cite{Tu2002}. For instance, in \cite{Ochiai2005} all nodes acquire their relative locations from the beacon of a nearby reference point that doesn't have to be the destination node. This allows open-loop synchronization, but each node requires knowledge of its relative position from a predetermined reference point within the cluster. In \cite{Barriac2004}, frequency and phase synchronization is achieved by locking the secondaries' oscillators to the reference beacon sent by the controller node after pre-compensating the phase mismatch and the propagation delay.

A solution more robust to link and node failures than Master-slave architectures is the \acrshort{DCA} proposed in \cite{Rahman2012b}, where each node broadcasts its carrier signal to all of its neighbors. Thus, the total received signal at any node is the superposition of its neighbors' carrier signals distorted by the channel. The goal of the \acrshort{DCA} is to use this received signal to adjust the instantaneous frequency at each distributed node in such a way that all the distributed nodes eventually become frequency locked to a common carrier \cite{Rahman2012b}.

Another Open-loop algorithm based in the retrodirective principle is \acrshort{2WS}, proposed in \cite{Preuss2011}. The main idea of this method is to implement a retrodirective beamforming between the base stations in a mobile network without the need of \acrshort{CSI} or feedback from the mobiles. To this end, a sinusoidal beacon is transmitted in a forward and backward propagation through all the base stations allowing them to compute a common carrier frequency and phase used to perform beamforming. The most significant drawback of retrodirective beamforming methods is that generally, the channel and the \acrshort{RF} chains at each transceiver are not reciprocal. A relative calibration method to compensate for this drawback is proposed in \cite{Bidigare2015}. In  \cite{NingXie2013a}, it is proposed a faster version of \acrshort{2WS} by exploiting the broadcast nature of the wireless links. The difference with \cite{Preuss2011} is that in the Fast Open-loop protocol, each base station only considers the signal from its adjacent two neighbors and ignores any signal from other base stations. In this way, the algorithm minimizes the latency by reducing the required non-overlapping time slots. 

Finally, the blind method or \acrshort{0F} is a synchronization algorithm that intends to synchronize the network without any feedback from the target or the rest of the distributed nodes. According to \cite{Bletsas2010}, the offsets of the oscillator drift cause intermittent coherent addition of signals at the receiver, which can result in significant beamforming gains. Even though the complexity of the \acrshort{0F} algorithm is simple, the data rate that can be achieved with this method is very low as the coherent beamforming is not stable. In addition, the statistical analysis reported in \cite{Bletsas2011} shows that \acrshort{0F} only works efficiently for a small network. 

\begin{table*}[]
\centering
\caption{Summary of synchronization methods}
\resizebox{\textwidth}{!}{%
%\begin{tabular}{m{0.1\textwidth}>{\centering}m{0.15\textwidth}>{\centering}m{0.15\textwidth}m{0.2\textwidth}m{0.35\textwidth}}
\begin{tabular}{ |m{0.1\textwidth} | m{0.16\textwidth} | m{0.13\textwidth} | m{0.19\textwidth} | m{0.37\textwidth} |}
\hline
\textbf{Classification} & \textbf{Method} & \textbf{Sync target} &  \textbf{Proposed application} & \textbf{Feasibility of its use in \acrshort{DSS}} \\ \hline
\multirow{5}{0.1\textwidth}{\centering Closed-loop: Iterative Bit Feedback} 
& \acrshort{1BF} \cite{Mudumbai2006} / \acrshort{2BF} \cite{Thibault2010} & Phase & Terrestrial wireless networks & \\ \cline{2-4}
& \acrshort{OCB} \cite{Pun2009} & Phase & \acrshort{WSN} & \\ \cline{2-4}
& \cite{Lee2018} & Phase & \acrshort{WSN} & Not suitable for synchronization of \acrshort{DSS} due to their slow\\ \cline{2-4}
& \acrshort{BioRARSA} \cite{Tseng2014} & Phase & Wireless sensor/relay network &convergence. \\ \cline{2-4}
& \acrshort{D1BF} \cite{Thibault2010} / \acrshort{SDDB} \cite{Thibault2013} & Phase & \acrshort{WSN} &  \\ \hline
\multirow{32}{0.1\textwidth}{\centering Closed-loop: Rich Feedback} 
& \acrshort{TPSN} \cite{Rucksana2015} and \acrshort{FTSP} \cite{Maroti2004,Shi2020}  & Time & \acrshort{WSN} & The basic ideas of these methods can be applied in \acrshort{DSS}, specially for large distributed networks. However, some refinement should be considered in order to achieve the accuracy needed for communication applications.\\ \cline{2-5}
& \acrshort{E1BF} \cite{Tushar2012} & Frequency and phase & Wireless networks & It could be used in some scenarios, depending on the distance between the \acrshort{DSS} and the receiver. However, the convergence time, 50\% smaller in this method than the original \acrshort{1BF}, can still be an problem. An advantage of this algorithm is that only one element of the \acrshort{DSS} processes the feedback from the receiver. \\ \cline{2-5}
& \cite{Kumar2014} & Phase & Wireless networks & Even, when the possibility to jointly achieve null and beamforming is a very desirable characteristic in satellite systems, the slow convergence of this method may make it unsuitable for synchronization of \acrshort{DSS} \\ \cline{2-5}
& \acrshort{PA} \cite{Jeevan2008} & Frequency and phase & Wireless communication networks & It can be used in a \acrshort{DSS} as long as a feedback channel is guarantee. This method presents scalability limitations. \\ \cline{2-5}
& \cite{Goguri2016a} & Phase & Wireless networks & This method allows simultaneously null and beamforming, which a desirable characteristic in satellite system. However, some limitations could be that (1) most of the calculation are performed by the distributed nodes and, (2) it can produce some latency for large \acrshort{DSS}. \\ \cline{2-5}
& \cite{Brown2012}\cite{Brown2015a} & Frequency and phase & Wireless networks & Allows beam and null-forming considering significant feedback latency, which implies that this method could be used for long baseline communications. However, it can present scalability problems. \\ \cline{2-5}
& \acrshort{F-RT} \cite{Brown} and \acrshort{T-RT} \cite{Brown2008} & Frequency and phase & Wireless communication networks & These methods can be used to synchronize \acrshort{DSS} as long as a feedback channel is guarantee. Another constraint related   \\ \cline{2-4}
& \acrshort{R-RT} \cite{Manosas-Caballu2011} & Phase, frequency and time & \acrshort{WSN} & to these methods is that all the satellites in the distributed system have to receive the reference signal broadcasted by the destination node.\\ \cline{2-5}
& \cite{Amor2019} & Phase, frequency and time & Cooperative decode-and-forward communication system & To implement this method in a \acrshort{DSS}, all the spacecraft have to simultaneously receive a common training sequence broadcasted by the destination node. In addition, some latency constraints must be taken into account. \\ \hline
\multirow{12}{0.1\textwidth}{\centering Open-loop: Intra-node communication} 
& \acrshort{RBS} \cite{10.1145/844128.844143} and \acrshort{PBS} \cite{Noh2008} & Time & \acrshort{WSN} & The basic ideas of these methods can be applied in \acrshort{DSS}, specially for large distributed networks. However, some refinement should be considered in order to achieve the accuracy needed for communication applications.\\ \cline{2-5}
& \acrshort{TWTT} \cite{Gun2009,Kirchner1991,Shengkang2013,Xiaochun2010,Jiuling2018,Pan2008,Ma2013,Yaowei2021} & Time & Satellite communications & It is used in \acrshort{DSS} \\ \cline{2-5}
& Master-Slave \cite{Ochiai2005,Barriac2004} & Frequency and phase & \acrshort{WSN} & Suitable for \acrshort{DSS} with the constraint that it requires accurate inter-satellite ranging. \\ \cline{2-5}
& \acrshort{DCA} \cite{Rahman2012b} & Frequency & Wireless networks & This method can be used to synchronize \acrshort{DSS} as long as all the distributed nodes are inside the range of the rest. \\ \cline{2-5}
& \acrshort{2WS} \cite{Preuss2011,Bidigare2015,NingXie2013a}  & Frequency and phase & Mobile networks & Despite some scalability problems, it can be used to synchronize \acrshort{DSS}. \\ \cline{2-5}
& \cite{Alvarez2018} & Phase, frequency and time & Dense and compact wireless networks & This method is suitable for \acrshort{DSS} where each node transmission can be received by almost all the satellites in the distributed system. However, algorithms with a central coordinator node acting as a reference for the delay compensation can be a simpler solution for less-dense networks. \\ \hline
\centering Open-loop: Blind synchronization & \acrshort{0F} \cite{Bletsas2010,Bletsas2011} & Frequency and phase & Wireless networks & Not recommended for \acrshort{DSS}, since it is limited to small networks, and it allows very low data rate. \\ \hline
\end{tabular}%
}
\label{tab:summary_synchronization_methods}
\end{table*}

\subsection{Summary and Lessons Learnt}
\begin{itemize}
    \item The synchronization algorithms can be classified as Closed-loop or Open-loop methods based on the use of feedback from a node external to the \acrshort{DSS}.
    \item Another classification considers the communication between the elements of the \acrshort{DSS}. It can be Closed-loop when the exchange of information among the distributed satellites is done as a two-way message exchange or; Open-loop when it is done as a broadcast or one-way communication.
    \item The time synchronization of \acrshort{DSS} is mainly based on the \acrshort{TWTT} algorithm. However, recent publications refer to the use of pseudo-random noise code and other techniques to achieve time synchronization and inter-satellite ranging simultaneously. 
    \item The time synchronization methods used in \acrshort{WSN} do not guarantee the accuracy required by communications applications. However, they could be applied to \acrshort{DSS} after some refinement.
    \item Some of the methods used for time synchronization in terrestrial distributed wireless networks can be applied to synchronize \acrshort{DSS}.
    \item Among the Closed-loop synchronization algorithms that use the feedback from a node external to the \acrshort{DSS}, the rich feedback methods are more suitable to implement in \acrshort{DSS}. Specifically, the \acrshort{PA} algorithm for \acrshort{DTB} and the reciprocity-based methods are the most recommended.
    \item Most Open-loop synchronization algorithms without any external aid, but with intra-node communication in the form of two-way message exchange are suitable for the synchronization of \acrshort{DSS}.
    \item Higher synchronization accuracy could be achieved by combining the use of intra-node communication in the form of two-way message exchange and the feedback from a node external to the \acrshort{DSS}. However, this configuration is not feasible for all applications.  
\end{itemize}

\section{Ranging and Relative Positioning in Distributed Satellite Systems}
\label{section::Ranging}

There are other operations closely related to the synchronization in \acrshort{DSS}, such as the inter-node ranging and relative position. For these operations, the requirements and accuracy of the coherent operation depend on the performance of the ranging and relative positioning algorithms. Several synchronization methods are based on these measurements \cite{RichardBrown2012}\cite{Comberiate2016}\cite{Hodkin2015a}. For example, in \cite{RichardBrown2012}, a dynamic model describes the stochastic kinematics and the clock evolution of each distributed node relative to the frame of the receiver of a \acrshort{DTB} system. Other examples are the remote sensing \acrshort{DSS} missions, such as \acrshort{GRACE} and PROBA-3, where knowing the relative position among the distributed nodes is fundamental to combining their measurements. For that reason, this section compiles recent advances in inter-node ranging and relative positioning for \acrshort{DSS}.

\subsection{Inter-Satellite Ranging}
\label{section::ranging_subsection}

For many \acrshort{DSS}, inter-satellite relative range measurement is a requirement for cooperative tasks. Generally, autonomous inter-satellite measurements and communications can reduce the dependence on ground stations, signal transmission delays, and improve the resilience and maneuverability of the \acrshort{DSS}. To this end, the distributed nodes must have the capability of inter-satellite ranging. 

\begin{figure}
    \begin{tikzpicture}[edge from parent fork down, sibling distance=35mm, level distance=15mm,every node/.style = {shape=rectangle, rounded corners,
            draw, align=center}] 
            \node {Inter-satellite Ranging Methods}
            child { node {Optical Inter-satellite \\ Ranging} }
            child { node {Ranging based on \\ \acrshort{RF} Signals}
                child { node {Carrier-ranging}}
                child { node {pseudo-random code ranging}} };
    \end{tikzpicture}
    \caption{Classification of the inter-satellite ranging methods}
    \label{fig::ranging_class}
\end{figure}
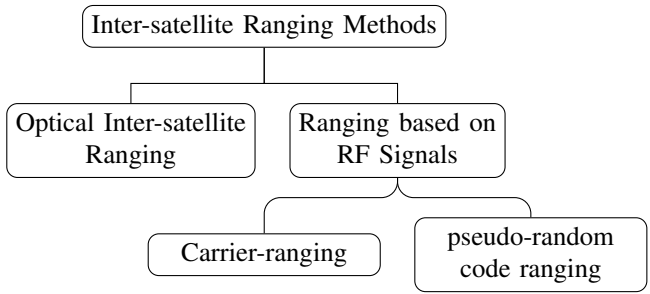

The methods to do inter-satellite ranging can be classified into two main groups represented in Fig. \ref{fig::ranging_class}: those based on \acrshort{RF} and those based on optical signals. Measurements made using radio signals are the most mature technology, but optical measurements can achieve better ranging performance. However, the higher directivity of the laser beam, in comparison with the \acrshort{RF} antenna's patterns, can represent a limitation for specific applications. Besides, the sunshine can blind optical sensors.

\subsubsection{Optical Inter-Satellite Ranging}

\begin{figure}[!t]
\centering
\includegraphics[width = \columnwidth]{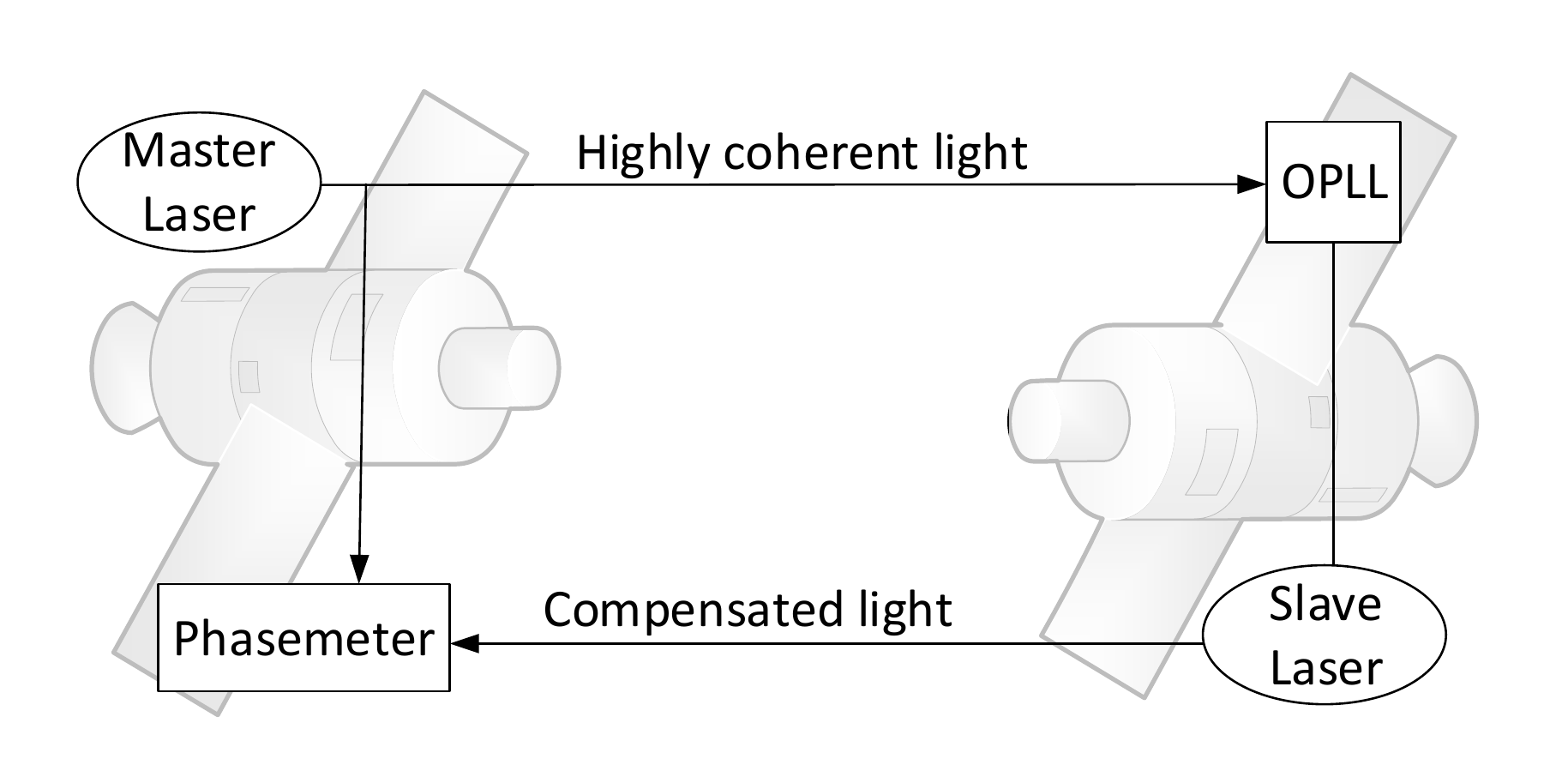}
\caption{Basic principle of optical inter-satellite ranging measurement}
\label{fig::Basic_optical_ranging}
\end{figure}

The basic principle of optical inter-satellite ranging measurement, also known as transponder laser interferometry, is represented in Fig. \ref{fig::Basic_optical_ranging} as explained in \cite{Ming2020}. The frequency-stabilized and power-stabilized master laser provides a highly coherent light source to meet interference requirements between the local and receiver laser lights. First, the master's laser light travels through the inter-satellite space and arrives at the slave satellite, where the \acrshort{OPLL} locks the phase of the slave laser to that of the weak light received. Then, the compensated slave laser light points and propagates back to the master satellite. The precision phasemeter measures the phase difference between the master laser and the received light to calculate the inter-satellite distance.

The authors in \cite{Ming2020} discussed the state of the art of inter-satellite laser interferometer technologies for spaceborne gravitational-wave detection. However, inter-satellite interferometric ranging has been previously implemented in missions such as \acrshort{GRACE} follow-on, where the ranging performance of the former mission was improved by a factor of 10 by including an interferometric laser ranging system \cite{Ales2014}. This improvement is due to the laser wavelength being 10,000 times shorter than the microwave wavelength. 

As a variant to the methods mentioned before, which required two phase-synchronized lasers on both satellites, a single-laser design was proposed in \cite{Bykhovsky2015}. This solution included a Mach-Zehnder interferometer on one of the satellites and a secondary satellite equipped with a retro-reflector. The work in \cite{Luo2016} overcame the need for highly stable lasers and complex phase-locking solutions of the previous ones. Though, it is required to maintain the coherence properties of the laser light in the reference arm, which is very technologically challenging for distances more significant than 500~km, i.e., twice the expected inter-satellite distance. However, simulations results in \cite{Bykhovsky2015} showed similar ranging resolution to the two phase-synchronized lasers method.

Recent publications proposed the use of a single optical \acrshort{ISL} for ranging and communications simultaneously \cite{Tian2019}\cite{Calvo2020}. For instance, \cite{Calvo2020} addressed the development of the optical transceivers to transmit ranging and frequency synchronization information through a coherent optical link between the spacecraft of the Kepler constellation. The Kepler system is based on a constellation of 24 Galileo-like \acrshort{MEO} and six \acrshort{LEO} satellites carrying stabilized lasers as optical frequency references and equipped with terminals for two-way optical links. This provides a means for broadcasting synchronization, and navigation messages across the \acrshort{DSS} without any communication with the ground segment \cite{Calvo2020}. A ranging precision of $300~\mu$m and $10^{-15}$ s/s stability  (Allan deviation at 1~s) is achieved through an \acrshort{OPLL} locked to a \acrshort{PRN} sequence with a chip rate of 25.6~GChip/s. An additional data channel at 50~Mbit/s is multiplexed with the ranging signal to transmit timestamps in a known repetition rate. This allows time alignment of both satellites with the resolution of the \acrshort{FPGA}' clock rate used at both ends.
Furthermore, the data link can be used for communication purposes by the constellation. Each \acrshort{MEO} satellite has four bidirectional transceivers, two for connecting to neighboring satellites in the same orbit, and two for connecting to \acrshort{LEO} satellites. The terminal aperture size is constrained to 75~mm due to the satellite's weight and size considered. However, to calculate the absolute ranging between terminals with an accuracy of 0.3~mm, the system requires aligning precision of $2.5\%$ the chip length or better \cite{Calvo2020}.

Another example of simultaneous ranging and communications inter-satellite optical link is the coherent optical receiver implemented in \cite{Tian2019}. In this article, the authors presented an \acrshort{FPGA}-based feedback-homodyne scheme as an alternative to the \acrshort{OPLL} to obtain a more flexible coherent optical receiver. Besides, a parallel \acrshort{FFT} wavelength drift estimation algorithm was proposed, aiming to improve the speed and range of wavelength drift tracking simultaneously. Simulation results showed that the wavelength drift tracking performance depends on the number of \acrshort{FFT} estimators used in parallel. However, using a real-time \acrshort{FPGA} implementation, the authors demonstrated that the design meets the needs of phase offset compensation when three \acrshort{FFT} estimators are used in parallel \cite{Tian2019}.

\subsubsection{Inter-Satellite Ranging Based on RF Signals}

The use of \acrshort{RF} for inter-satellite ranging is not a new concept. Articles such as \cite{4072332}, published in 1985, already proposed a design to measure satellite-to-satellite range-rate with a precision smaller than $1~\mu$m/s in distances between 100~km and 300~km. However, further analyses included the frequency instability of the oscillators \cite{Kim2003}, the channel noise conditions \cite{Yang2010}, and the requirements of low-cost small satellites \cite{Alawieh2016} among other specifications. Distances can be determined from either the signal's modulation (\acrshort{PRN} codes) or the carrier phase. 

\begin{figure}[!t]
\centering
\includegraphics[width = \columnwidth]{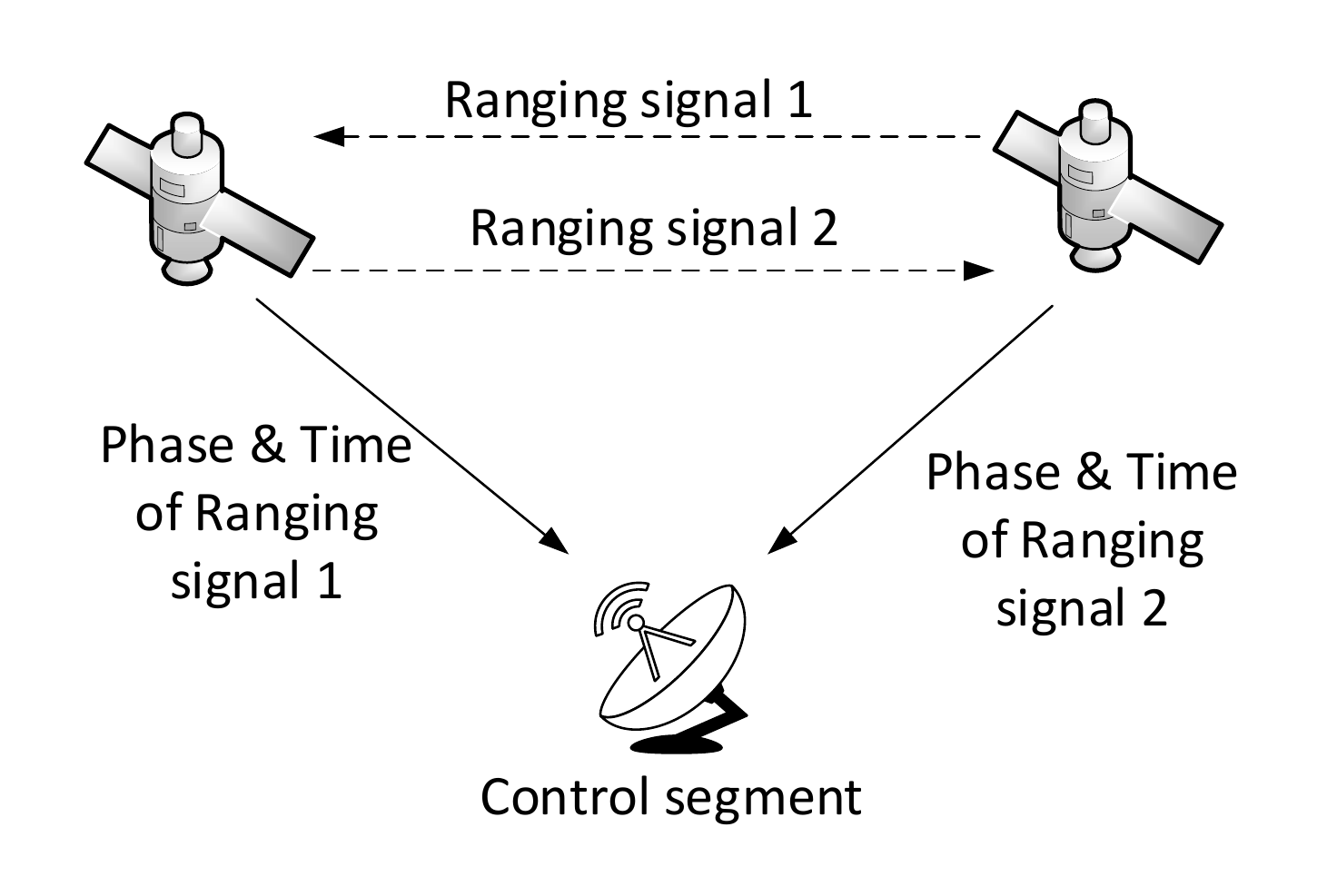}
\caption{Basic principle of \acrshort{RF} carrier-ranging by \acrshort{DOWR}.}
\label{fig::DOWR}
\end{figure}

One of the most popular carrier-phase based ranging methods is the \acrshort{DOWR} \cite{Gun2009}\cite{Kim2003}\cite{Yang2010}. By combining the one-way range measurements from two microwave-ranging devices, the method minimizes the effect of oscillator phase noise. Each satellite uses identical transmission and reception subsystems to send a carrier signal to the other. The recorded measurements are transmitted to a control segment for processing and calculating the inter-satellite range. \acrshort{DOWR} method is represented in Fig. \ref{fig::DOWR}. 

Some variations of this method have been proposed. For example, in \cite{Gun2009} the authors addressed the use of \acrshort{DOWT} synchronization and ranging (\acrshort{DOWT}\&\acrshort{DOWR}) in \acrshort{DSS}. Simulation results indicated less than 0.2~m and 0.5~ns precise ranging and time synchronization accuracy and error time synchronization smaller than $\pm10$~ns for inter-satellite distances up to 200~km in satellite formation flying \cite{Gun2009}. Another example is the Random Access Inter-satellite Ranging (RAISSR) system proposed in \cite{Xiaoyi2020}. In this system, each satellite transmits a signal to allow other satellites to perform a one-way distance measurement with respect to it. Even when the authors claimed no strict time synchronization, the method requires all the satellites to share a common time reference.  

In \cite{Wang2012}, an approach to solving the problem of autonomous timing is proposed. This method aims to achieve time synchronization and high precision ranging between onboard oscillators and clocks on Earth, with only scarce information exchange. The main idea is to calculate the satellites \acrshort{CFO} using a frequency tracking loop. Then, the inter-satellite range and the satellites' relative radial velocity are estimated on the ground \cite{Wang2012}. The reported results showed a precision of 0.1~ns for inter-satellite baseline measurements and the accuracy of velocity measurement around $10^{-3}$~m/s, for approximately 0.01~Hz carrier frequency tracking error \cite{Wang2012}. 

An essential aspect of \acrshort{DOWR} that limits its performance is its synchronization requirements. As this method requires the transmitter and receiver to be synchronized, its accuracy is directly related to the performance of the synchronization algorithm. Clock drifts and offsets between nodes can introduce errors in the order of microseconds which can be too large for specific applications. On the other hand, the \acrshort{TWR} method does not have this drawback. In \acrshort{TWR}, introduced in \cite{Alawieh2016}, one satellite sends the ranging signal and activates a counter to calculate the elapsed time until the reception of the other satellite replies. In this way, there is no need to synchronize both satellites' clocks \cite{Alawieh2016}. The evaluation of the algorithm on a \acrshort{SDR} platform is reported in \cite{Alawieh2016}, showing an accuracy of a few centimeters.

Another hardware demonstration was presented in \cite{Comberiate2016}. In this case, the authors developed a coherent distributed transmission system based on high-accuracy microwave ranging to perform \acrshort{DTB} for mobile platforms. The Two-tone ranging method used for the indoor and outdoor experiments was introduced in \cite{Hodkin2015a}.  

The most well-known example of pseudo-random code ranging is the method applied by \acrshort{GNSS} constellations. A variation of this method to measure the inter-satellite range in deep-space missions, where the navigation constellation is not available, was suggested in \cite{Rui2008}. The design combines the typical \acrshort{GPS} receiver with a \acrshort{RF} transceiver to use when the satellite does not receive the \acrshort{GPS} constellation' signals. The \acrshort{GPS} module is a flight-qualified design that performs orbit determination using a \acrshort{EKF} and the inter-satellite distance measurement is obtained by differencing both filtered absolute \acrshort{GPS} solutions. The core of the whole system is a \acrshort{CPU} core which controls the switch and coordination of the two ranging units \cite{Rui2008}. Similarly, an \acrshort{ARNS} for autonomous satellite formation flying in \acrshort{LEO} is proposed in  \cite{Yang2014}. The idea of this article is that inter-satellite ranging systems can provide additional observation information, which can be used to increase the \acrshort{GPS} stand-alone observation dimension. Reported results indicated that the \acrshort{ARNS} improved the relative positioning accuracy by one order of magnitude in comparison to the \acrshort{GPS} stand-alone solution \cite{Yang2014}.

In addition, other authors addressed ranging methods based on \acrshort{PRN} codes. For instance, in \cite{Wang2019} it is presented an algorithm to determine the initial value of smooth pseudo-range for the smoothing method proposed by Hatch \cite{Hatch}. The algorithm uses the least-squares straight line fitting technique to decrease the convergence time and improve the ranging accuracy \cite{Wang2019}. On the other hand, the authors of \cite{Tang2017} analyzed how it is affected the accuracy of inter-satellite ranging in a \acrshort{DSSS} by the used bandwidth and \cite{Xue2020} introduced a new ranging scheme combining continuous phase modulation and a \acrshort{PRN} ranging code. The chip pulse used in this method is based on a normally distributed signal, which improves the ranging accuracy and reliability of the system. 

Another example of a \acrshort{RF} ranging technique using the signal's modulation was proposed in \cite{Crisan2018}. This paper analyzed a hybrid \acrshort{OFDM} communication-metrology system for a two satellites formation flying. The system uses \acrshort{OFDM} signals for \acrshort{ISL} communications and measurements functions. The inter-satellite distance is estimated using a training symbol, which is also used for time and frequency synchronizations and estimation of the channel impulse response. The accuracy of the technique depends mainly on the bandwidth of the transmitted signal \cite{Crisan2018}.

Most recent works focus on carrier ranging methods assisted by pseudo-random code ranging, which allows higher precision range values. For example, \cite{Xiang2019a} proposed a pseudo-code-assisted carrier ranging algorithm, which is the combination of pseudo-range and dual-frequency carrier phase ranging methods. In \cite{Xiang2019}, the impact of the frequency selection on the ranging accuracy was analyzed. Too high frequency leads to problems in the resolution of integer ambiguity, but too low frequency impacts ranging accuracy. Both articles reported range errors on the order of 1~cm for inter-satellite distances around 100~km \cite{Xiang2019a}\cite{Xiang2019}. 

Even though the accuracy of carrier-ranging methods is higher than pseudo-random code ranging methods, the integer ambiguity is more difficult to resolve for the former ones. The solution of the phase integer ambiguity problem has been dealt in \cite{Yang2014}, \cite{Sun2012} and \cite{Yue2017}. For instance, \cite{Sun2012} presents a robust integer-cycle ambiguity resolution method by modifying the well-known \acrshort{LAMBDA} method \cite{Teunissen1995}. The authors derived the validation threshold in closed-form as a function of the phase noise variance and the antenna baseline geometry. As a result, the success rate of the original \acrshort{LAMBDA} is improved. However, the method requires that the spacecraft of which the \acrshort{LOS} will be estimated is equipped with a body-fixed array of at least three antennas in different planes \cite{Sun2012}. This could be a limitation for small satellites missions. Another procedure based on the \acrshort{LAMBDA} method was proposed in \cite{Yang2014}. This paper contains two approaches to combine the inter-satellite ranging measurements with \acrshort{GPS}. Besides, it proposes a feedback scheme to convert the ambiguity float solutions obtained from the \acrshort{EKF} into pseudo-perfect measurements through the \acrshort{LAMBDA} method. After the possible integer solutions are verified, they are adopted as pseudo-perfect measurements to update the \acrshort{EKF} for the next epoch. 

Another approach to the phase integer ambiguity problem was proposed in \cite{Yue2017}. In this case, the authors used the simplified time-differenced technique \cite{Wendel2006} to eliminate the influence of ambiguity in a \acrshort{GPS}/\acrshort{INS} integrated navigation algorithm \cite{Yue2017}. The simplified time-differenced method consists of processing time differences of successive carrier phase measurements at a \acrshort{GPS} base station, which increases the velocity and attitude accuracy. It reduces noise in the position information compared to the traditional tightly coupled systems. Unfortunately, the time-differenced carrier phase measurements do not fit easily in the framework of a \acrshort{KF} measurement equation, which has adverse effects on the derivation of an appropriate measurement matrix \cite{Wendel2006}.

\subsection{Relative Positioning}
\label{section::relative_posit}

Spacecraft relative positioning in formation flying is a crucial enabler of new space missions and of paramount importance for distributed satellites architectures. Clearly, two subcategories can be distinguished: in-orbit autonomous and assisted from the ground. Relative positioning can be seen as a coordinated operation of the ranging systems described in the section above, which can occur either in-orbit or on-ground.

\subsubsection{In-Orbit Autonomous}
The primary technology used for precise relative positioning of autonomous formation flying satellites is \acrshort{GNSS}.

If the relative position between two \acrshort{GNSS} receivers is required, the \acrshort{GNSS} data differences from two receivers can be used. This method reduces common data errors, such as the \acrshort{GPS} satellite clock offset. The difference is typically calculated using the pseudo-range estimation between the receivers and a \acrshort{GNSS} satellite or more. Once resolved the code ambiguity, sub-metric precision is obtained by using this method \cite{Kroes2006}.

In order to enable, cm to mm positioning accuracy levels, the handling of carrier differential \acrshort{GNSS} (CD\acrshort{GNSS}) estimations needs to happen \cite{DAmico2009}. Within the setting of the \acrshort{GRACE} formation flying mission, the possibility of 1~mm level relative navigation over a 200~km separation has been illustrated by utilizing carrier \acrshort{GPS} estimations from a high-grade dual-frequency BlackJack receiver \cite{Kroes2005}.

A major problem in using CD\acrshort{GNSS} is tackling for the unknown integer number of cycles (integer ambiguity) \cite{Renga2013}. Especially when the resilience and precision of the integer solution are affected by variations in the number of common-in-view satellites, and ephemeris and ionospheric differential errors. This is enhanced when the inter-satellite distances are highly variable due to the relative orbital trajectory.

A few strategies based on arrays of \acrshort{GNSS} sensors have been proposed to unravel the phase ambiguity problem. Since antennas are unbendingly mounted on the stage, the relative antenna position within the local body frame is known in advance. It can be used to improve the accuracy of the estimated integer ambiguity. In \cite{Nadarajah2016}, the use of the Multivariate Constrained Least-squares AMBiguity Decorrelation Adjustment (MC-\acrshort{LAMBDA}) strategy \cite{Teunissen2007} successfully utilizing nonlinear geometrical constraints. By consolidating the known antenna geometry into its ambiguity objective function, this strategy has been appeared to illustrate reliable and immediate single-frequency integer ambiguity determination.

\acrshort{DSS} operating above \acrshort{LEO} loses coverage of the \acrshort{GNSS} infrastructure and has to rely on other means for tracking and navigation. For those missions, a trilateration scheme is proposed in \cite{Cheung2017} that evaluates the \acrshort{3D} relative position between a reference spacecraft and a target spacecraft using raw-range measurements from a distance baseline of known locations, which is called “anchors”. The anchors can be antennas of a ground-based network or satellites of a space-based network (e.g., \acrshort{GPS}). The method assumes the clocks of the anchors to be perfectly synchronized and requires some synchronization between the anchors and the reference spacecraft. However, the synchronization errors between the reference spacecraft and the rest of the satellites in the \acrshort{DSS} is compensated by using an additional anchor. This method achieves sub-meter accuracy for a \acrshort{GEO} \acrshort{DSS} with two spacecraft, using a baseline network of three ground stations as anchors. 

Another example of very accurate tracking and control of the relative position is the PROBA-3 mission. PROBA-3 is a European Space Agency (ESA) mission to obtain highly accurate formation flying. A couple of satellites, the Coronagraph Spacecraft (CSC) and the Occulter Spacecraft (OSC) will work together as an externally occulted solar coronagraph. The CSC hosts the optical assembly of the coronagraph as the primary payload, while the OSC carries the coronagraph external occulter disk. The mission requirements are a longitudinal accuracy better than 1~mm for the inter-satellite distance of 144~m \cite{wwwProba3}. To make this feasible, the formation flight system is distributed between both spacecraft: the CSC hosts the formation flying sensors, while the OSC performs the data processing \cite{Casti2019}. The formation flight task requires the use of several metrology subsystems, from coarse accuracy and large scale range determination to very accurate and shorter scale range measurement and absolute positioning. All the data generated by these subsystems are processed in real-time by the Guidance and Navigation Control system, obtaining an unprecedented accuracy \cite{Capobianco2019}.

IRASSI is an interferometry-based mission concept composed of five free-flying telescopes orbiting the Sun-Earth/Moon second Lagrangian point, L2. It focuses on observing specific regions of the sky to study star formation, evolution processes, and early planetary origins. The study of these processes requires a telescope with angular resolution lower than 0.1 arcsec \cite{Buinhas2016}. The interferometer relies on dynamically changing baselines obtained through the physical separation during scientific observations to achieve such strict resolution values. For example, the baseline vectors between the telescope reference points must be determined with an accuracy of $5~\mu$m to guarantee the precise correlation of the detected signals. The navigation concept capable of achieving these strict requirements consist of two components: the absolute position estimation concerning Earth and the relative position estimation, which determines the satellite's positions to each other \cite{Buinhas2018}. The description of two autonomous relative positioning algorithms based on a geometric snapshot approach can be found in \cite{Linz2020}.

For formation flying missions, in addition to the inter-satellites position, it is imperative to estimate and control the satellites' attitudes accurately. In \cite{Weiqing2015}, the relationship between the precision estimation against the ranging accuracy, ranging distance, and satellite relative position were analyzed. As a result, it was concluded that obtaining accurate measurements required a balanced number of transmitting and receiving antennas with suitable configurations. It is a necessity to install more antennas to improve the precision estimation of the attitude angles \cite{Weiqing2015}. 

Another critical requirement of formation flying is accurate relative navigation. Significant research results have been published about this topic. For example, \cite{Carrillo2016} used the nonlinear dynamics describing the relative positioning of multiple spacecraft for formation flying trajectory tracking control. Using Lyapunov-based control design and stability analysis techniques, the authors developed a nonlinear adaptive higher-order sliding mode control commonly known as adaptive super twisting sliding mode control. On the other hand, \cite{Dan2006} develops an efficient approach of autonomous relative orbit determination for satellite formation flying. The proposed solution uses the inter-satellite local measurements by the microwave radar and laser devices on board the satellites to perform the relative navigation. The design uses a decentralized Schmidt \acrshort{KF} \cite{Dan2006} to estimate the state of relative orbit between the satellites and proves that this approach is immune to the single satellite failure. Simulation results showed that the relative position estimation might achieve centimeter-level accuracy, and relative velocity estimation may achieve mm/s-level accuracy for a circular spatial formation consisting of three satellites where the chief satellite is at the center. The radius is about l~km \cite{Dan2006}.
Similarly, in \cite{Wang2018b} a method for autonomous orbit determination combining X-ray pulsar measurements and inter-satellite ranging during Mars orbiting phase was presented. The method calculates an observability index reflecting the measurement information quality and optimizes the observable target selection and the observation scheduling. Then, the Unscented \acrshort{KF} is used to estimate the autonomous pulsar assisted orbit determination \cite{Wang2018b}.

\subsubsection{Assisted from Ground}
Determining the orbital position of satellites from the ground requires obtaining azimuth and elevation view angles of the satellite and the distance from the ground station to the satellite (range). 

One of the most precise \acrshort{POD} measurements assisted from the ground can be achieved by Radio Interferometry methods. Interferometry is a technique for passive tracking. A simple interferometer consists of two antennas. Using the phase difference between the signals received by the two antennas, the direction of the target can be determined. Connected-Element Interferometry (CEI) is a technique for determining the phase offset from the \acrshort{TOA} difference of a downlink radio signal to two antennas on a short baseline. In \cite{Liu2019}, the authors used a small-scale CEI system of two orthogonal baselines (75 m $\times$ 35 m) to track a \acrshort{GEO} satellite. Reported results showed accuracies smaller than 1~km in the radial and the cross-track directions, and \acrshort{3D} position accuracy in the order of 2~km. Another example of the positioning of \acrshort{GEO} satellites by Radio Interferometry was proposed in \cite{Sadeghi2019}. In this case, the authors focused on determining the view angles of \acrshort{GEO} satellites with an estimation accuracy of 0.001\textdegree~in Ku band. 

In addition, \acrshort{ISL} can improve the accuracy of orbit determination for \acrshort{GNSS} constellations. One example of this was presented in \cite{Michalak2020}, where it was demonstrated that the \acrshort{POD} of the Kepler system could be performed with just one ground station achieving orbit accuracies of 5~cm in \acrshort{3D} and 0.24~cm in radial direction \cite{Michalak2020}. On the other hand, in \cite{Chen2019}, it was proven that the use of \acrshort{ISL} ranging measurements could reduce the first positioning time about six times in conventional navigation receivers.

\subsection{Summary and Lessons Learnt}

\begin{itemize}
    \item In a \acrshort{DSS}, the inter-node ranging and relative position are closely related to the synchronization algorithm. Generally, the requirements and accuracy of the coherent operation depend on the performance of the ranging and relative positioning algorithms as much as the synchronization itself. 
    \item Optical inter-satellite ranging methods can achieve higher accuracy, but the higher directivity of the laser beam can represent a limitation for specific applications.
    \item The use of a single optical \acrshort{ISL} to simultaneously perform ranging and communication is a trending topic in this field.
    \item Inter-satellite ranging based on \acrshort{RF} can use measurements of the carrier's phase or use the signal modulation. Recent publications considered the combination of both: carrier ranging assisted by pseudo-random code ranging methods.
    \item Both optical and \acrshort{RF}-based ranging algorithms have to deal with the phase integer ambiguity problem.
    \item Spacecraft relative positioning can be seen as a coordinated operation of the ranging systems described above, which can take place either in-orbit autonomous or assisted from the ground.
    \item The leading technology used for precise relative positioning of autonomous formation flying satellites is \acrshort{GNSS}.
    \item \acrshort{DSS} operating above \acrshort{LEO} loses coverage of the \acrshort{GNSS} infrastructure and has to rely on more complex schemes. Examples of these schemes were analyzed in this section.
    \item One of the most precise \acrshort{POD} measurements assisted from the ground can be achieved by Radio Interferometry methods.
\end{itemize}

\section{Applications of Distributed Satellite Systems Synchronization}
\label{section::App}

As stated above, carrier and time synchronization are critical and very challenging requirements for \acrshort{CDSS}. Consequently, the research in the area is very prolific. In this section, the most significant articles about synchronization for \acrshort{DSS} found in the literature are summarized. They are arranged in two groups attending to its application: Communications or Remote Sensing.

\subsection{Communications}
\label{section::Comm_app}

\begin{figure}[!t]
\centering
\includegraphics[width = \columnwidth]{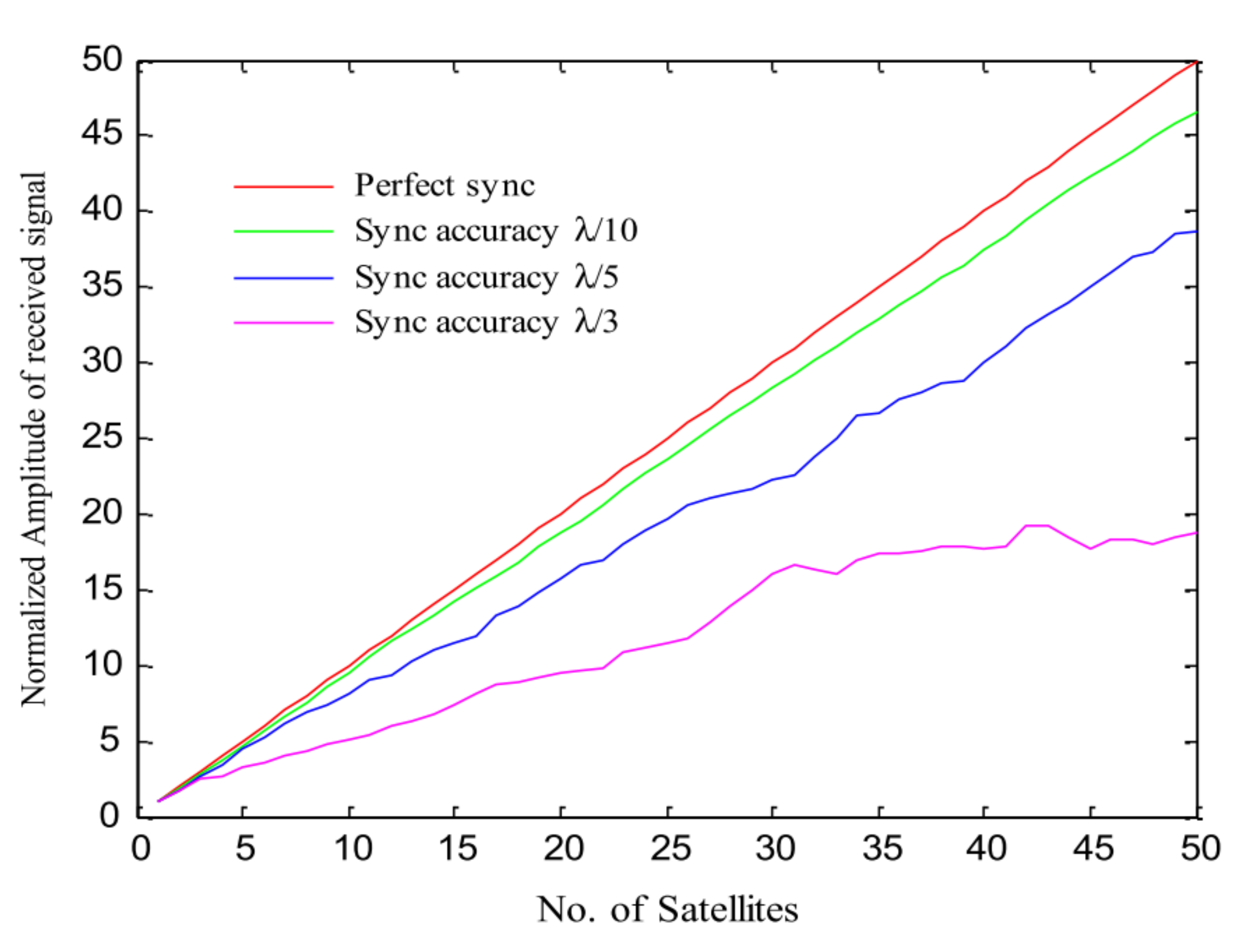}
\caption{Normalized amplitude of the received signal vs. the number of satellites for different values of phase synchronization accuracy \cite{Sundaramoorthy2013}}
\label{fig::Phase_synch_req}
\end{figure}

Small satellites, organized in distributed systems, are envisioned to be the future of space communications \cite{Kodheli2020}. Forming a dynamic phased array in space with the nodes of a \acrshort{DSS} can improve the communications capabilities between the network and the Earth. However, implementing such an array requires effective open-loop carrier synchronization \cite{Radhakrishnan2016}. In \cite{Sundaramoorthy2013}, the phase synchronization constraints of a \acrshort{DSS} of very simple, resource-limited femto-satellites communicating with Earth were derived and discussed \cite{Sundaramoorthy2013}. Fig. \ref{fig::Phase_synch_req} from \cite{Sundaramoorthy2013} shows the normalized amplitude of the received signal as a function of the number of transmitters for different accuracies of phase synchronization. As can be appreciated in the figure for a high number of nodes, the effects of phase synchronization inaccuracies are more evident.

In \cite{Sundaramoorthy2016}, a time and phase synchronization solution to perform beamforming in a \acrshort{LEO}-\acrshort{DSS} is proposed. The mathematical framework and analysis of the synchronization scheme is developed, which relays in an external beacon transmitted from a \acrshort{GEO} spacecraft above the intended ground receiver. This open-loop method achieves subcentimeter-level (subnanosecond-level) phase synchronization with localization accuracy on the order of meters \cite{Sundaramoorthy2016}. However, these results are restrained to a particular array geometry, which is not always feasible.  

In \cite{Gun2009}, the simulation and performance evaluation of the two-way range and time synchronization method for a \acrshort{DSS} were presented. The \acrshort{DSS} consisted of four satellites in a formation flying mission, with 50~km of inter-satellite distance and a carrier frequency of 400~MHz simulated in the Systems Tool Kit (STK) platform from Analytical Graphics, Inc. According to the simulation results, the error of the time synchronization was less than 10~ns \cite{Gun2009}.

\subsection{Remote Sensing}
\label{section::Remote_sensing_app}

Remote sensing and Earth observation from \acrshort{DSS} have become attractive for the community in the last years. New concepts, like distributed synthetic aperture radiometers or fractionated radars, offer the potential to significantly reduce the costs of future multistatic \acrshort{SAR} missions \cite{Krieger2018}\cite{Xiao2020}. Besides, the low radar power budget and high spatial resolution requirements of future remote sensing satellite missions seem to be only possible to achieve through multistatic configurations \cite{Querol2020}. However, the frequency, phase, and time synchronization is still the major challenge that slows down the launch of new missions. For that reason, new synchronization methods for Remote Sensing \acrshort{DSS} are published daily.

For instance, an inter-satellite time synchronization algorithm for a micro-satellite cluster was proposed in \cite{Jiuling2018}. The authors proposed a time control loop that dynamically adjusts each satellite reference frequency according to the time difference with the time benchmark of the cluster. Consequently, time synchronization is achieved by locking the clocks of all the satellites to a chosen one. Similarly, \cite{Ubolkosold2005} presents a method to relatively synchronize the clocks in a \acrshort{DSS} without the use of \acrshort{GNSS} signals. A reference pulse transmitted by a master satellite is received and time-tamped by all the secondary satellites in its broadcast domain. The method estimates the clock offsets relative to the other secondary satellites by calculating the receive-time differences between two secondary satellites. Simulation results reported in \cite{Ubolkosold2005} showed excellent convergence rates. However, this method requires that all the distributed nodes are in the same broadcast domain of a single master satellite, and the \acrshort{LOS} from one satellite to all others must always be available. Besides, the algorithm is affected by the difference in propagation time between two receivers. This is not a problem for relatively small networks but, for \acrshort{DSS} with longer inter-satellite distances, these propagation delays have to be compensated.

Other authors have studied the effects of time and frequency synchronization on \acrshort{DSS} \acrshort{SAR} \cite{Zhang2006}\cite{Younis2006}. In \cite{Zhang2006}, a model considering the time and frequency synchronization errors in \acrshort{InSAR} were presented. The performance of the range and azimuth compression with the synchronization errors was analyzed to obtain the time and frequency synchronization requirements for parasitic \acrshort{InSAR} system design. On the other hand, \cite{Younis2006} studied the performance degradation in bi-static and multi-static \acrshort{SAR} due to the oscillators' phase noise. Using a dedicated synchronization link to quantify and compensate the oscillators phase noise was proposed considering three different synchronization schemes: continuous duplex, pulsed duplex and, pulsed alternated. The analysis included additional factors such as receiver noise and the Doppler effect and contributions known from sampling theory like aliasing and interpolation errors. According to the reported results, successful oscillator phase noise compensation is possible if the compensation algorithm and the signal timing are adapted to the link hardware and \acrshort{SAR} parameters \cite{Younis2006}.

More advanced works have reached the implementation and test steps, which are fundamental to using a synchronization method in a mission. For example, \cite{He2011} proposed a hardware-in-the-loop simulation and evaluation approach for \acrshort{DSS} \acrshort{SAR}. The proposal was used to model a typical bi-satellite formation spaceborne distributed \acrshort{SAR} system in the X band. The \acrshort{SAR}' central electronic equipment was implemented in hardware, whereas the echo generation and the processing and evaluation of the results were performed in software. Another example is the testbed for coherent distributed remote sensing systems proposed in \cite{Querol2020}. This platform is composed of two satellites, the channel emulator, and two targets, all implemented in \acrshort{USRP}. The authors proposed and validated a dual carrier point-to-point synchronization loop through the testbed. The synchronization algorithm is based on a Master-Slave architecture to autonomously synchronize both satellite clocks with a common reference using \acrshort{ISL} \cite{Querol2020}. Generally, the hardware simulations include features that can not be easily or accurately considered with computer-based simulations. First, there are coupling relationships among the channel mismatch, the time and phase synchronization errors, and other error sources. The single error analysis method used in the computer-based simulation can't meet this requirement. Besides, the hardware implementation of the synchronization algorithm gives a more accurate description of its performance. Computer-based simulations cannot precisely measure features like processing time and resources. On the other hand, the error source characteristic has to be known for data analysis and evaluation, which is more complicated for hardware simulations.

Several remote sensing distributed satellite missions have been proposed and deployed during the last years. Section \ref{section::Examples} will refer to already deployed missions, but the proposed synchronization algorithms for future missions are of interest to this section.
\begin{enumerate}
    \item ARGOS:  The Advanced Radar Geosynchronous Observation System (ARGOS) will be a \acrshort{MIMO} \acrshort{SAR} system hosted on a swarm of mini-satellites in quasi-geostationary orbits \cite{MontiGuarnieri2015}. It consists of: a swarm of active and passive spacecraft in a zero inclination quasi-\acrshort{GEO} orbit; a data-link to a telecommunication satellite that is used for both data download and synchronization; a ground segment and; a network of active calibrators for precise estimation of sensors clocks and orbits. According to \cite{Guarnieri2015}, the link with the telecommunication satellite can be used to synchronize the \acrshort{DSS} to a common clock and support the estimation of the precise orbits, in a similar way to a cloud \acrshort{PLL}. Besides, the oscillators' phase noise can be compensated by exploiting a network of Compact Active Transponders or through an on-ground synchronization scheme capable of estimating the phase on very bright point targets.
    
    \item LuTan-1: LuTan-1 (LT-1), also known as TwinSAR-L mission, is an innovative spaceborne bistatic \acrshort{SAR} mission based on the use of two radar satellites operating in the L-band with flexible formation flying, to generate the global digital terrain models in the bistatic interferometry mode \cite{Liang2019}. Several articles about the LT-1 phase synchronization scheme have been published \cite{Liang2019}\cite{Jin2020}\cite{Zhang2020}\cite{Liang2020}. The algorithm, proposed in \cite{Jin2020}, considers the time multiplexing of the synchronization pulses with the radar pulse-repetition intervals avoiding any interference in the \acrshort{SAR} data acquisition. Besides, \cite{Liang2019} proposed a \acrshort{KF} phase-error estimation and compensation method, which improves the synchronization accuracy for low \acrshort{SNR}. On the other hand, \cite{Zhang2020} evaluated the performance of the LT-1's synchronization scheme for multipath effects, and \cite{Liang2020} presents its experimental verification.
    
    \item ConGaLSAR: A Constellation of Geostationary and \acrshort{LEO} \acrshort{SAR} (ConGaLSAR) radars has been proposed in \cite{Xiao2020}. The phase and time synchronization of this constellation would be supported by a novel transponding mode known as MirrorSAR, where the \acrshort{LEO} satellites work as relay nodes for the ground echoes to the illuminating \acrshort{GEO} satellite. The key idea of MirrorSAR is to redirect the radar echoes to the spacecraft, where they can be coherently demodulated with the same clock references previously used to generate the radar pulses \cite{Krieger2018}. The authors introduce two alternatives for synchronization. In the most simple one, the \acrshort{LEO} subsystem functionality is limited to amplify and forward the received \acrshort{RF} signal. The second possible configuration, named Double Mirror Synchronization, requires the transmission of a very stable reference signal from the illuminator to the \acrshort{LEO}-satellites by using a dedicated low-gain antenna. Even though this method increases the synchronization processing considerably and requires more complex \acrshort{LEO}-satellites, it still assumes that the distance between the transmitter and receiver satellites is almost constant, and it can be accurately estimated \cite{Krieger2018}.
\end{enumerate}

\subsection{Summary and Lessons Learnt}
\begin{itemize}
    \item Not many publications consider \acrshort{CDSS} for communications applications. However, there are some simulations for \acrshort{LEO} \acrshort{DSS}, synchronized with the aid of a \acrshort{GEO} spacecraft. In addition, the two-way time and range synchronization method has been considered for the synchronization of a \acrshort{DSS} of four satellites in a formation flying mission.
    \item Remote sensing and Earth observation from \acrshort{DSS} has become attractive for the community in the last years. However, the frequency, phase, and time synchronization is still the major challenge that slows down the launch of new missions. For that reason, new synchronization methods for Remote Sensing \acrshort{DSS} are published constantly.
    \item Time and phase synchronization by master-slave architecture using intra-node communications are the preferred methods for remote sensing \acrshort{DSS}. However, the complexity of the system increases with the number of distributed nodes to be synchronized due to the increment of \acrshort{ISL} required.
    \item Information about the synchronization algorithms for future remote sensing distributed satellites missions was discussed in this section, specifically ARGOS, LuTan-1, and ConGaLSAR missions.
\end{itemize}

\section{Examples of Synchronization for Distributed Satellite Systems}
\label{section::Examples}

The development of new synchronization techniques in \acrshort{DSS} as in any other wireless network requires three main steps: design, prototyping, and deployment. The new synchronization method is theoretically analyzed, tested, and improved using software simulators during the design step. However, no matter how thorough the design step, new algorithms should comply with the prototyping before deployment. Theoretical design - software simulation - prototyping is a closed-loop that must be executed repeatedly to improve a synchronization technique before its deployment in \acrshort{DSS}. Theoretical design and simulation of synchronization algorithms suitable for \acrshort{DSS} were already covered in previous sections. This section summarizes the most relevant hardware implementations and prototyping of such algorithms.

The most relevant and advanced examples of synchronization techniques for cohesive distributed systems are analyzed in the following. Unlike the previous section, where concepts and theoretical methods were discussed, this section presents synchronization algorithms that have been implemented and tested at least as a hardware \acrshort{PoC}. Similar to the previous section, the examples are classified according to their application as Synchronization Examples in Distributed Communications Systems and Synchronization Examples in Remote Sensing \acrshort{DSS}.     

\subsection{Synchronization Examples in Distributed Communications Systems}
\label{section::Comm_examples}

Hardware development for communication satellite systems has been pointing toward reconfigurable \acrshort{SDR} System-on-a-Chip ground receivers, and to the ultimate extreme of Satellite-on-a-chip during the last years, \cite{Kodheli2020}. In general, recent trends on prototyping and deployment of spacecraft are based on \acrshort{COTS} \cite{Lovascio2019}, \acrshort{FPGA} \cite{Yu2022} and \acrshort{SDR} \cite{Maheshwarappa2015} designs. Mostly due to its reconfigurable characteristics and because they allow less expensive and faster development processes. Several examples of the use of \acrshort{SDR} for inter-satellite communication in small satellite systems can be found in \cite{Radhakrishnan2016}.

However, there are not many publications about the hardware prototypes of synchronization algorithms, specifically in Communications \acrshort{CDSS}. Mainly because the use of \acrshort{CDSS} for communications is a very recent topic that does not have any launched missions yet. Nevertheless, it is helpful to analyze the \acrshort{PoC} of the synchronization algorithms used in distributed terrestrial communications that \acrshort{DSS} could apply. This can support the selection of the synchronization technique and the hardware platform during the design of a prototype for Communications \acrshort{CDSS}. The examples are presented in order of increasing complexity, and some technical details are listed in Table \ref{tab::summary_synch_examples} at the end of the section.

\subsubsection{Mobile Communications Networks}

One example of distributed coherent radio system that shows some similarities with the \acrshort{DSS} is the recently proposed \acrshort{CF} massive \acrshort{MIMO} concept. In \acrshort{CF} operation, the network infrastructure performs beamforming towards the \acrshort{UE} using multiple distributed radio nodes. The beamforming coordination relies on the \acrshort{TDD} operation and uses a densely distributed network topology \cite{Interdonato2019,9120654}. Although there is a \acrshort{CPU} in \acrshort{CF} Massive \acrshort{MIMO} system, the exchange of information happens only in the form of payload data while the power control coefficients change slowly. Neither is there a provision of sharing the instantaneous \acrshort{CSI} among the \acrshort{AP} or the \acrshort{CPU}. Channel estimation is performed only at the \acrshort{AP} using uplink pilot signals and further used for decoding uplink symbols and precoding the downlink symbols. \cite{363e19c89514405e8e37ba23474c2b64}.

A crucial requirement of the \acrshort{CF} Massive \acrshort{MIMO} is the global synchronization of \acrshort{RF} carriers and information symbols across the geographically distributed service nodes network. In \cite{Jeong2020}, \acrshort{CFO} estimation for Distributed Massive \acrshort{MIMO} and \acrshort{CF} Massive \acrshort{MIMO} is studied. Specifically, the article evaluates the performance of distributed large-scale \acrshort{MIMO} systems with two collaborative (successive and joint), and one non-collaborative (independent) \acrshort{CFO} estimation techniques \cite{Jeong2020}. According to \cite{Borg2018}, the global synchronization of \acrshort{RF} carriers and information symbols across a distributed \acrshort{MIMO} network can be done with a common 1~pps (pulse per second) \acrshort{GPS} timing signal used to provide \acrshort{GPS} disciplined oscillators \cite{Borg2018}. This solution was demonstrated for two 50~MHz carriers tracked by frequency lock loops implemented on \acrshort{FPGA} development platforms. Results showed a locking time of 500~s and a coherence time of 10~s, which is enough for the \acrshort{MIMO} processing \cite{Borg2018}.

The \acrshort{TDD} operation implemented in \acrshort{CF} Massive \acrshort{MIMO} is not very appealing for a \acrshort{DSS} used for communication applications due to the long latency of the satellite orbits. However, an extrapolation of the method to \acrshort{FDD} scenarios can be applied. In such a case, more robust synchronization in time and phase is required between the distributed satellite nodes. One example of \acrshort{FDD} standardized technology in mobile networks is the \acrshort{CoMP} transmission and/or reception used in Long Term Evolution Advanced (LTE-A) standard. 

In \acrshort{CoMP} operation, multiple transmission points coordinate to combine the user's information signal from neighboring evolved Node B (eNB) to improve the received signal quality. To this end, \acrshort{CoMP} transmission requires sharing coordination information, usually consisting of user's feedback \acrshort{CSI}, through the backhaul links \cite{Qamar2017}. It can be implemented in four different types depending on the degree of coordination among cells: (1) Coordinated Scheduling and Beamforming and, (2) Joint Processing, both in the downlink; (3) Coordinated Scheduling and, (4) Joint Reception and Processing, both in the uplink. Even though multiple types of \acrshort{CoMP} can be used together \cite{Qamar2017}. For example, in \cite{Gu2018} proposed to switch the \acrshort{CoMP} transmission mode between Coordinated Beamforming and Joint Processing adaptively to maximize the average achievable rate, considering that the last one is more sensitive to phase synchronization errors than Coordinated Beamforming \cite{Gu2018}. 

\acrshort{CoMP} requires the transmission points to be tight time and frequency synchronized. This constraint mostly affects coordination between \acrshort{BS} because some additional solution is needed to provide accurate phase synchronization. Large distances between the \acrshort{BS} result in unavoidable differences in \acrshort{TOA} between the users' signals which in turn lead to inter-symbol interference in \acrshort{OFDM} systems if the cyclic prefix length is exceeded. Moreover, \acrshort{CFO} which are caused by imperfect oscillators, lead to inter-carrier interference and are a problem, particularly in the downlink due to the feedback delay of the measured \acrshort{CSI}. The most extended approach to synchronization in \acrshort{CoMP} at outdoor \acrshort{BS} is to obtain a precise clock reference from \acrshort{GNSS} \cite{Chaloupka2018}. This could be extended to \acrshort{DSS} in \acrshort{LEO} \acrshort{MEO}. However, indoor \acrshort{BS} require to be synchronized over the backhaul network using standard protocols such as \acrshort{PTP} or proprietary solutions like synchronous Ethernet \cite{Jungnickel2013} \cite{Tian2017} \cite{Song2018}. 

In any typical cellular system, the users perform synchronization with a single \acrshort{BS} where the \acrshort{BS} transmits predefined signals to the \acrshort{UE} to adjust their transmit timings so that signals received from many \acrshort{UE}s to the \acrshort{BS} are time-aligned. However, a concern in \acrshort{CoMP} is that the \acrshort{UE} needs to be synchronized to multiple \acrshort{BS} simultaneously. Hence, the previously discussed technique is not going to work in the \acrshort{CoMP} case. Due to this, the usage of \acrshort{CoMP} becomes limited to deployments where the inter-site distances between \acrshort{BS} are small \cite{Hamza2012}. 

Different approaches to the time synchronization problem stated before have been proposed. The most straightforward is to use a larger cyclic prefix, but it reduces the system's throughput. Another widely used approach is adjusting the time advance based on the nearest \acrshort{BS}. However, such a solution is not capable of fully solving the problem since some of the users may not be able to use the \acrshort{CoMP}. In \cite{Hamza2012}, the authors propose a linear combination of samples corresponding to the two consecutively received \acrshort{OFDM} symbols to identify 
multiple transmitted signals. Such a scheme enables the mitigation of Multi-user Interference by applying Successive Interference Cancellation, as also proposed in \cite{Huang2014} and \cite{Zhao2014}. Another approach considers the positions of the serving and interfering \acrshort{BS} and the estimated location of the \acrshort{UE} to timely relate the synchronization signals received from adjacent \acrshort{BS} \cite{Dammann2013}. Several other approaches which consider asynchronous interference mitigation are discussed in \cite{Pilaram2015} and \cite{Hamza2018}. All these methods could be extrapolated to time synchronization in \acrshort{DSS}.

Joint processing of the carrier phase (along-with frequency) is one of the significant challenges in \acrshort{CoMP} as in \acrshort{DSS}. The time difference of arrivals imposes a phase offset on the transmit covariance matrix leading to improper precoding matrix index selection and therefore limiting the promised performance of \acrshort{CoMP} \cite{Iwelski2014}. It is a two-fold problem: (a) The \acrshort{RF}-\acrshort{LO} of the participating \acrshort{BS} in \acrshort{CoMP} are required to be synchronized among them, (b) the \acrshort{LO} of the \acrshort{UE} are also required to be synchronized with the corresponding \acrshort{BS}. Since the phase noise of the oscillator evolves quickly and only with the help of backhaul information, it becomes challenging to establish synchronization between the participating \acrshort{BS}. For achieving an acceptable level of synchronization, the phase noise needs to be very low in the \acrshort{LO}, which makes such deployment cost-inefficient. In another approach, the \acrshort{UE} can also keep tracking the phase noise. The effect of \acrshort{BS} coordination and synchronization on the phase noise estimation at the user receiver was studied in \cite{Khanzadi2013}, where it is derived the data-aided and non-data-aided \acrshort{CRLB} for the phase noise estimation in a \acrshort{CoMP} system. Another approach presented in \cite{Chang2012} addresses the data-aided joint synchronization and channel estimation for the uplink of \acrshort{CoMP} systems based on the \acrshort{MMSE} criterion. This proposal considers an alternating minimization method to simplify the \acrshort{MMSE} cost function such that the multiuser \acrshort{CFO}, sampling timing offset, and channel frequency response can be estimated in an iterative approach more feasible for practical systems. Another method that addresses the estimation and compensation of time and frequency offsets in \acrshort{CoMP} was proposed in \cite{Koivisto2013}. In this case, the authors compare two solutions to the time-frequency offset estimation problem, one in which the \acrshort{UE} is not aware of the transmitting point, and one in which the \acrshort{UE} is provided with assistance information.

\subsubsection{Implementation Examples of Closed-Loop Synchronization Methods}
\begin{itemize}
    \item \textbf{One Bit Feedback}: In \cite{Mudumbai2006}, an \acrshort{FPGA} (Spartan-3) based hardware prototype was developed to demonstrate \acrshort{1BF}. Phase synchronization is achieved \acrshort{OTA} whereas the frequency synchronization is attained via \acrshort{RF} cables, and a frequency distribution source Octoclock \cite{OctoCloc26:online}. The implementation used three single transmitters (beam-formers) and a single antenna receiver. At a feedback rate of 300~ms, convergence within 90\% of the theoretical limit is achieved after 60 iterations. The experiments were performed inside the laboratory, i.e., static environment. Hence, the performance under the realistic time-varying channels was not investigated. Besides, frequency synchronization was also required to be achieved \acrshort{OTA}.  
    
    A hardware prototype of the \acrshort{1BF} method, using \acrshort{USRP} \acrshort{SDR} \cite{USRP2Ett93:online} was presented in \cite{Quitin2012a}. This prototype functionality is "all-wireless", i.e., no wired connections are used to attain frequency and timing synchronization. This experimental setup used three single-antenna transmitters and one single antenna receiver. With a feedback interval of 50~ms, nine times beamforming gain was achieved. However, like the implementation in \cite{Mudumbai2006}, the experiments were performed indoors. Then, the performance of the system against the time-varying channel was not investigated. Besides, only phase synchronization is achieved \acrshort{OTA} whereas frequency synchronization is achieved through coaxial cables and frequency distribution source.
    
    Another prototype described in \cite{MunkyoSeo2008} performs both frequency and phase synchronization using \acrshort{1BF}, which has been derived from \cite{Mudumbai2006}. Individual components of the hardware prototype have been developed by the authors themselves and the beamforming algorithms have been programmed on \acrshort{DSP} cores. At a test frequency of 60~GHz, using three elements distributed phased array, 9.2~dB of distributed beamforming gain is achieved. In contrast to the methods discussed previously, i.e., \cite{Mudumbai2006} and \cite{Quitin2012a}, both frequency and phase synchronization are achieved \acrshort{OTA}. However, the feedback interval of 0.5~ms is excessively high as compared to the other \acrshort{1BF} methods discussed (\cite{Mudumbai2006} and \cite{Quitin2012a}). Besides, the prototype was tested against static channel environments; hence the performance of the system against the time-varying channel was not investigated.
    
    \item \textbf{Full-feedback}: Very few \acrshort{PoC} demonstrators are available for Full-Feedback implementations. For example, in \cite{Scherber2013} a notable range extension by performing outdoor \acrshort{OTA} tests was shown using \acrshort{COTS} handheld radio (based on \acrshort{FPGA} and \acrshort{DSP}) equipment. Also, the \acrshort{KF} is used to predict the offset in frequency and phase, which further assists in tackling the time variation of the channels. Thus making the system perform satisfactorily even when the nodes are mobile. It is shown that as compared to a \acrshort{COTS} radio transmitter (approximate range 2~km), by performing distributed beamforming, the range can be extended up to 6~km using ten such radio nodes. Further extension of range up to 10~km can be achieved by using 30 such radio nodes. In \cite{Bidigare2012}, using the same experimental setup and beamforming method, 0.1 dB distributed beamforming gain was demonstrated using ten \acrshort{COTS} radio nodes. \acrshort{OTA} operation and application of \acrshort{KF} for prediction makes this implementation suitable for \acrshort{DSS} scenarios. 
\end{itemize}

\subsubsection{Implementation Examples of Open-Loop Synchronization Methods}
Among the open-loop algorithms, in \cite{Mghabghab2019} an \acrshort{SDR} implementation (\acrshort{USRP}~X310 \cite{USRPX31043:online}) using transmitter nodes was demonstrated. \acrshort{OTA} tests confirmed reliable frequency lock even at very low power. In another open-loop \acrshort{SDR} implementation (\acrshort{USRP} N210 \cite{USRPN2105:online}), a solution for \acrshort{OFDM} based frames is developed by designing specific preambles to estimate the frequency and timing offset. The system is capable to operate reliably under frequency selective and slow fading channels. \acrshort{OTA} showed performance gains from 2.5~dB to 2.8~dB.

\subsubsection{Implementation Examples of Master-Slave Synchronization Methods}
In \cite{Alemdar}, an \acrshort{SDR} implementation of Master-Slave closed-loop synchronization procedure named as NetBeam was demonstrated. NetBeam involves a network of radios that broadcast information locally and jointly transmit it to a remote receiver. This is equivalent to a Master-Slave closed-loop synchronization. The performance is further enhanced by the creation of a cluster of beamformers based on \acrshort{CSI}. As for the hardware implementation: a dual antenna configuration of \acrshort{USRP}~X310 is used as a receiver, and a single antenna configuration of \acrshort{USRP} B210 is used as transmitters. A significant aspect of this implementation is that the \acrshort{3D} distribution of radio nodes is considered, which is crucial for \acrshort{DSS} (full-dimensional beamforming). Besides, a faster convergence is attained through machine learning techniques.
    
A \acrshort{SDR} (\acrshort{USRP}) implementation for parallel frequency and phase synchronization was demonstrated in \cite{Rahman2012}. In this case, the Master-Slave method was used for frequency synchronization whereas \acrshort{1BF} was used for phase synchronization. This is an all wireless implementation, where no wired medium is used for \acrshort{CSI} feedback or clock/frequency distribution. Convergence time for the beamforming is of the order of several milliseconds. Another example of frequency and phase synchronization by the Master-Slave method is described in \cite{Pooler2018b} using a \acrshort{SDR} (\acrshort{USRP} X310). In this case, coherent pulses are received with more than 90\% of the ideal coherent energy. Similarly, \cite{Schmid2017} provided a comparison between wired and wireless clock distribution to achieve frequency and phase synchronization by Master-Slave. \acrshort{OTA} tests showed a near-ideal 6~dB gain from a two-transmitter system at a distance of 85~m and coherent gain up to 90\% of the ideal signal summation is achieved. 
    
On the other hand, in \cite{Peiffer2016} a \acrshort{SDR} (\acrshort{USRP} N210) implementation of a Master-Slave architecture for phase synchronization is presented. The external frequency source is used for stable clock distribution whereas phase synchronization is achieved through the Master-Slave method. Channel reciprocity is exploited to obtain \acrshort{CSI} without the need for feedback. Besides, the nodes cooperate in disseminating the \acrshort{CSI}. Beamforming gain as close as 90\% to the ideal beamforming gain is achieved.

Besides, in \cite{Yan2018} the authors analyze the wireless carrier frequency and sample timing synchronization by performing an exchange of \acrshort{RF} signals between the master and slave nodes. Further, the proposed solution was implemented in GNU Radio, and wireless tests were performed using Ettus \acrshort{USRP} N210 \acrshort{SDR}. Besides, the estimation of the fractional clock phase was performed using matched filter bank consisting of sixteen fractionally delayed Zadoff-Chu sequences \cite{Frank1962} which were capable of estimating residual timing offsets as small as 1/16 of the sample duration. However, the system did not address propagation delays, and also the system bandwidth was limited to only 1~MHz. The method yields a residual timing precision of approximately 500~ns which exceeds the expectation of \acrshort{CRLB}.   

In addition, authors in \cite{Overdick2017} used a \acrshort{SDR} implementation (using USRP-E312) to perform synchronization, which did not require time-stamps. The observed accuracy of the method was less than 1~$\mu$s (0.8~$\mu$s) for 150~kHz sampling rate with precision limited to 1/10 of the sample rate. The authors reported the limited real-time processing capability of the USRP-E312 as the reason behind using such a low sampling rate. Three E310s were used in the over-the-wire setting: one as the master node, a second as the slave node, and a third as the measurement device for determining the resulting clock offset. 

\subsubsection{Implementation Examples of Consensus-Based Synchronization Methods} 
The \acrshort{PoC} presented in \cite{Alvarez2015} demonstrates consensus-based synchronization through \acrshort{SDR} (\acrshort{USRP} N210) implementation. GNU Radio was used for software-based signal processing. Both time delay and \acrshort{CFO} estimation, as well as tracking, are addressed in the implementation. Convergence time is of the order of seconds when different nodes have different \acrshort{CFO}. \acrshort{OTA} tests showed carrier frequency offsets to be within 100 parts per billion (ppb) while timing offset were correctly estimated. Besides, the implementation also showed tracking capabilities. 
    
Another notable consensus-based implementation can be found in \cite{Yan2018}. For \acrshort{OFDM} type frames, a specific frame structure was designed to facilitate frequency synchronization. A residual timing offset within 1/16 of symbol duration and a residual frequency offset of 5~Hz is achieved. With such residual frequency and timing offset, a near-optimal received signal power gain is shown when distributed beamforming is performed. 

\begin{table*}[]
\centering
\caption{Summary of \acrshort{PoC} hardware implementations of distributed synchronization techniques}
\resizebox{\textwidth}{!}{%
\begin{tabular}{|c|c|c|c|c|c|c|c|}
\hline
\multicolumn{1}{|c|}{} &  \multicolumn{1}{c|}{\textbf{Synch. Method}} & \multicolumn{1}{c|}{\begin{tabular}[c]{@{}c@{}}\textbf{Frequency} \\ \textbf{Synch}\end{tabular}} &  \multicolumn{1}{c|}{\textbf{Phase Synch.}}  & \multicolumn{1}{c|}{\begin{tabular}[c]{@{}c@{}}\textbf{Feedback} \\ \textbf{rate (ms)}\end{tabular}} &  \multicolumn{1}{c|}{\begin{tabular}[c]{@{}c@{}}\textbf{Beamforming Gain /} \\ \textbf{Convergence}\end{tabular}} & \multicolumn{1}{c|}{\begin{tabular}[c]{@{}c@{}}\textbf{Static /} \\ \textbf{Time-Varying}\end{tabular}} & \multicolumn{1}{|c|}{\textbf{Platform}} \\
\hline
\multicolumn{1}{|l|}{\cite{Mudumbai2006}}   & \multicolumn{1}{l|}{Closed-Loop}  & \multicolumn{1}{l|}{Wired}    & \multicolumn{1}{l|}{Wireless} & \multicolumn{1}{l|}{300} & \multicolumn{1}{l|}{90\% of theoretical limit}        & \multicolumn{1}{l|}{Static}       & \multicolumn{1}{l|}{\acrshort{FPGA}} \\ \hline
\multicolumn{1}{|l|}{\cite{Quitin2012a}}     & \multicolumn{1}{l|}{Closed-Loop}  & \multicolumn{1}{l|}{Wired}    & \multicolumn{1}{l|}{Wireless} & \multicolumn{1}{l|}{50}  & \multicolumn{1}{l|}{9x}                              & \multicolumn{1}{l|}{Static}       & \multicolumn{1}{l|}{\acrshort{FPGA}}         \\ \hline
\multicolumn{1}{|l|}{\cite{MunkyoSeo2008}}    & \multicolumn{1}{l|}{Closed-Loop}  & \multicolumn{1}{l|}{Wireless} & \multicolumn{1}{l|}{Wireless} & \multicolumn{1}{l|}{0.5} & \multicolumn{1}{l|}{9.2 dB}                           & \multicolumn{1}{l|}{Static}       & \multicolumn{1}{l|}{\acrshort{DSP}}          \\ \hline
\multicolumn{1}{|l|}{\cite{Scherber2013}}  & \multicolumn{1}{l|}{Closed-Loop}  & \multicolumn{1}{l|}{Wireless} & \multicolumn{1}{l|}{Wireless} & \multicolumn{1}{l|}{NA}  & \multicolumn{1}{l|}{0.1 dB}                           & \multicolumn{1}{l|}{Time varying} & \multicolumn{1}{l|}{\acrshort{FPGA} and \acrshort{DSP}} \\ \hline
\multicolumn{1}{|l|}{\cite{Mghabghab2019}} & \multicolumn{1}{l|}{Open-Loop}    & \multicolumn{1}{l|}{Wireless} & \multicolumn{1}{l|}{Wireless} & \multicolumn{1}{l|}{NA}  & \multicolumn{1}{l|}{2.5 - 2.8 dB}                       & \multicolumn{1}{l|}{Time Varying} & \multicolumn{1}{l|}{\acrshort{FPGA}}         \\ \hline
\multicolumn{1}{|l|}{\cite{Alemdar}}         & \multicolumn{1}{l|}{Master-Slave} & \multicolumn{1}{l|}{Wireless} & \multicolumn{1}{l|}{Wireless} & \multicolumn{1}{l|}{NA}  & \multicolumn{1}{l|}{10x}                             & \multicolumn{1}{l|}{Time Varying} & \multicolumn{1}{l|}{\acrshort{FPGA}}         \\ \hline
\multicolumn{1}{|l|}{\cite{Rahman2012}}       & \multicolumn{1}{l|}{Master-Slave} & \multicolumn{1}{l|}{Wireless} & \multicolumn{1}{l|}{Wireless} & \multicolumn{1}{l|}{NA}  & \multicolumn{1}{l|}{NA}                              & \multicolumn{1}{l|}{Static}       & \multicolumn{1}{l|}{\acrshort{FPGA}}         \\ \hline
\multicolumn{1}{|l|}{\cite{Schmid2017}}       & \multicolumn{1}{l|}{Master-Slave} & \multicolumn{1}{l|}{Wireless} & \multicolumn{1}{l|}{NA}            & \multicolumn{1}{l|}{NA}  & \multicolumn{1}{l|}{90\% of theoretical limit}        & \multicolumn{1}{l|}{Time Varying} & \multicolumn{1}{l|}{\acrshort{COTS}}         \\ \hline
\multicolumn{1}{|l|}{\cite{Peiffer2016}}     & \multicolumn{1}{l|}{Master-Slave} & \multicolumn{1}{l|}{Wireless} & \multicolumn{1}{l|}{Wireless} & \multicolumn{1}{l|}{NA}  & \multicolumn{1}{l|}{90\% of theoretical limit}        & \multicolumn{1}{l|}{Static}       & \multicolumn{1}{l|}{\acrshort{FPGA}}         \\ \hline
\multicolumn{1}{|l|}{\cite{Pooler2018b}}      & \multicolumn{1}{l|}{Master-Slave} & \multicolumn{1}{l|}{NA}            & \multicolumn{1}{l|}{Wireless} & \multicolumn{1}{l|}{NA}  & \multicolumn{1}{l|}{90\% of theoretical limit}        & \multicolumn{1}{l|}{Time Varying} & \multicolumn{1}{l|}{\acrshort{FPGA}}         \\ \hline
\multicolumn{1}{|l|}{\cite{Yan2018}}      & \multicolumn{1}{l|}{Master-Slave} & \multicolumn{1}{l|}{Wireless}            & \multicolumn{1}{l|}{Wireless} & \multicolumn{1}{l|}{NA}  & \multicolumn{1}{l|}{NA}        & \multicolumn{1}{l|}{Static} & \multicolumn{1}{l|}{\acrshort{USRP}}         \\ \hline
\multicolumn{1}{|l|}{\cite{Overdick2017}}      & \multicolumn{1}{l|}{Master-Slave} & \multicolumn{1}{l|}{Wired}            & \multicolumn{1}{l|}{Wired} & \multicolumn{1}{l|}{NA}  & \multicolumn{1}{l|}{NA}        & \multicolumn{1}{l|}{Static} & \multicolumn{1}{l|}{\acrshort{USRP}}         \\ \hline
\multicolumn{1}{|l|}{\cite{Alvarez2015}}      & \multicolumn{1}{l|}{Consensus-Based} & \multicolumn{1}{l|}{Wired}            & \multicolumn{1}{l|}{Wireless} & \multicolumn{1}{l|}{NA}  & \multicolumn{1}{l|}{NA}        & \multicolumn{1}{l|}{Time Varying} & \multicolumn{1}{l|}{\acrshort{FPGA}}         \\ \hline
\end{tabular}%
}
\label{tab::summary_synch_examples}
\end{table*}

\subsection{Synchronization Examples in Remote Sensing Distributed Satellite Systems}
\label{section::remote_sensing_examples}

This section analyzes the synchronization methods used by \acrshort{DSS} missions to perform Earth observation and remote sensing tasks. At the end of the section, Table \ref{tab:summary_EO_RA} summarizes the technical details of these missions.

\subsubsection{\acrshort{GRACE}}
The \acrfull{GRACE} was a mission devoted to monitoring changes of the Earth's gravity field irregularities from its dispatch in March 2002 to the conclusion of its science mission in October 2017 \cite{Tapley2004}. \acrshort{GRACE} comprised of two equal satellites in near‐circular orbits at ~500 km elevation and 89.5$^\circ$ inclination, detached from each other by around 220 km along‐track, and connected by an exceedingly precise inter‐satellite, \acrshort{KBR} system. The satellites were nominally held in a 3-axis stabilized, nearly Earth-pointed orientation, such that the \acrshort{KBR} antennas were pointed accurately at each other. The \acrshort{KBR} gives a micron-level precision (10~$\mu$m) using carrier phase estimations within the K (26 GHz) and Ka (32 GHz) frequencies \cite{Kim2009}. A single horn serves as the K/Ka antenna for both transmitting and receiving the inter-satellite dual-band wave signals \cite{Thomas1999}. Each satellite transmits two sinusoidal signals (at K and Ka bands) with a frequency offset (nominally set to 0.5~MHz). The two 0.5~MHz down-converted \acrshort{RF} signals are sampled at approximately 19 MHz and passed to the digital signal processing part of the receiver. Dedicated digital signal processing channels are used to digitally counter-rotate the phase of each down-converted signal, track phase with a digital phase-locked loop, and extract phase. 

\subsubsection{\acrshort{LISA}}
\acrshort{LISA} stands for Laser Interferometer Space Antenna \cite{LISAPage}. It is a distributed satellite mission formed of three spacecraft operating in formation flying at a 5,000,000 km distance, being the three peaks of an equilateral triangle. This arrangement composes a huge interferometer to monitor the gravitational waves coming from galactic and out of the galaxy sources. Each spacecraft contains two verification masses and laser bars that measure the separation between its masses and those from the other nodes with a required precision of 20~pm. This is the most extreme example of accurate ranging between flying payloads ever seen before. The spacecraft use inertial sensors and micro-newton thrusters to determine and control their orbits, whereas laser interferometry is used to measure the distance with the required accuracy.

\subsubsection{\acrshort{OLFAR}}
The \acrfull{OLFAR} is a space-based low-frequency radio telescope that investigates the universe's so-called dark ages, maps the interstellar medium, and finds planetary and sun-powered bursts in other solar systems. The telescope, which is composed of a swarm of fifty satellites, was sent to an area distant from Earth to maintain a strategic distance from the high Radio Frequency Interference found at frequencies underneath 30 MHz, coming from Earth \cite{Budianu2011OLFAR:Swarms}. The satellites can be maintained in a \acrshort{3D} configuration with a maximum diameter of 100 km \cite{Rajan2015JointNodes} by using round-trip pulse-based synchronization. 

\subsubsection{TanDEM-X}
TanDEM-X was a scientific mission that comprised of two X-band \acrshort{SAR} satellites following an orbit in near arrangement, with variable separation between them between 500~and~1100~m \cite{Fiedler2008TheOverview}. The mission produces high accuracy \acrshort{SAR} snapshots at X-band in both monostatic and bistatic setups. For the bistatic design, one of the two satellites works as a transmitter and the other one as a receiver, and the mission performs a closed-loop synchronization scheme to get the required coherence.

\subsubsection{GRAIL}
The \acrfull{GRAIL} was a NASA mission to outline the gravity field of the Moon to a remarkable level of detail \cite{Enzer2010}. Twin shuttles were propelled on 10 September 2012 and were embedded into lunar orbit on 31 December 2011 and 01 January 2012 \cite{Oudrhiri2014} correspondingly. The instrument for this mission was based on \acrshort{GRACE}. Nevertheless, there were a few contrasts between both missions. The principal difference between the \acrshort{GRAIL} and \acrshort{GRACE} instruments emerged because \acrshort{GRAIL} was not suffering the Earth's atmosphere drag, nor \acrshort{GPS} navigation system was accessible. \acrshort{GRACE} compensated for air disturbance basically by having an accelerometer to measure non-gravitational speed variations and by using two independent microwave ranging frequencies instead of only a single one.

\acrshort{GRAIL} was streamlined by excluding the K-Band frequency (26~GHz) as well as the accelerometer (non-gravitational strengths are small enough to be modeled rather than measured). Furthermore, for \acrshort{GRACE}, \acrshort{GPS} was utilized to estimate the relative delay between the two spacecraft, calibrate on-board ultra-stable oscillators, and track the two satellite orbits. Without \acrshort{GPS} accessible at the Moon, \acrshort{GRAIL} included an extra S-Band (2~GHz) Time Transfer System to supplant the \acrshort{GPS} timing measurements, and an extra X-Band (8~GHz) Radio Science Signal used for Doppler following of the shuttle and ultra-stable oscillator frequency calibrations through the Deep Space Network \cite{Enzer2010}.

\begin{table*}[]
\centering
\caption{Summary of synchronization requirements for Remote Sensing \acrshort{DSS} missions}
\resizebox{\textwidth}{!}{%
\begin{tabular}{|c|c|c|c|c|c|c|c|}
\hline
\multicolumn{1}{|c|}{\textbf{Mission}} & \multicolumn{1}{c|}{\textbf{Application}} & \multicolumn{1}{c|}{\begin{tabular}[c]{@{}c@{}}\textbf{Distance}\\ \textbf{between}\\ \textbf{nodes}\end{tabular}} & \multicolumn{1}{c|}{\begin{tabular}[c]{@{}c@{}}\textbf{Band}\\ \textbf{of}\\ \textbf{operation}\end{tabular}} & \multicolumn{1}{c|}{\begin{tabular}[c]{@{}c@{}}\textbf{Required}\\ \textbf{accuracy}\end{tabular}} &  \multicolumn{1}{c|}{\begin{tabular}[c]{@{}c@{}}\textbf{Number}\\ \textbf{of}\\ \textbf{nodes}\end{tabular}} & \multicolumn{1}{c|}{\begin{tabular}[c]{@{}c@{}}\textbf{Synch.}\\ \textbf{signal}\end{tabular}} & \multicolumn{1}{c|}{\begin{tabular}[c]{@{}c@{}}\textbf{Phase (P)}\\ \textbf{Timing (T)}\\  \textbf{Ranging (R)}\end{tabular}} \\
\hline
\acrshort{GRACE} & Gravity & 220 km & \begin{tabular}[c]{@{}c@{}}K (26 GHz)\\ Ka (32 GHz)\end{tabular} & 10 $\mu$m & 2 & CW & R \\ \hline
\acrshort{LISA} & \begin{tabular}[c]{@{}c@{}}Gravitational\\ waves\end{tabular} & $5\cdot10^6$ km & Optical & 20 pm & 3 & CW & R \\ \hline
\acrshort{OLFAR} & \begin{tabular}[c]{@{}c@{}}Radio\\ telescope\end{tabular} & 100 km & TBD & $<$ 1 m & 10 - 50 & Pulsed & PT \\ \hline
TanDEM-X & \acrshort{SAR} & 0.5 - 1.1 km & X  & $\pm 0.1^\circ$ & 2 & \begin{tabular}[c]{@{}c@{}}Pulsed\\ chirp\end{tabular} & PT \\ \hline
\acrshort{GRAIL} & Lunar gravity & 50 - 225 km & \begin{tabular}[c]{@{}c@{}}S (2 GHz)\\ X (8 GHz)\\ Ka (32 GHz)\end{tabular} & 1 $\mu$m & 2 & CW & PTR \\ \hline
\end{tabular}%
}
\label{tab:summary_EO_RA}
\end{table*}

\subsection{Summary and Lessons Learnt}

\begin{itemize}
    \item Hardware development for communication satellite systems has been pointing toward re-configurable \acrshort{SDR} System-on-a-Chip ground receivers and to the ultimate extreme of Satellite-on-a-chip during the last years. In general, recent trends for prototyping and deployment of spacecraft are based on \acrshort{COTS}, \acrshort{FPGA} and \acrshort{SDR} designs.
    \item Synchronization algorithms from Mobile Networks performing cohesive distributed communications, such as \acrshort{CF} Massive \acrshort{MIMO} and \acrshort{CoMP}, can be extrapolated to synchronize \acrshort{CDSS}.
    \item Most prototypes of synchronization methods for distributed wireless communication networks implemented Master-Slave architecture and \acrshort{FPGA} platforms.
    \item \acrshort{DSS} missions performing Earth observation and remote sensing tasks were discussed in this section. Most of them do not include more than three distributed nodes, and the synchronization is achieved using \acrshort{RF} signals. Missions included in the section are: \acrshort{GRACE}, \acrshort{LISA}, \acrshort{OLFAR}, TanDEM-X and, \acrshort{GRAIL}.
\end{itemize}

\section{Synchronization Using Machine Learning}
\label{section::ML}

\acrshort{ML} techniques can be significantly useful to address various synchronization problems associated with the end-to-end communication system. Some applications of \acrshort{ML} in facilitating the synchronization process include frame synchronization, compensation of the errors caused by sampling frequency/time offsets, carrier synchronization, and characterization of phase noise. While modeling an end-to-end communication system, there may arise synchronization problems between transmitter and receiver due to various reasons including, sampling frequency offset, sampling timing error, and mismatch about the beginning of each frame, i.e., frame header, at the receiver. To address these problems, a \acrshort{CNN} could be promising to build an additional synchronization model in order to achieve better frame synchronization and also to compensate for the impairments caused due to sampling timing error and sampling time offset. To this end, a \acrshort{CNN}-based synchronization model with softmax activation function proposed in \cite{Wu2019DL} has demonstrated 2 dB better detection than the direct correlation detection in terms of correctly detecting the actual position of a frame header. Furthermore, the traditional frame synchronization techniques based on maximum likelihood and correlation may fail due to hardware implementation constraints or frequency deviation problems. To address this, \acrshort{ML} techniques such as multi-instance learning can be used to solve the frame synchronization problem under different frequency ranges without additional modifications \cite{Wang2013learning}.  

Moreover, \acrshort{ML} techniques can be used to accurately characterize amplitude and phase noises, which are essential parameters in the synchronization process of distributed satellite systems. In this regard, in \cite{Zibar2015} a Bayesian filtering-based framework was used in combination with the expectation-maximization to characterize the amplitude and phase noise characterization of the lasers. The carrier synchronization has been experimentally demonstrated using the proposed framework, and the Bayesian filtering has been shown as an efficient method to estimate laser phase noise even in the presence of low \acrshort{SNR}.   

 Due to limited computing power, the complexity of the \acrshort{ML} model, and the increased data volume, a single learning node/machine is generally not able to execute an \acrshort{ML} model, and this will require the distributed implementation of \acrshort{ML} models across several distributed nodes in a network. In distributed learning, a local node computes a data subset's local updates/models. It updates the local updates/model parameters to the centralized server, which then calculates the global parameters \cite{Nguyen2021IoT}. The computed global parameters are then distributed to the local nodes to all distributed nodes. This distributed approach significantly reduces the communication burden required to communicate all the local raw data to the centralized server and utilize the local computational power to generate local models.  
 
 However, the application of \acrshort{ML} techniques in distributed systems may result in relatively low accuracies and poor convergence rate due to differences in the transmission delays and computational capabilities of distributed nodes/clusters. Therefore, it is crucial to have a proper synchronization among different nodes/clusters of distributed systems to enhance the accuracy of the \acrshort{ML} training model and to accelerate the training time. The synchronization cost may result in a significant performance loss of a distributed \acrshort{ML} model \cite{Zhang2018}. In general, to parallelize the data across distributed clusters, \acrshort{ML} techniques utilize \acrshort{BSP} strategy \cite{Gerbessiotis1994}, in which all computational nodes need to commit and receive new global parameters before starting the next iteration, resulting in a load imbalance problem. This load imbalance problem can be addressed with an asynchronous strategy as it enables the distributed learning nodes to utilize local model parameters for the next iteration \cite{DeepRL2016}. The main problem with this asynchronous method is that the model may not provide the accuracy guarantee since the model can be trapped in a local optimum without converging to a globally optimum solution.   Another approach is \acrshort{SSP} strategy \cite{Qirong2013}, in which the modes can utilize the stale global parameters to train the local model; however, this may not guarantee convergence due to the limitations of stale global parameters. A promising approach to address the drawbacks of the methods above is to dynamically adapt the communications method between the centralized/parameter server and the distributed nodes based on the performance of each node. In this regard, the \acrshort{ASP} strategy proposed in \cite{Zhang2018} has been shown to achieve higher convergence speed and provide better accuracy than the \acrshort{SSP} methods. However, the applicability of the \acrshort{ASP} method in a large-size distributed \acrshort{ML} framework remains an open problem. 
  
The emerging cyber-physical systems requiring time synchronization are vulnerable to the threat of time synchronization attacks, which mainly focus on modifying the measurements' sampling time/time stamps without modifying the system measurements. Such time synchronization attacks may lead to serious consequences in cyber-physical systems, such as incorrect voltage stabilization in smart grid networks. The existing attack detection techniques such as residual-based bad data detection and the conventional supervised \acrshort{ML}-based detectors may not be able to effectively detect such attacks, leading to the need for innovative \acrshort{ML}-based solutions \cite{Wang2017}. In this regard, "first difference aware" \acrshort{ML} classifier proposed in \cite{Wang2017} could be promising to detect two types of time synchronization attacks, namely, direct time synchronization, which only modifies some time stamps, and stealth time synchronization attack, which modifies all the timestamps at a certain time. The First Difference \acrshort{ML} (FD\acrshort{ML}) techniques utilize the backward first difference of the time-series data in order to process the input data stream before employing an \acrshort{ML} method. 

For a satellite system with a \acrshort{DRT} working in the \acrshort{FH-FDMA} mode, it is common to use different hopping sequences for the uplink and downlink communications to avoid possible jamming or interference in both the links \cite{Bae2016}. However, the synchronization between \acrshort{DRT} and ground equipment with different hopping sequences becomes very complicated, and it is a crucial challenge to investigate an efficient synchronization method. In this regard, authors in \cite{Lee2019} proposed an \acrshort{ML}-based novel method to carry out synchronization with the Frequency Hopping signal for tactical SatCom system by utilizing serial search for coarse acquisition and \acrshort{LSTM} network for fine acquisition. The main objective of the proposed work is to reduce the synchronization time. It has been shown that the proposed \acrshort{LSTM}-based method enables the fast and quick fine acquisition in comparison to the existing methods in the literature and provides benefits in saving the overall synchronization time by allowing the fine acquisitions for both downlink and uplink and learning the temporal trend of the signal.   

In the context of 5G integrated SatCom utilizing \acrshort{OFDM}, authors in \cite{Tong2020} proposed a Cyclic Prefix based multi-symbol merging blind timing algorithm to enhance the timing accuracy. Also, an improved synchronization method has been submitted to realize more accurate time-frequency error correction in the considered 5G integrated SatCom system. With the help of simulations results, it has been depicted that the proposed multi-symbol merging method provides better performance than the single-symbol method in low \acrshort{SNR} conditions, and the proposed synchronization method can provide a good Bit Error Ratio (BER) performance. 

Due to the availability of massive volumes of high-resolution videos and images taken by \acrshort{LEO} satellites and \acrshort{UAV}, many data sources have become available for \acrshort{SIN}. However, the application of \acrshort{ML} becomes challenging due to limited computing and communication resources and also the low frequency and orbit resources in \acrshort{SIN}. Also, the unstable connections of  \acrshort{SIN} further create the challenges of employing \acrshort{ML} among a swarm of satellites and \acrshort{UAV}. To address this, authors in \cite{ Guo2021} considered the application of distributed \acrshort{ML} in \acrshort{SIN}, called SpaceDML, to effectively reduce the communication overhead among the \acrshort{SIN} devices. The proposed SpaceDML utilizes adaptive loss-aware quantization and partial weight averaging algorithms to compress models without sacrificing their quality and selectively average the active agents' partial model updates, respectively. The evaluation with public dataset and realistic model presented in  \cite{ Guo2021} has demonstrated that the proposed SpaceDML can enhance the model accuracy by about 2-3 \% and reduce the communication burden among \acrshort{SIN} devices can be up to 60 \% as compared to the baseline algorithm.

\subsection{Summary and Lessons Learnt}

\begin{itemize}
\item \acrshort{ML} can be considered as a promising technique to address several 
synchronization problems related to frame synchronization, compensation of the errors caused due to sampling frequency/time offsets, carrier synchronization, and characterization of phase noise in communication systems, including a SatCom network. 
\item For distributed satellite systems, \acrshort{ML} techniques can be used to accurately characterize amplitude and phase noises, which are essential parameters in the synchronization process of a distributed satellite system.
\item Distributed learning techniques find significant importance in distributed satellite systems as they enable the distributed implementation of \acrshort{ML} models across several distributed nodes in a network.
\item  Synchronization cost may result in a significant performance loss in a distributed \acrshort{ML} implementation, and it is imperative to have a proper synchronization among different nodes/clusters of a distributed system to enhance the accuracy of the \acrshort{ML} training model and to accelerate the training time.
\item Time synchronization attack is one important problem to be considered in the emerging cyber-physical systems involving a distributed satellite system as it may lead to severe consequences in the overall system operation. The conventional supervised \acrshort{ML}-based detectors may not be able to effectively detect such attacks, leading to the need for innovative \acrshort{ML}-based solutions such as "first difference aware" \acrshort{ML} classifier proposed in \cite{Wang2017}.
\item Synchronization between \acrshort{DRT} and ground equipment with different hopping sequences in an \acrshort{FH-FDMA}-based satellite system becomes very complicated, and ML-based synchronization techniques such as \acrshort{LSTM} seem to be promising as illustrated in \cite{Lee2019}.
\item \acrshort{ML} techniques could also be promising to enhance timing accuracy in 5G integrated satellite system \cite{Tong2020} and to reduce the communication burden among different distributed \acrshort{SIN} nodes in emerging \acrshort{SIN} with the help of distributed \acrshort{ML} \cite{ Guo2021}.
\end{itemize}

\section{Research Challenges and Opportunities}
\label{section::Challenges}

This section summarizes some of the critical research challenges and opportunities identified while conducting this survey.

\subsection{Synchronization through inter-satellite links in Next Generation GNSS}
\label{section::chall_gnss}

Multiple research results have indicated significant performance improvements on navigation constellations, thanks to the introduction of inter-satellite ranging and communication links \cite{Xue2020}\cite{Sun2019}. At the same time, a consolidation of the technologies enabling this improvement is essential to reduce further the complexity of the inter-satellite ranging and communication payload and simplify the ground-based orbit determination and clock synchronization algorithms. \acrshort{ISL} currently perform navigation and communication functions through independent low-rate telemetry and high-rate data channels, respectively. The integration of both functionalities into one common channel would simplify the onboard equipment and improve the electromagnetic compatibility, which implies the reduction of power consumption and the required frequency resources. 

Current \acrshort{GNSS} employ a network of ground stations to track the satellite clocks and estimate their deviations for a single time scale. The estimated time offsets are included in the navigation messages sent to the final user, which use the information to correct the satellite clock delays in the received signals. This procedure would be simplified if the satellites of the \acrshort{GNSS} could synchronize their carriers without the intervention of the ground segment. The use of optical \acrshort{ISL} could be a viable solution to achieve this goal enabling a novel \acrshort{GNSS} architecture in which, rather than independently maintaining a time scale, each navigation satellite would synchronize the emitted broadcast navigation signals to a common system time \cite{Giorgi2019}\cite{Henkel2020}.

\subsection{Federated Satellite Systems}
\label{section::chall_fss}

To fully achieve the potential that \acrshort{FSS} represents for space systems, several challenges need to be addressed to achieve its successful implementation, and operation \cite{Ruiz-De-Azua2018}. Some of the technical challenges that need to be considered in the development of \acrshort{FSS} are \cite{Ruiz-de-Azua22}:

\begin{itemize}
    \item Coordination/cooperation between heterogenous satellites belonging to the same network. This requires the definition of \textbf{compatible communication standards}, including allocation of the appropriate frequency bands, bandwidths, and the definition of the modulations. Inter-satellite links can be either radio or optical. Both impose different requirements on the spacecraft in terms of pointing accuracy and jitter and have different ranges and transmission speeds.
    \item The different layers of the \textbf{protocol stack} also need to be defined. Routing protocols have to be able to properly manage the fast switches/handoffs and bottlenecks that for polar \acrshort{LEO} satellites typically occur in the polar regions, where the satellite density is larger. Quality of service (QoS) must also be maintained to keep latency bounded. Some works have proposed the use of \acrshort{MANET} routing protocols to autonomously determine the optimum route in a context where all nodes are constantly moving. In the \acrshort{IoSat} context \acrshort{MANET} can provide the self-organization, self-configuration, and flexibility required by \acrshort{FSS} \cite{Ruiz-De-Azua2018}.
    \item Additionally, all the \textbf{backbone technologies} to support \acrshort{FSS} operations so as to interconnect seamlessly in-situ sensors and user terminals from ground to/from the space infrastructure (e.g. \cite{rs13194014}), to become the \acrshort{IoSat} paradigm \cite{DeAzua2018}, will also have to be developed.
    \item Depending on the above requirements and technologies, this can be achieved by \textbf{multi-layer constellations} formed by \acrshort{LEO} satellites at different heights and/or \acrshort{MEO} satellites, and/or possibly \acrshort{GEO} satellites.
    \item Last, but not least, \textbf{privacy and security problems} must be addressed to prevent potentially malicious users from operating and taking control of the \acrshort{FSS} \cite{Shi2020} or preventing third parties from sniffing the information gathered by or transmitted through the satellites forming the \acrshort{FSS}. In this line, the use of distributed keys and blockchain offers a new field of research.
\end{itemize}

\subsection{Distributed Beamforming as an Enabler for Distributed Satellite Systems with Small Satellites}
\label{section::chall_DB_small_sat}

Despite the well-known advantages of small satellites over traditional ones, small satellites present some limitations in mass and volume that lead to restrictions in power consumption and the antennas' location. These constraints could be overcome by their use in \acrshort{DSS} configurations \cite{Querol2020}\cite{Ghasempour2017}. Key challenges to achieving this potential benefit include distributed timing, carrier frequency, and phase synchronization \cite{Radhakrishnan2016}. Besides, the limited kinematic capabilities and the small cross-section of these satellites affect the accurate measurement of inter-satellite range and precise orbit determination and control, making the synchronization tasks even more difficult. However, several research results have proved the advantages of \acrshort{DBF} for distributed wireless networks in terrestrial communications. Results such as the increment of the transmission range \cite{Sriploy2018} and the \acrshort{SNR} \cite{Diao2019}, as well as the experimental demonstration of \acrshort{DBF} by a swarm of \acrshort{UAV} \cite{Mohanti2019} could be extrapolated to \acrshort{DSS}. Future research in the field should focus on the synchronization algorithms to perform \acrshort{DBF} with small satellites systems.

\subsection{Synchronization Algorithms Suitable for Communications Cohesive Distributed Satellite Systems}
\label{section::chall_synch_comm_CDSS}

Generally, closed-loop synchronization methods are not the ideal solution for space-ground or long distances communication links due to the intrinsic delay between the \acrshort{DSS} and the target nodes. In addition, other constraints such as the slow convergence of the iterative bit feedback algorithms \cite{Tseng2014}; the high power consumption and scalability problems of \acrshort{T-RT} \cite{Brown2008} and; the self-interference problem of \acrshort{F-RT} algorithm \cite{Merlano-Duncan2021} makes them not suitable for synchronization in \acrshort{CDSS}. Maybe the most promising solution among the closed-loop methods for \acrshort{CDSS} could be the consensus synchronization algorithm proposed in \cite{Alvarez2018}. However, it is still required to analyze the effect on this algorithm of the transmission delay between the \acrshort{CDSS} and the target node, especially when the design proposed in \cite{Alvarez2018} considered the channel time invariant. 

Unlike the previous group, open-loop methods do not require feedback from the target node. That makes them the preferable algorithms for \acrshort{CDSS} dedicated to communications and the only available solution for remote sensing applications. However, these algorithms require accurate inter-satellite ranging measurement \cite{Ochiai2005}\cite{Barriac2004}. In most remote sensing missions, it is necessary to know precisely the exact position of each spacecraft in the \acrshort{CDSS} for the primary mission goal. For that reason, using this measurement for synchronization purposes does not represent an additional cost. However, implementing these synchronization methods in Communications \acrshort{CDSS} would require including both subsystems: the accurate inter-satellite ranging and the synchronization algorithm, which may incur excessive resource consumption.

Therefore, a synchronization method suitable for Communications \acrshort{CDSS} using small satellites is still missing. The most promising solution seems to be related to the development of open-loop synchronization methods independent of the \acrshort{DSS} geometry. Another possibility is the development of very efficient (in volume and resources consumption) ranging and synchronization methods.

\subsection{Hardware Implementation of Synchronization Algorithms}
\label{section::chall_hardware}

As stated before, one of the most relevant open problems in Synchronization for \acrshort{DSS} is the development of algorithms capable of working in small satellites. To this end, the hardware implementation of these methods has to consider the low power, volume, and computational resources available in these spacecraft. Some general solutions could include distributed computation and wireless power transfer. However, the development of new hardware components for micro and nanosatellites is a critical research field that highly impacts the implementation of the synchronization algorithms for \acrshort{DSS}.

Another approach that has not been taken into account sufficiently is to consider the hardware limitations during the design of the synchronization algorithms. For example, the distortions produced by the power amplification and the phase noise introduced by the local oscillators are some of the hardware impairments that are obviated in most of the synchronization methods proposed. These physical phenomena are treated as random errors during the hardware implementation. However, including their description as part of the design can improve the performance of the synchronization method implemented. 

Finally, there is another practical limitation related to the hardware implementation of \acrshort{F-RT} synchronization algorithms. In this case, the accuracy of the method relies on the assumption that all the synchronization signals exchanged by the distributed nodes are in the same frequency band \cite{Brown}. However, this assumption implies that the nodes should simultaneously transmit and receive signals in the same frequency band, which is not possible in practice due to self-interference. Assigning different frequencies affects the method's accuracy; performing time division multiple access introduces delays and scalability problems. The feasible solution to this problem can be the use of In-band Full Duplex techniques, as suggested in \cite{Querol2020}.  

\subsection{Efficient Inter-Satellite Ranging Methods}
\label{section::chall_ranging}

The methods for Inter-Satellite Ranging are still an open research topic. Advanced \acrshort{DSS} will require an accurate inter-satellite ranging and baseline determination for sensing and communication applications. These accuracy requirements become ever stringent when no external or ground-based aids are used for synchronization. 

The required accuracy also depends on the final mission goals. For instance, Remote Sensing \acrshort{DSS} missions without any feedback from external aids, as is the case of a spaceborne distributed radiometer or radio-telescopes, will require a stringent level of accuracy because the image quality will depend directly on the accuracy of the inter-satellite ranging \cite{Lan2018}. The most promising advances in precise ranging methods for Remote Sensing missions point to using inter-satellite optical links as the most accurate alternatives \cite{Ming2020}. However, solving the integer ambiguity problem remains an open research question for ranging methods based on optical or \acrshort{RF} technologies either for communications or remote sensing applications \cite{Yue2017}. 

On the other hand, Communications \acrshort{DSS} require more efficient algorithms in terms of resource consumption. In this case, one possible solution could be the implementation of methods to simultaneously perform range measurements and communications over \acrshort{ISL} in a similar way to the solutions proposed in \cite{Xue2020}, and \cite{Calvo2020}.

\subsection{Machine Learning-Enabled Synchronization}
\label{section::chall_ML}
%\cite{Dreifuerst2020} Deep Learning-based Carrier Frequency Offset Estimation with One-Bit ADCs

As described in section \ref{section::ML},  \acrshort{ML} techniques could be promising in distributed satellite systems for various synchronization aspects such as frame synchronization carrier synchronization, characterization of phase noise, and the compensation of errors caused due to time/frequency offsets. The estimation of carrier frequency offset with low-resolution \acrshort{ADC} is necessary for the synchronization process as low-resolution \acrshort{ADC} can enable the operation of advanced \acrshort{MIMO} operations with the help of full digital architectures. However, this becomes challenging at the higher carrier frequencies due to the higher channel bandwidth. To address this problem, \acrshort{ML} techniques based on various deep Neural Network (NN) architectures could be used in finding the best estimate of the involved nonlinear function towards performing carrier frequency offset estimation \cite{Dreifuerst2020}. Some of the critical design aspects to be considered in employing \acrshort{ML}-based carrier frequency offset estimation include the sensitivity to noise, length of the training sequence, and computational complexity of the \acrshort{ML} model. Also, the convergence rate and accuracy of the \acrshort{ML} models should be carefully considered while employing \acrshort{ML} techniques in distributed satellite systems due to differences in the propagation delays and heterogeneous capabilities of the distributed nodes. 

Due to the distributed nature of the considered satellite applications, analyzing the feasibility of distributed \acrshort{ML} techniques such as federated learning in distributed satellite systems could be a promising future research direction. One of the main problems to be considered while employing \acrshort{ML} in distributed satellite systems is the synchronization cost. Although some synchronization techniques such as \acrshort{BSP} \cite{Gerbessiotis1994}, \acrshort{SSP} \cite{Qirong2013}, and \acrshort{ASP} \cite{Zhang2018}  have been proposed in the literature to deal with the performance loss caused due to distributed \acrshort{ML}, application of distributed \acrshort{ML} in large-scale distributed satellite systems remains an open problem. Moreover, another future research direction is to investigate suitable \acrshort{ML} techniques to address time synchronization attacks \cite{Wang2017} in distributed satellite systems caused by the modification of time/time stamps of the underlying measurements.  

\section{Conclusions}
\label{section::Concl}

\acrshort{CDSS} is a core technological paradigm for future remote sensing and communication missions. However, achieving effective and accurate carrier synchronization in these systems is still an open challenge. For that reason, many researchers are currently working on more precise synchronization methods for \acrshort{DSS}. In addition, satellite systems have significant time and resource constraints, especially for small satellites such as the spacecraft that are envisioned to be part of the future \acrshort{DSS}. This implies an additional challenge for the scientific community working on the topic.

In this context, this article has captured the latest advances in architectures and synchronization methods for \acrshort{DSS}. A brief survey of the \acrshort{DSS} architectures was provided, classifying them into five general groups: Constellations, Clusters, Swarms, Fractionated, and Federated spacecraft. Other parameters considered for the classification of \acrshort{DSS} were the \acrshort{ISL}: Ring, Star, Mesh, or Hybrid topology and their synchronization scheme. Generally, Distributed synchronization algorithms are more robust than centralized synchronization algorithms but more complex.

The distributed time, phase, and frequency wireless synchronization methods reported in the literature were summarized and compared, analyzing their feasibility for \acrshort{DSS}. The synchronization algorithms were classified as Closed-loop or Open-loop methods based on the use of
feedback from a node external to the \acrshort{DSS}. In addition, another classification considered the communication between the elements of the \acrshort{DSS} as Closed-loop, when the exchange of information among the distributed satellites was done as a two-way message exchange or;
Open-loop when it was done as a broadcast or one-way communication. 

The time synchronization of \acrshort{DSS} is mainly based on the TWTT algorithm. However, recent publications refered to the use of pseudo-random noise code and other techniques to achieve time synchronization and inter-satellite ranging simultaneously. Among the Closed-loop synchronization algorithms that use the feedback from a node external to the \acrshort{DSS}, the rich feedback methods are more suitable to implement in \acrshort{DSS}. Specifically, the \acrshort{PA} algorithm for \acrshort{DTB} and the reciprocity-based methods are the most recommended. On the other hand, most of the Open-loop synchronization algorithms are suitable for the synchronization of \acrshort{DSS}. As a conclusion, higher synchronization accuracy could be achieved by combining the use of intra-node communication in the form of two-way message exchange and the feedback from a node external to the \acrshort{DSS}. 

Other operations, closely related to synchronization in \acrshort{DSS} such as inter-satellite ranging and relative positioning were also analyzed. Generally, the requirements and accuracy of the coherent operation depend on the performance of the ranging and relative positioning algorithms as much as the synchronization itself. These operations can be performed by optical or \acrshort{RF} inter-satellite ranging methods. The former one can achieve higher accuracy, but the higher directivity of the laser beam can represent a limitation for specific applications. The use of a single \acrshort{ISL} to simultaneously perform ranging and communication and the solution to the phase integer ambiguity problem are trending topics in this field.

An extensive compilation of missions and proof of concept implementations have been included. Besides, the use of \acrshort{ML} as a promising technique to address several synchronization problems was considered. For instance, frame synchronization, compensation of the errors caused due to sampling frequency/time offsets, carrier synchronization, and characterization of phase noise in communication systems, including a SatCom network. Distributed learning techniques as enablers of the distributed implementation of \acrshort{ML} models across several distributed nodes in \acrshort{DSS} was analyzed. Finally, a collection of current research activities and potential research topics was proposed, identifying problems and open challenges that can be useful for researchers in the field.

\bibliographystyle{ieeetr}
\bibliography{DSS_survey}

\begin{thebibliography}{100}

\bibitem{Graziano2013}
M.~D. Graziano, ``{Overview of Distributed Missions},'' in {\em Distributed
  Space Missions for Earth System Monitoring}, pp.~375--386, New York, NY:
  Springer New York, 2013.

\bibitem{Radhakrishnan2016}
R.~Radhakrishnan, W.~Edmonson, F.~Afghah, R.~Rodriguez-Osorio, F.~Pinto, and
  S.~Burleigh, ``{Survey of Inter-satellite Communication for Small Satellite
  Systems: Physical Layer to Network Layer View},'' {\em IEEE Communications
  Surveys and Tutorials}, vol.~18, pp.~2442--2473, sep 2016.

\bibitem{Savazzi}
P.~Savazzi and A.~Vizziello, ``{Carrier Synchronization in Distributed MIMO
  Satellite Links},''

\bibitem{Barton2014}
R.~J. Barton, ``{Distributed MIMO Communication Using Small Satellite
  Constellations},'' in {\em IEEE International Conference on Wireless for
  Space and Extreme Environments (WiSEE)}, IEEE, 2014.

\bibitem{Chen2020}
Y.~Chen, D.~Liang, H.~Yue, D.~Liu, X.~Wu, H.~Zhang, Y.~Jiao, K.~Liu, and
  R.~Wang, ``{Implementation of a Phase Synchronization Scheme Based on Pulsed
  Signal at Carrier Frequency for},'' {\em Sensors}, pp.~1--14, 2020.

\bibitem{Merlano-Duncan2021}
J.~C. Merlano-Duncan, L.~Martinez-Marrero, J.~Querol, S.~Kumar, A.~Camps,
  S.~Chatzinotas, and B.~Ottersten, ``{A Remote Carrier Synchronization
  Technique for Coherent Distributed Remote Sensing Systems},'' {\em IEEE
  Journal of Selected Topics in Applied Earth Observations and Remote Sensing},
  vol.~14, pp.~1909--1922, 2021.

\bibitem{Gun2009}
L.~Gun and H.~Feijiang, ``{Precise two way time synchronization for distributed
  satellite system},'' in {\em 2009 IEEE International Frequency Control
  Symposium Joint with the 22nd European Frequency and Time forum},
  pp.~1122--1126, IEEE, apr 2009.

\bibitem{Sundaramoorthy2013}
P.~P. Sundaramoorthy, E.~Gill, and C.~J. Verhoeven, ``{Enhancing ground
  communication of distributed space systems},'' {\em Acta Astronautica},
  vol.~84, pp.~15--23, 2013.

\bibitem{Sundaramoorthy2016}
P.~Sundaramoorthy, E.~Gill, and C.~Verhoeven, ``{Beamforming with spacecraft
  under reduced localization and clock constraints},'' {\em IEEE Transactions
  on Aerospace and Electronic Systems}, vol.~52, no.~3, pp.~1197--1208, 2016.

\bibitem{Nasir2016}
A.~A. Nasir, S.~Durrani, H.~Mehrpouyan, S.~D. Blostein, and R.~A. Kennedy,
  ``{Timing and carrier synchronization in wireless communication systems: a
  survey and classification of research in the last 5 years},'' {\em Eurasip
  Journal on Wireless Communications and Networking}, vol.~2016, no.~1, 2016.

\bibitem{SundararamanClockSurvey}
B.~Sundararaman, U.~Buy, and A.~D. Kshemkalyani, ``{Clock synchronization for
  wireless sensor networks: a survey},''

\bibitem{Newman2008}
R.~Newman and M.~Hammoudeh, ``{Pennies from Heaven: A retrospective on the use
  of wireless sensor networks for planetary exploration},'' {\em Proceedings of
  the 2008 NASA/ESA Conference on Adaptive Hardware and Systems, AHS 2008},
  pp.~263--270, 2008.

\bibitem{Selva2017}
D.~Selva, A.~Golkar, O.~Korobova, I.~L. Cruz, O.~L. de~Weck, and P.~Collopy,
  ``{Distributed Earth Satellite Systems : What Is Needed to Move Forward?},''
  {\em Journal of Aerospace Information Systems}, no.~August, 2017.

\bibitem{Kodheli2020}
O.~Kodheli, E.~Lagunas, N.~Maturo, S.~K. Sharma, B.~Shankar, J.~F.~M. Montoya,
  J.~C.~M. Duncan, D.~Spano, S.~Chatzinotas, S.~Kisseleff, J.~Querol, L.~Lei,
  T.~X. Vu, and G.~Goussetis, ``{Satellite Communications in the New Space Era:
  A Survey and Future Challenges},'' pp.~1--45, 2020.

\bibitem{wwwGlobalstar}
``{OneWeb - eoPortal Directory - Satellite Missions}.''

\bibitem{Maine1995}
K.~Maine, C.~Devieux, and P.~Swan, ``{Overview of IRIDIUM satellite network},''
  {\em Wescon Conference Record}, pp.~483--490, 1995.

\bibitem{wwwOneWeb}
``{OneWeb - eoPortal Directory - Satellite Missions}.''

\bibitem{Tapley2004}
B.~D. Tapley, S.~Bettadpur, M.~Watkins, and C.~Reigber, ``{The gravity recovery
  and climate experiment: Mission overview and early results},'' {\em
  Geophysical Research Letters}, vol.~31, pp.~1--4, may 2004.

\bibitem{Racca2010}
G.~D. Racca and P.~W. McNamara, ``{The LISA Pathfinder Mission},'' {\em Space
  Science Reviews}, vol.~151, pp.~159--181, mar 2010.

\bibitem{wwwPrisma}
``{PRISMA (Prototype) - eoPortal Directory - Satellite Missions}.''

\bibitem{wwwProba3}
``{Technologies / Proba Missions / Space Engineering and Technology / Our
  Activities / ESA}.''

\bibitem{Maurer2016}
E.~Maurer, R.~Kahle, F.~Mrowka, G.~Morfill, A.~Ohndorf, and S.~Zimmermann,
  ``{Operational aspects of the TanDEM-X Science Phase},'' in {\em SpaceOps
  2016 Conference}, (Reston, Virginia), American Institute of Aeronautics and
  Astronautics, may 2016.

\bibitem{wwwQB50}
``{The von Karman Institute for Fluid Dynamics}.''

\bibitem{Engelen2010}
S.~Engelen, C.~Verhoeven, and M.~Bentum, ``{Olfar, A Radio Telescope Based on
  Nano-Satellites in Moon Orbit},'' {\em AIAA/USU Conference on Small
  Satellites}, aug 2010.

\bibitem{Brown2006a}
O.~Brown and P.~Eremenko, ``{Fractionated Space Architectures: A Vision for
  Responsive Space},'' 2006.

\bibitem{Ruiz-De-Azua2020}
J.~A. Ruiz-De-Azua, A.~Calveras, and A.~Camps, ``{A Novel Dissemination
  Protocol to Deploy Opportunistic Services in Federated Satellite Systems},''
  {\em IEEE Access}, vol.~8, pp.~142348 -- 142365, 2020.

\bibitem{Camps2021}
A.~Camps, J.~Munoz-Martin, J.~Ruiz-de Azua, L.~Fernandez, A.~Perez-Portero,
  D.~Llaveria, C.~Herbert, M.~Pablos, A.~Golkar, A.~Gutierrrez, C.~Antonio,
  J.~Bandeiras, J.~Andrade, D.~Cordeiro, S.~Briatore, N.~Garzaniti, F.~Nichele,
  R.~Mozzillo, A.~Piumatti, M.~Cardi, M.~Esposito, B.~C. Dominguez, M.~Pastena,
  G.~Filippazzo, and A.~Reagan, ``{FSSCat Mission Description and First
  Scientific Results of the FMPL-2 Onboard 3CAT-5/A},'' in {\em 2021 IEEE
  International Geoscience and Remote Sensing Symposium IGARSS},
  pp.~1291--1294, 2021.

\bibitem{Prager2020}
S.~Prager, M.~S. Haynes, and M.~Moghaddam, ``{Wireless Subnanosecond RF
  Synchronization for Distributed Ultrawideband Software-Defined Radar
  Networks},'' {\em IEEE Transactions on Microwave Theory and Techniques},
  vol.~68, no.~11, pp.~4787--4804, 2020.

\bibitem{Poincare1898}
H.~Poincar{\'{e}}, ``{La mesure du temps},'' {\em Revue de M{\'{e}}taphysique
  et de Morale}, vol.~6, no.~1, pp.~1--13, 1898.

\bibitem{Einsten1905}
A.~Einsten, ``{Zur elektrodynamik bewegter k{\"{o}}rper},'' {\em Annalen der
  Physik}, vol.~322, no.~10, pp.~891--921, 1905.

\bibitem{Kirchner1991}
D.~Kirchner, ``{Two-Way Time Transfer Via Communication Satellites},'' {\em
  Proceedings of the IEEE}, vol.~79, no.~7, pp.~983--990, 1991.

\bibitem{Shengkang2013}
Z.~Shengkang and Z.~Li, ``{Ultra-Short Term Clock Offset Prediction for Two-
  Way Satellite Time Synchronization},'' in {\em 2013 Joint European Frequency
  and Time Forum and International Frequency Control Symposium (EFTF/IFC)},
  pp.~335--338, IEEE, 2013.

\bibitem{Xiaochun2010}
L.~Xiaochun, ``{Autonomous Time Synchronization Algorithm and Time
  Synchronization Link Performance Analysis in the Satellite Constellation},''
  in {\em 6th International Conference on Wireless Communications Networking
  and Mobile Computing (WiCOM)}, no.~10673011, pp.~15--18, IEEE, 2010.

\bibitem{Jiuling2018}
X.~Jiuling, Z.~Chaojie, W.~Chunhui, and J.~Xiaojun, ``{Approach to
  inter-satellite time synchronization for micro-satellite cluster},'' {\em
  Journal of Systems Engineering and Electronics}, vol.~29, no.~4,
  pp.~805--815, 2018.

\bibitem{Pan2008}
L.~J. Pan, T.~Jiang, L.~Y. Zhou, H.~Xu, and W.~D. Chen, ``{A research on
  high-precision time-synchronization and ranging system between satellites},''
  {\em 2008 International Conference on Microwave and Millimeter Wave
  Technology Proceedings, ICMMT}, vol.~2, no.~1, pp.~926--929, 2008.

\bibitem{Ma2013}
H.~J. Ma, H.~Wu, J.~Wu, M.~Li, K.~Wang, Z.~He, and D.~Zhao, ``{Design and
  implementation of dual one-way precise ranging and time synchronization
  system},'' in {\em 2013 Joint European Frequency and Time Forum and
  International Frequency Control Symposium, EFTF/IFC 2013}, vol.~3,
  pp.~831--834, IEEE, 2013.

\bibitem{Yaowei2021}
Z.~H.~U. Yaowei, X.~U. Zhaobin, J.~I.~N. Xiaojun, G.~U.~O. Xiaoxu, and J.~I.~N.
  Zhonghe, ``{Integrated method for measuring distance and time difference
  between small satellites},'' in {\em Journal of Systems Engineering and
  Electronics}, vol.~32, pp.~596--606, 2021.

\bibitem{Mills2006}
D.~L. Mills, {\em {Computer Network Time Synchronization: The Network Time
  Protocol on Earth and in Space}}.
\newblock second~ed., 2006.

\bibitem{IEEE_Standard}
{\em {IEEE Standard for a Precision Clock Synchronization Protocol for
  Networked Measurement and Control Systems}}.
\newblock IEEE Standard 1588-2008, 2008.

\bibitem{Shi2015}
B.~Shi, D.~Zhang, and J.~Hu, ``{Preliminary Investigation in Wide Area
  Protection Protocol},'' in {\em 2015 IEEE International Symposium on
  Precision Clock Synchronization for Measurement, Control, and Communication
  (ISPCS)}, IEEE, 2015.

\bibitem{Dierikx2016}
E.~F. Dierikx, A.~E. Wallin, T.~Fordell, J.~Myyry, P.~Koponen, M.~Merimaa,
  T.~J. Pinkert, J.~C. Koelemeij, H.~Z. Peek, and R.~Smets, ``{White rabbit
  precision time protocol on long-distance fiber links},'' {\em IEEE
  Transactions on Ultrasonics, Ferroelectrics, and Frequency Control}, vol.~63,
  no.~7, pp.~945--952, 2016.

\bibitem{Noh2007}
K.~L. Noh, Q.~M. Chaudhari, E.~Serpedin, and B.~W. Suter, ``{Novel clock phase
  offset and skew estimation using two-way timing message exchanges for
  wireless sensor networks},'' {\em IEEE Transactions on Communications},
  vol.~55, no.~4, pp.~766--777, 2007.

\bibitem{10.1145/844128.844143}
J.~Elson, L.~Girod, and D.~Estrin, ``Fine-grained network time synchronization
  using reference broadcasts,'' {\em SIGOPS Oper. Syst. Rev.}, vol.~36,
  p.~147–163, Dec. 2003.

\bibitem{Noh2008}
K.~L. Noh, E.~Serpedin, and K.~Qaraqe, ``{A new approach for time
  synchronization in wireless sensor networks: Pairwise broadcast
  synchronization},'' {\em IEEE Transactions on Wireless Communications},
  vol.~7, no.~9, pp.~3318--3322, 2008.

\bibitem{Rucksana2015}
S.~Rucksana, C.~Babu, and S.~Saranyabharathi, ``{Efficient timing-sync protocol
  in wireless sensor network},'' in {\em 2015 International Conference on
  Innovations in Information, Embedded and Communication Systems (ICIIECS)},
  (Coimbatore, India), IEEE, 2015.

\bibitem{Maroti2004}
M.~Mar{\'{o}}ti, B.~Kusy, G.~Simon, and {\'{A}}.~L{\'{e}}deczi, ``{The flooding
  time synchronization protocol},'' in {\em SenSys'04 - Proceedings of the
  Second International Conference on Embedded Networked Sensor Systems},
  no.~January, pp.~39--49, 2004.

\bibitem{Shi2020}
F.~Shi, S.~Member, X.~Tuo, S.~X. Yang, S.~Member, and J.~Lu, ``{Rapid-Flooding
  Time Synchronization for Large-Scale Wireless Sensor Networks},'' {\em IEEE
  Transactions on Industrial Informatics}, vol.~16, no.~3, pp.~1581--1590,
  2020.

\bibitem{Roehr2007}
S.~Roehr, P.~Guldeny, and M.~Vossiek, ``{Method for high precision clock
  synchronization in wireless systems with application to radio navigation},''
  in {\em Proceedings - 2007 IEEE Radio and Wireless Symposium, RWS},
  pp.~551--554, 2007.

\bibitem{Fiedler2008TheOverview}
H.~Fiedler, G.~Krieger, M.~Zink, M.~Younis, M.~Bachmann, S.~Huber, I.~Hajnsek,
  and A.~Moreira, ``{The TanDEM-X mission: an overview},'' in {\em 2008
  International Conference on Radar}, pp.~60--64, IEEE, 9 2008.

\bibitem{Zachariah2017}
D.~Zachariah, S.~Dwivedi, P.~Handel, and P.~Stoica, ``{Scalable and Passive
  Wireless Network Clock Synchronization in LOS Environments},'' {\em IEEE
  Transactions on Wireless Communications}, vol.~16, no.~6, pp.~3536--3546,
  2017.

\bibitem{Segura2015}
M.~Segura, S.~Niranjayan, H.~Hashemi, and A.~F. Molisch, ``{Experimental
  demonstration of nanosecond-accuracy wireless network synchronization},'' in
  {\em IEEE International Conference on Communications}, vol.~2015-Septe,
  pp.~6205--6210, IEEE, 2015.

\bibitem{Denis2006}
B.~Denis, J.~B. Pierrot, and C.~Abou-Rjeily, ``{Joint distributed
  synchronization and positioning in UWB Ad Hoc networks using TOA},'' {\em
  IEEE Transactions on Microwave Theory and Techniques}, vol.~54, no.~4,
  pp.~1896--1910, 2006.

\bibitem{Dongare2017}
A.~Dongare, P.~Lazik, N.~Rajagopal, and A.~Rowe, ``{Pulsar: A wireless
  propagation-aware clock synchronization platform},'' in {\em Proceedings of
  the IEEE Real-Time and Embedded Technology and Applications Symposium, RTAS},
  pp.~283--292, IEEE, 2017.

\bibitem{Knappe2004}
S.~Knappe, V.~Shah, P.~D.~D. Schwindt, L.~Hollberg, and J.~Kitching, ``{A
  microfabricated atomic clock},'' {\em Applied Physics Letters}, vol.~85,
  no.~9, 2004.

\bibitem{Exel2012}
R.~Exel, ``{Receiver design for time-based ranging with IEEE 802.11b
  signals},'' {\em International Journal of Navigation and Observation},
  vol.~2012, 2012.

\bibitem{1986Gardner}
F.~Gardner, ``A bpsk/qpsk timing-error detector for sampled receivers,'' {\em
  IEEE Transactions on Communications}, vol.~34, no.~5, pp.~423--429, 1986.

\bibitem{ELG_Schmith}
T.~Schmidl and D.~Cox, ``Robust frequency and timing synchronization for
  ofdm,'' {\em IEEE Transactions on Communications}, vol.~45, no.~12,
  pp.~1613--1621, 1997.

\bibitem{Jayaprakasam2017}
S.~Jayaprakasam, S.~K.~A. Rahim, and C.~Y. Leow, ``{Distributed and
  Collaborative Beamforming in Wireless Sensor Networks: Classifications,
  Trends, and Research Directions},'' {\em IEEE Communications Surveys and
  Tutorials}, vol.~19, no.~4, pp.~2092--2116, 2017.

\bibitem{Mudumbai2006}
U.~M. {Raghuraman Mudumbai, Ben Wild} and K.~Ramchandran, ``{Distributed
  Beamforming using 1 Bit Feedback: from Concept to Realization},'' in {\em
  Proc. 44th Allerton Conf. Commun., Control, Comp.,}, (Monticello, IL),
  pp.~1020--1027, Springer, 2006.

\bibitem{Thibault2010}
I.~Thibault, G.~E. Corazza, and L.~Deambrogio, ``{Random, deterministic, and
  hybrid algorithms for distributed beamforming},'' in {\em 2010 5th Advanced
  Satellite Multimedia Systems Conference and the 11th Signal Processing for
  Space Communications Workshop}, pp.~221--225, IEEE, sep 2010.

\bibitem{Pun2009}
M.-O. Pun, D.~Brown, and H.~Poor, ``{Opportunistic collaborative beamforming
  with one-bit feedback},'' {\em IEEE Transactions on Wireless Communications},
  vol.~8, pp.~2629--2641, may 2009.

\bibitem{Lee2018}
J.~Lee, S.~Lee, and J.~Park, ``{Fast Phase Synchronization with Clustering and
  One-Bit Feedback for Distributed Beamforming in a Wireless Sensor Network},''
  in {\em 2018 IEEE 87th Vehicular Technology Conference (VTC Spring)},
  pp.~1--4, IEEE, jun 2018.

\bibitem{Song2012}
S.~Song, J.~S. Thompson, P.-J. Chung, and P.~M. Grant, ``{Exploiting Negative
  Feedback Information for One-Bit Feedback Beamforming Algorithm},'' {\em IEEE
  Transactions on Wireless Communications}, vol.~11, pp.~516--525, feb 2012.

\bibitem{Tseng2014}
C.-S. Tseng, J.~Denis, and C.~Lin, ``{On the Robust Design of Adaptive
  Distributed Beamforming for Wireless Sensor/Relay Networks},'' {\em IEEE
  Transactions on Signal Processing}, vol.~62, pp.~3429--3441, jul 2014.

\bibitem{Xie2018}
N.~Xie, K.~Xu, and J.~Chen, ``{Exploiting Cumulative Positive Feedback
  Information for One-Bit Feedback Synchronization Algorithm},'' {\em IEEE
  Transactions on Vehicular Technology}, vol.~67, pp.~5821--5830, jul 2018.

\bibitem{Thibault2013}
I.~Thibault, A.~Faridi, G.~E. Corazza, A.~V. Coralli, and A.~Lozano, ``{Design
  and Analysis of Deterministic Distributed Beamforming Algorithms in the
  Presence of Noise},'' {\em IEEE Transactions on Communications}, vol.~61,
  pp.~1595--1607, apr 2013.

\bibitem{Jeevan2008}
P.~Jeevan, S.~Pollin, A.~Bahai, and P.~P. Varaiya, ``{Pairwise Algorithm for
  Distributed Transmit Beamforming},'' in {\em 2008 IEEE International
  Conference on Communications}, pp.~4245--4249, IEEE, 2008.

\bibitem{Tushar2012}
W.~Tushar, D.~B. Smith, A.~Zhang, T.~A. Lamahewa, and T.~Abhayapala,
  ``{Distributed transmit beamforming: Phase convergence improvement using
  enhanced one-bit feedback},'' in {\em 2012 IEEE Wireless Communications and
  Networking Conference (WCNC)}, pp.~528--532, IEEE, apr 2012.

\bibitem{Kumar2014}
A.~Kumar, R.~Mudumbai, and S.~Dasgupta, ``{Scalable algorithms for joint beam
  and null-forming using distributed antenna arrays},'' in {\em 2014 IEEE
  Global Communications Conference}, pp.~4042--4047, IEEE, dec 2014.

\bibitem{Tu2002}
Y.-S. Tu and G.~Pottie, ``{Coherent cooperative transmission from multiple
  adjacent antennas to a distant stationary antenna through AWGN channels},''
  in {\em Vehicular Technology Conference. IEEE 55th Vehicular Technology
  Conference. VTC Spring 2002 (Cat. No.02CH37367)}, vol.~1, pp.~130--134, IEEE,
  2002.

\bibitem{Brown2012}
D.~R. Brown, P.~Bidigare, S.~Dasgupta, and U.~Madhow, ``{Receiver-coordinated
  zero-forcing distributed transmit nullforming},'' in {\em 2012 IEEE
  Statistical Signal Processing Workshop (SSP)}, pp.~269--272, IEEE, aug 2012.

\bibitem{Brown2015a}
D.~R. Brown, R.~David, and P.~Bidigare, ``{Improving coherence in distributed
  MISO communication systems with local accelerometer measurements},'' in {\em
  2015 49th Annual Conference on Information Sciences and Systems (CISS)},
  pp.~1--6, IEEE, mar 2015.

\bibitem{Goguri2016a}
S.~Goguri, B.~Peiffer, R.~Mudumbai, and S.~Dasgupta, ``{A class of scalable
  feedback algorithms for beam and null-forming from distributed arrays},'' in
  {\em 2016 50th Asilomar Conference on Signals, Systems and Computers},
  pp.~1447--1451, IEEE, nov 2016.

\bibitem{Amor2019}
S.~B. Amor, S.~Affes, F.~Bellili, U.~Vilaipornsawai, L.~Zhang, and P.~Zhu,
  ``{Multi-Node ML Time and Frequency Synchronization for Distributed
  MIMO-Relay Beamforming Over Time-Varying Flat-Fading Channels},'' {\em IEEE
  Transactions on Communications}, vol.~67, pp.~2702--2715, apr 2019.

\bibitem{Brown}
D.~Brown, G.~Prince, and J.~McNeill, ``{A method for carrier frequency and
  phase synchronization of two autonomous cooperative transmitters},'' in {\em
  IEEE 6th Workshop on Signal Processing Advances in Wireless Communications,
  2005.}, pp.~260--264, IEEE, 2005.

\bibitem{Brown2008}
D.~Brown and H.~Poor, ``{Time-Slotted Round-Trip Carrier Synchronization for
  Distributed Beamforming},'' {\em IEEE Transactions on Signal Processing},
  vol.~56, pp.~5630--5643, nov 2008.

\bibitem{Manosas-Caballu2011}
M.~Ma{\~{n}}osas-Caball{\'{u}} and G.~Seco-Granados, ``{Robust time-slotted
  round-trip carrier and timing synchronization for distributed beamforming},''
  in {\em Proceedings of the ... European Signal Processing Conference
  (EUSIPCO).}, (Barcelona, Spain), 2011.

\bibitem{Alvarez2018}
M.~A. Alvarez and U.~Spagnolini, ``{Distributed Time and Carrier Frequency
  Synchronization for Dense Wireless Networks},'' {\em IEEE Transactions on
  Signal and Information Processing over Networks}, vol.~4, pp.~683--696, dec
  2018.

\bibitem{Ochiai2005}
H.~Ochiai, P.~Mitran, H.~Poor, and V.~Tarokh, ``{Collaborative beamforming for
  distributed wireless ad hoc sensor networks},'' {\em IEEE Transactions on
  Signal Processing}, vol.~53, pp.~4110--4124, nov 2005.

\bibitem{Barriac2004}
G.~Barriac, R.~Mudumbai, and U.~Madhow, ``{Distributed Beamforming for
  Information Transfer in Sensor Networks},'' in {\em IPSN 2004 : Third
  International Symposium on Information Processing in Sensor Networks},
  (Berkeley, California, USA), p.~452, Association for Computing Machinery,
  2004.

\bibitem{Rahman2012b}
M.~M. Rahman, S.~Dasgupta, and R.~Mudumbai, ``{A distributed consensus approach
  to synchronization of RF signals},'' in {\em 2012 IEEE Statistical Signal
  Processing Workshop, SSP 2012}, pp.~281--284, IEEE, 2012.

\bibitem{Preuss2011}
R.~D. Preuss and D.~R. Brown, ``{Two-Way Synchronization for Coordinated
  Multicell Retrodirective Downlink Beamforming},'' {\em IEEE Transactions on
  Signal Processing}, vol.~59, pp.~5415--5427, nov 2011.

\bibitem{Bidigare2015}
T.~P. Bidigare, U.~Madhow, D.~R. Brown, R.~Mudumbai, A.~Kumar, B.~Peiffer, and
  S.~Dasgupta, ``{Wideband distributed transmit beamforming using channel
  reciprocity and relative calibration},'' in {\em 2015 49th Asilomar
  Conference on Signals, Systems and Computers}, pp.~271--275, IEEE, nov 2015.

\bibitem{NingXie2013a}
{Ning Xie}, {Xu Bao}, {Hui Wang}, and {Xiaohui Lin}, ``{Fast open-loop
  synchronization for distributed downlink beamforming},'' in {\em 2013 47th
  Annual Conference on Information Sciences and Systems (CISS)}, pp.~1--6,
  IEEE, mar 2013.

\bibitem{Bletsas2010}
A.~Bletsas, A.~Lippman, and J.~Sahalos, ``{Simple, zero-feedback, distributed
  beamforming with unsynchronized carriers},'' {\em IEEE Journal on Selected
  Areas in Communications}, vol.~28, pp.~1046--1054, sep 2010.

\bibitem{Bletsas2011}
A.~Bletsas, A.~Lippman, and J.~N. Sahalos, ``{Zero-Feedback, Collaborative
  Beamforming for Emergency Radio: Asymptotic Analysis},'' {\em Mobile Networks
  and Applications}, vol.~16, pp.~589--599, oct 2011.

\bibitem{RichardBrown2012}
D.~{Richard Brown}, P.~Bidigare, and U.~Madhow, ``{Receiver-coordinated
  distributed transmit beamforming with kinematic tracking},'' in {\em 2012
  IEEE International Conference on Acoustics, Speech and Signal Processing
  (ICASSP)}, pp.~5209--5212, IEEE, mar 2012.

\bibitem{Comberiate2016}
T.~M. Comberiate, K.~S. Zilevu, J.~E. Hodkin, and J.~A. Nanzer, ``{Distributed
  transmit beamforming on mobile platforms using high-accuracy microwave
  wireless positioning},'' vol.~9829, p.~98291S, International Society for
  Optics and Photonics, may 2016.

\bibitem{Hodkin2015a}
J.~E. Hodkin, K.~S. Zilevu, M.~D. Sharp, T.~M. Comberiate, S.~M. Hendrickson,
  M.~J. Fitch, and J.~A. Nanzer, ``{Microwave and millimeter-wave ranging for
  coherent distributed RF systems},'' {\em IEEE Aerospace Conference
  Proceedings}, vol.~2015-June, pp.~1--7, 2015.

\bibitem{Ming2020}
M.~Ming, Y.~Luo, Y.~R. Liang, J.~Y. Zhang, H.~Z. Duan, H.~Yan, Y.~Z. Jiang,
  L.~F. Lu, Q.~Xiao, Z.~Zhou, and H.~C. Yeh, ``{Ultraprecision intersatellite
  laser interferometry},'' {\em International Journal of Extreme
  Manufacturing}, vol.~2, no.~2, 2020.

\bibitem{Ales2014}
F.~Ales, P.~F. Gath, U.~Johann, and C.~Braxmaier, ``{Modeling and simulation of
  a laser ranging interferometer acquisition and guidance algorithm},'' {\em
  Journal of Spacecraft and Rockets}, vol.~51, no.~1, pp.~226--238, 2014.

\bibitem{Bykhovsky2015}
D.~Bykhovsky, D.~Kedar, and S.~Arnon, ``{Fiber-ring delay line for
  high-resolution intersatellite ranging},'' {\em IEEE Photonics Technology
  Letters}, vol.~27, no.~6, pp.~673--676, 2015.

\bibitem{Luo2016}
Y.~Luo, H.~Li, Y.~Liang, H.~Z. Duan, J.~Zhang, and H.~C. Yeh, ``{A preliminary
  prototype of laser frequency stabilization for spaceborne interferometry
  missions},'' in {\em 2016 European Frequency and Time Forum, EFTF 2016},
  IEEE, 2016.

\bibitem{Tian2019}
Y.~Tian, J.~Zhong, X.~Lin, H.~Yang, and D.~Kang, ``{Inter-Satellite Integrated
  Laser Communication/Ranging Link with Feedback-Homodyne Detection and
  Fractional Symbol Ranging},'' in {\em 2019 IEEE International Conference on
  Space Optical Systems and Applications, ICSOS 2019}, 2019.

\bibitem{Calvo2020}
R.~M. Calvo, J.~Poliak, J.~Surof, and R.~Wolf, ``{Evaluation of optical ranging
  and frequency transfer for the Kepler system : PPreliminary laboratory
  tests},'' in {\em 2020 European Navigation Conference, ENC 2020}, pp.~1--9,
  2020.

\bibitem{4072332}
J.~L. Macarthur and A.~S. Posner, ``{Satellite-to-satellite range-rate
  measurement},'' {\em Geoscience and Remote Sensing, IEEE Transactions on},
  vol.~GE-23, no.~4, pp.~517--523, 1985.

\bibitem{Kim2003}
J.~Kim and B.~D. Tapley, ``{Simulation of Dual One-Way Ranging Measurements},''
  {\em Journal of Spacecraft and Rockets}, vol.~40, pp.~419--425, may 2003.

\bibitem{Yang2010}
J.~Yang, Y.~Yang, L.~F. Liang, and L.~Liu, ``{Research on digital phase-locked
  loop about K/Ka-band high precision receiver},'' in {\em Proceedings - 2010
  International Conference on Intelligent System Design and Engineering
  Application, ISDEA 2010}, vol.~2, pp.~185--188, IEEE, 2010.

\bibitem{Alawieh2016}
M.~Alawieh, N.~Hadaschik, N.~Franke, and C.~Mutschler, ``{Inter-satellite
  ranging in the Low Earth Orbit},'' in {\em 2016 10th International Symposium
  on Communication Systems, Networks and Digital Signal Processing, CSNDSP
  2016}, IEEE, 2016.

\bibitem{Xiaoyi2020}
X.~Xiaoyi, W.~Chunhui, and J.~Zhonghe, ``{Design, analysis and optimization of
  random access inter-satellite ranging system},'' {\em Journal of Systems
  Engineering and Electronics}, vol.~31, no.~5, pp.~871--883, 2020.

\bibitem{Wang2012}
L.~Wang, X.~Li, Y.~Yang, and L.~Liu, ``{Research on inter-satellite ranging and
  velocity measurement method based on Doppler frequency measurement},'' in
  {\em 2012 2nd International Conference on Consumer Electronics,
  Communications and Networks, CECNet 2012 - Proceedings}, no.~60901017,
  pp.~3446--3450, IEEE, 2012.

\bibitem{Rui2008}
S.~Rui, X.~Guodong, and L.~Shengchang, ``{Inter-satellite ranging system design
  in formation},'' in {\em 2008 2nd International Symposium on Systems and
  Control in Aerospace and Astronautics, ISSCAA 2008}, pp.~1--5, IEEE, 2008.

\bibitem{Yang2014}
Y.~Yang, Y.~Li, C.~Rizos, A.~G. Dempster, and X.~Yue, ``{Inter-satellite
  ranging augmented GPS relative navigation for satellite formation flying},''
  {\em Journal of Navigation}, vol.~67, no.~3, pp.~437--449, 2014.

\bibitem{Wang2019}
Z.~Wang, H.~Huan, and M.~Wang, ``{Determining the Initial Value of Carrier
  Phase Smoothing Pseudorange by Least Squares Straight Line Fitting
  Technique},'' {\em Proceedings of 2019 IEEE International Conference on
  Artificial Intelligence and Computer Applications, ICAICA 2019},
  pp.~303--307, 2019.

\bibitem{Hatch}
R.~Hatch, ``{The synergism of GPS code and carrier measurements},'' in {\em 3rd
  International Geodetic Symposium on Satellite Doppler Positioning}, (Las
  Cruces, NM), pp.~1213--1231, 1982.

\bibitem{Tang2017}
Y.~Tang, Y.~Wang, and J.~Chen, ``{Impact of finite bandwidth for
  inter-satellite ranging using direct sequence spread spectrum signal},'' in
  {\em 2016 IEEE International Conference on Signal and Image Processing, ICSIP
  2016}, pp.~488--492, IEEE, 2017.

\bibitem{Xue2020}
R.~Xue, T.~Wang, and H.~Tang, ``{A Novel Chip Pulse Employed by Ranging Code
  Based on Simultaneous Transmitting CPM Modulation and PN Ranging in
  Inter-Satellite Links of GNSS},'' {\em IEEE Access}, vol.~8,
  pp.~132860--132870, 2020.

\bibitem{Crisan2018}
A.~M. Crisan, A.~Martian, and D.~Coltuc, ``{Inter-satellite radio frequency
  ranging in a hybrid OFDM communication-metrology system},'' {\em 2018 15th
  Workshop on Positioning, Navigation and Communications, WPNC 2018}, pp.~3--7,
  2018.

\bibitem{Xiang2019a}
B.~Xiang, F.~Yan, Y.~Zhang, F.~Shen, W.~Xia, and L.~Shen, ``{A New
  Inter-Satellite Ranging Method Based on Pseudo-Range and Dual-Frequency
  Carrier Phase},'' in {\em 2019 11th International Conference on Wireless
  Communications and Signal Processing (WCSP)}, IEEE, 2019.

\bibitem{Xiang2019}
B.~Xiang, F.~Yan, F.~Shen, W.~Xia, S.~Xing, Y.~Wu, and L.~Shen, ``{Selection of
  frequency pairs and accuracy analysis in inter-satellite carrier ranging},''
  in {\em 2019 IEEE Globecom Workshops, GC Wkshps 2019 - Proceedings},
  pp.~0--4, 2019.

\bibitem{Sun2012}
R.~Sun, J.~Guo, and E.~Gill, ``{Antenna array based line-of-sight estimation
  using a GNSS-like inter-satellite ranging system},'' in {\em 6th ESA Workshop
  on Satellite Navigation Technologies: Multi-GNSS Navigation Technologies
  Galileo's Here, NAVITEC 2012 and European Workshop on GNSS Signals and Signal
  Processing}, IEEE, 2012.

\bibitem{Yue2017}
Z.~Yue, B.~Lian, and C.~Tang, ``{GPS/INS integrated navigation algorithm based
  on the double-difference of time/inter-satellite with carrier phase},'' in
  {\em 2017 IEEE International Conference on Signal Processing, Communications
  and Computing, ICSPCC 2017}, vol.~2017-Janua, pp.~1--6, 2017.

\bibitem{Teunissen1995}
P.~J. Teunissen, ``{The least-squares ambiguity decorrelation adjustment: a
  method for fast GPS integer ambiguity estimation},'' {\em Journal of
  Geodesy}, vol.~70, pp.~65--82, 1995.

\bibitem{Wendel2006}
J.~Wendel, O.~Meister, R.~M{\"{o}}nikes, and G.~F. Trommer, ``{Time-differenced
  carrier phase measurements for tightly coupled GPS/INS integration},'' in
  {\em Record - IEEE PLANS, Position Location and Navigation Symposium},
  vol.~2006, pp.~54--60, 2006.

\bibitem{Kroes2006}
R.~Kroes, {\em {Precise Relative Positioning of Formation Flying Spacecraft
  using GPS}}.
\newblock Doctoral thesis, 2006.

\bibitem{DAmico2009}
O.~{D'Amico Simone and Montenbruck}, ``{Differential GPS: An Enabling
  Technology for Formation Flying Satellites},'' in {\em Small Satellite
  Missions for Earth Observation} (H.-P. {Sandau Rainer and Roeser} and
  V.~Arnoldo, eds.), (Berlin, Heidelberg), pp.~457--465, Springer Berlin
  Heidelberg, 2010.

\bibitem{Kroes2005}
R.~Kroes, O.~Montenbruck, W.~Bertiger, and P.~Visser, ``{Precise GRACE baseline
  determination using GPS},'' {\em GPS Solutions}, vol.~9, pp.~21--31, apr
  2005.

\bibitem{Renga2013}
A.~Renga, M.~Grassi, and U.~Tancredi, ``{Relative Navigation in LEO by
  Carrier-Phase Differential GPS with Intersatellite Ranging Augmentation},''
  {\em International Journal of Aerospace Engineering}, vol.~2013, pp.~1--11,
  2013.

\bibitem{Nadarajah2016}
P.~J.~G. {Nadarajah Nandakumaran and Teunissen} and V.~Sandra, ``{Attitude
  Determination and Relative Positioning for LEO Satellites Using Arrays of
  GNSS Sensors},'' in {\em IAG 150 Years} (P.~{Rizos Chris and Willis}, ed.),
  (Cham), pp.~743--749, Springer International Publishing, 2016.

\bibitem{Teunissen2007}
P.~Teunissen, ``{A General Multivariate Formulation of the Multi-Antenna GNSS
  Attitude Determination Problem},'' {\em Artificial Satellites}, vol.~42, jan
  2007.

\bibitem{Cheung2017}
K.~M. Cheung and C.~Lee, ``{A trilateration scheme for relative positioning},''
  {\em IEEE Aerospace Conference Proceedings}, pp.~1--10, 2017.

\bibitem{Casti2019}
M.~Casti, S.~Fineschi, A.~Bemporad, G.~Capobianco, F.~Landini, D.~Loreggia,
  V.~Noce, M.~Romoli, C.~Thizy, and D.~Galano, ``{Fine positioning algorithms
  for the ESA/PROBA-3 formation flying mission},'' {\em 2019 IEEE International
  Workshop on Metrology for AeroSpace, MetroAeroSpace 2019 - Proceedings},
  pp.~121--125, 2019.

\bibitem{Capobianco2019}
G.~Capobianco, S.~Fineschi, D.~Loreggia, A.~Bemporad, F.~Landini, M.~Casti,
  V.~Noce, M.~Romoli, D.~Galano, and C.~Thizy, ``{The in-flight calibration
  procedures of the shadow position sensors (SPS), a very accurate optical
  metrology system of the ESA/PROBA-3 formation flying mission},'' in {\em 2019
  IEEE International Workshop on Metrology for AeroSpace, MetroAeroSpace 2019 -
  Proceedings}, pp.~479--483, 2019.

\bibitem{Buinhas2016}
L.~Buinhas, E.~Ferrer-Gil, and R.~Forstner, ``{IRASSI: InfraRed astronomy
  satellite swarm interferometry - Mission concept and description},'' in {\em
  IEEE Aerospace Conference Proceedings}, vol.~2016-June, 2016.

\bibitem{Buinhas2018}
L.~Buinhas, M.~Philips-Blum, K.~Frankl, T.~Pany, B.~Eissfeller, and
  R.~Forstner, ``{Formation operations and navigation concept overview for the
  IRASSI space interferometer},'' {\em IEEE Aerospace Conference Proceedings},
  vol.~2018-March, pp.~1--16, 2018.

\bibitem{Linz2020}
H.~Linz, D.~Bhatia, L.~Buinhas, M.~Lezius, E.~Ferrer, R.~F{\"{o}}rstner,
  K.~Frankl, M.~Philips-Blum, M.~Steen, U.~Bestmann, W.~H{\"{a}}nsel,
  R.~Holzwarth, O.~Krause, and T.~Pany, ``{InfraRed Astronomy Satellite Swarm
  Interferometry (IRASSI): Overview and study results},'' {\em Advances in
  Space Research}, vol.~65, no.~2, pp.~831--849, 2020.

\bibitem{Weiqing2015}
M.~Weiqing, R.~Liu, Y.~Xinxin, and E.~Kamel, ``{Analysis of precision
  estimation of RF metrology in satellite formation flying},'' in {\em 2015
  International Conference on Wireless Communications and Signal Processing,
  WCSP 2015}, pp.~0--4, IEEE, 2015.

\bibitem{Carrillo2016}
L.~R.~G. Carrillo, F.~M. Palacios, E.~S.~E. Quesada, and K.~Alexis, ``{Adaptive
  high order sliding mode control for relative positioning and trajectory
  tracking of spacecraft formation flying},'' {\em 24th Mediterranean
  Conference on Control and Automation, MED 2016}, pp.~1095--1101, 2016.

\bibitem{Dan2006}
X.~Dan, C.~Xibin, and W.~Yunhua, ``{Decentralized determination of relative
  orbit for formation flying satellite},'' in {\em 1st International Symposium
  on Systems and Control in Aerospace and Astronautics}, vol.~2006,
  pp.~338--343, 2006.

\bibitem{Wang2018b}
S.~Wang and P.~Cui, ``{Autonomous orbit determination using pulsars and
  inter-satellite ranging for Mars orbiters},'' in {\em 2018 IEEE Aerospace
  Conference}, (Big Sky, MT, USA), 2018.

\bibitem{Liu2019}
Z.~Liu, L.~Du, Y.~Zhu, Z.~Qian, J.~Wang, and S.~Liang, ``{Investigation on GEO
  satellite orbit determination based on CEI measurements of short
  baselines},'' {\em The Journal of Navigation}, vol.~72, no.~6, pp.~1585 --
  1601, 2019.

\bibitem{Sadeghi2019}
M.~Sadeghi, F.~Behnia, and H.~Haghshenas, ``{Positioning of Geostationary
  Satellite by Radio Interferometry},'' {\em IEEE Transactions on Aerospace and
  Electronic Systems}, vol.~55, no.~2, pp.~903--917, 2019.

\bibitem{Michalak2020}
G.~Michalak, K.~H. Neumayer, and R.~Konig, ``{Precise orbit determination of
  the kepler navigation system - A simulation study},'' in {\em 2020 European
  Navigation Conference, ENC 2020}, pp.~1--10, 2020.

\bibitem{Chen2019}
L.~Chen, H.~Lin, Z.~Lu, J.~Li, and G.~Ou, ``{High Orbital Spacecraft Fast
  Positioning Algorithm Assisted by Inter-Satellite Links},'' in {\em 2019 2nd
  International Conference on Information Systems and Computer Aided Education,
  ICISCAE 2019}, pp.~598--602, 2019.

\bibitem{Krieger2018}
G.~Krieger, M.~Zonno, J.~Mittermayer, A.~Moreira, S.~Huber, and
  M.~Rodriguez-Cassola, ``{MirrorSAR: A fractionated space transponder concept
  for the implementation of low-cost multistatic SAR missions},'' in {\em
  Proceedings of the European Conference on Synthetic Aperture Radar, EUSAR},
  vol.~2018-June, pp.~1359--1364, 2018.

\bibitem{Xiao2020}
P.~Xiao, B.~Liu, and W.~Guo, ``{ConGaLSAR: A Constellation of Geostationary and
  Low Earth Orbit Synthetic Aperture Radar},'' {\em IEEE Geoscience and Remote
  Sensing Letters}, vol.~17, no.~12, pp.~2085--2089, 2020.

\bibitem{Querol2020}
J.~C. Merlano-Duncan, J.~Querol, L.~Martinez-Marrero, J.~Krivochiza, A.~Camps,
  S.~Chatzinotas, and B.~Ottersten, ``{SDR Implementation of a Testbed for
  Synchronization of Coherent Distributed Remote Sensing Systems},'' in {\em
  IGARSS 2020 - 2020 IEEE International Geoscience and Remote Sensing
  Symposium}, (Waikoloa, HI, USA), pp.~6588--6591, IEEE, 2020.

\bibitem{Ubolkosold2005}
P.~Ubolkosold, S.~Knedlik, and O.~Loffeld, ``{Clock synchronization protocol
  for distributed satellite networks},'' in {\em Proceedings. 2005 IEEE
  International Geoscience and Remote Sensing Symposium, 2005. IGARSS '05.},
  vol.~1, (Seoul, South Korea), pp.~681--684, IEEE, 2005.

\bibitem{Zhang2006}
Y.~Zhang, D.~Liang, and Z.~Dong, ``{Analysis of time and frequency
  synchronization errors in spaceborne parasitic InSAR system},'' in {\em
  International Geoscience and Remote Sensing Symposium (IGARSS)},
  pp.~3047--3050, IEEE, 2006.

\bibitem{Younis2006}
M.~Younis, R.~Metzig, and G.~Krieger, ``{Performance Prediction of a Phase
  Synchronization Link for Bistatic SAR},'' {\em IEEE Geoscience and Remote
  Sensing Letters}, vol.~3, pp.~429--433, jul 2006.

\bibitem{He2011}
Z.~He, F.~He, H.~Huang, and D.~Liang, ``{A hardware-in-loop simulation and
  evaluation approach for spaceborne distributed SAR},'' in {\em International
  Geoscience and Remote Sensing Symposium (IGARSS)}, pp.~886--889, IEEE, 2011.

\bibitem{MontiGuarnieri2015}
A.~{Monti Guarnieri}, A.~Broquetas, A.~Recchia, F.~Rocca, and J.~Ruiz-Rodon,
  ``{Advanced Radar Geosynchronous Observation System: ARGOS},'' {\em IEEE
  Geoscience and Remote Sensing Letters}, vol.~12, pp.~1406--1410, jul 2015.

\bibitem{Guarnieri2015}
A.~M. Guarnieri, A.~Broquetas, F.~L{\'{o}}pez-dekker, and F.~Rocca, ``{A
  GEOSTATIONARY MIMO SAR SWARM FOR QUASI-CONTINUOUS OBSERVATION},'' in {\em
  IEEE International Geoscience and Remote Sensing Symposium (IGARSS) 2015},
  pp.~2785--2788, 2015.

\bibitem{Liang2019}
D.~Liang, K.~Liu, H.~Zhang, Y.~Deng, D.~Liu, Y.~Chen, C.~Li, H.~Yue, and
  R.~Wang, ``{A High-Accuracy Synchronization Phase-Compensation Method Based
  on Kalman Filter for Bistatic Synthetic Aperture Radar},'' {\em IEEE
  Geoscience and Remote Sensing Letters}, vol.~17, no.~10, pp.~1722--1726,
  2019.

\bibitem{Jin2020}
G.~Jin, R.~Wang, K.~Liu, D.~Liu, D.~Liang, H.~Zhang, N.~Ou, Y.~Zhang, Y.~Deng,
  and C.~Li, ``{An Advanced Phase Synchronization Scheme for LT-1},'' {\em IEEE
  Transactions on Geoscience and Remote Sensing}, vol.~58, no.~3,
  pp.~1735--1746, 2020.

\bibitem{Zhang2020}
Y.~Zhang, H.~Zhang, N.~Ou, K.~Liu, D.~Liang, Y.~Deng, and R.~Wang, ``{First
  Demonstration of Multipath Effects on Phase Synchronization Scheme for
  LT-1},'' {\em IEEE Transactions on Geoscience and Remote Sensing}, vol.~58,
  no.~4, pp.~2590--2604, 2020.

\bibitem{Liang2020}
D.~Liang, K.~Liu, H.~Zhang, Y.~Chen, H.~Yue, D.~Liu, Y.~Deng, H.~Lin, T.~Fang,
  C.~Li, and R.~Wang, ``{The Processing Framework and Experimental Verification
  for the Noninterrupted Synchronization Scheme of LuTan-1},'' {\em IEEE
  Transactions on Geoscience and Remote Sensing}, pp.~1--11, 2020.

\bibitem{Lovascio2019}
A.~Lovascio, A.~D'Orazio, and V.~Centonze, ``{Design of COTS-based
  radio-frequency receiver for cubesat applications},'' in {\em 2019 IEEE
  International Workshop on Metrology for AeroSpace, MetroAeroSpace 2019 -
  Proceedings}, pp.~399--404, IEEE, 2019.

\bibitem{Yu2022}
L.~Yu, S.~Zhang, N.~Wu, and C.~Yu, ``{FPGA-based Hardware-in-the-Loop
  Simulation of User Selection Algorithms for Cooperative Transmission
  Technology over LOS Channel on Geosynchronous Satellites},'' {\em IEEE
  Access}, pp.~1--1, 2022.

\bibitem{Maheshwarappa2015}
M.~R. Maheshwarappa, M.~Bowyer, and C.~P. Bridges, ``{Software Defined Radio
  (SDR) architecture to support multi-satellite communications},'' in {\em IEEE
  Aerospace Conference Proceedings}, vol.~2015-June, pp.~1--10, IEEE, 2015.

\bibitem{Interdonato2019}
G.~Interdonato, E.~Bj{\"o}rnson, H.~Quoc~Ngo, P.~Frenger, and E.~G. Larsson,
  ``{Ubiquitous cell-free Massive MIMO communications},'' {\em EURASIP Journal
  on Wireless Communications and Networking}, vol.~2019, no.~1, p.~197, 2019.

\bibitem{9120654}
{\"{O}}.~T. {Demir} and E.~{Bj{\"{o}}rnson}, ``{Max-Min Fair Wireless-Powered
  Cell-Free Massive MIMO for Uncorrelated Rician Fading Channels},'' in {\em
  2020 IEEE Wireless Communications and Networking Conference (WCNC)},
  pp.~1--6, 2020.

\bibitem{363e19c89514405e8e37ba23474c2b64}
H.~Ngo, A.~Ashikhmin, H.~Yang, E.~Larsson, and T.~Marzetta, ``Cell-free massive
  mimo versus small cells,'' {\em IEEE Wireless Communications}, vol.~16,
  pp.~1834--1850, 1 2017.

\bibitem{Jeong2020}
S.~Jeong, A.~Farhang, and M.~Flanagan, ``{Collaborative Vs. Non-Collaborative
  CFO Estimation for Distributed Large-Scale MIMO Systems},'' in {\em IEEE
  Vehicular Technology Conference}, vol.~2020-Novem, 2020.

\bibitem{Borg2018}
G.~Borg, Z.~U.~D. Javaid, and A.~Khandaker, ``{The physical and engineering
  requirements of scalable, decentralised, distributed, large-scale MIMO},'' in
  {\em Proceedings of the 3rd International Conference on Informatics and
  Computing, ICIC 2018}, pp.~3--8, IEEE, 2018.

\bibitem{Qamar2017}
F.~Qamar, K.~B. Dimyati, M.~N. Hindia, K.~A.~B. Noordin, and A.~M. Al-Samman,
  ``{A comprehensive review on coordinated multi-point operation for LTE-A},''
  {\em Computer Networks}, vol.~123, no.~May, pp.~19--37, 2017.

\bibitem{Gu2018}
Z.~Gu and Z.~Zhang, ``{Mode selection for CoMP transmission with nonideal
  synchronization},'' {\em China Communications}, vol.~15, no.~12,
  pp.~132--146, 2018.

\bibitem{Chaloupka2018}
Z.~Chaloupka, L.~Ries, A.~Samperi, P.~Waller, and M.~Crisci, ``{Phase
  synchronization for 5G using mass market GNSS receivers},'' in {\em 2018
  European Frequency and Time Forum, EFTF 2018}, pp.~192--196, IEEE, 2018.

\bibitem{Jungnickel2013}
V.~Jungnickel, K.~Manolakis, S.~Jaeckel, M.~Lossow, P.~Farkas, M.~Schlosser,
  and V.~Braun, ``{Backhaul requirements for inter-site cooperation in
  heterogeneous LTE-Advanced networks},'' in {\em 2013 IEEE International
  Conference on Communications Workshops, ICC 2013}, pp.~905--910, 2013.

\bibitem{Tian2017}
Y.~Tian, K.~L. Lee, C.~Lim, and A.~Nirmalathas, ``{Performance evaluation of
  CoMP for downlink 60-GHz radio-over-fiber fronthaul},'' in {\em MWP 2017 -
  2017 International Topical Meeting on Microwave Photonics}, vol.~2017-Decem,
  pp.~1--4, 2017.

\bibitem{Song2018}
G.~Song, W.~Wang, D.~Chen, and T.~Jiang, ``{KPI / KQI-Driven Coordinated
  Multipoint in 5G : Measurements, Field Trials, and Technical Solutions},''
  {\em IEEE Wireless Communications}, no.~October, pp.~23--29, 2018.

\bibitem{Hamza2012}
A.~M. Hamza and J.~W. Mark, ``{A timing synchronization scheme in coordinated
  base-stations cooperative communications},'' in {\em 2012 International
  Conference on Wireless Communications and Signal Processing, WCSP 2012},
  pp.~3--8, IEEE, 2012.

\bibitem{Huang2014}
S.~Y. Huang, Y.~H. Lin, and J.~H. Deng, ``{Novel time offset pre-processing and
  interference cancellation for downlink OFDMA CoMP system},'' in {\em
  Proceedings, APWiMob 2014: IEEE Asia Pacific Conference on Wireless and
  Mobile 2014}, pp.~102--108, IEEE, 2014.

\bibitem{Zhao2014}
L.~Zhao, K.~Liang, G.~Cao, R.~Qian, and D.~L{\'{o}}pez-P{\'{e}}rez, ``{An
  enhanced signal-timing-offset compensation algorithm for coordinated
  multipoint-to-multiuser systems},'' {\em IEEE Communications Letters},
  vol.~18, no.~6, pp.~983--986, 2014.

\bibitem{Dammann2013}
A.~Dammann and R.~Raulefs, ``{Exploiting position information for
  synchronization in coordinated multipoint transmission},'' in {\em IEEE
  Vehicular Technology Conference}, IEEE, 2013.

\bibitem{Pilaram2015}
H.~Pilaram, M.~Kiamari, and B.~H. Khalaj, ``{Distributed synchronization and
  beamforming in uplink relay asynchronous OFDMA CoMP networks},'' {\em IEEE
  Transactions on Wireless Communications}, vol.~14, no.~6, pp.~3471--3480,
  2015.

\bibitem{Hamza2018}
A.~M. Hamza, J.~W. Mark, and E.~A. Sourour, ``{Interference analysis and
  mitigation for time-asynchronous OFDM CoMP systems},'' {\em IEEE Transactions
  on Wireless Communications}, vol.~17, no.~7, pp.~4780--4791, 2018.

\bibitem{Iwelski2014}
S.~Iwelski, B.~Badic, Z.~Bai, R.~Balraj, C.~Kuo, E.~Majeed, T.~Scholand,
  G.~Bruck, and P.~Jung, ``{Feedback generation for CoMP transmission in
  unsynchronized networks with timing offset},'' {\em IEEE Communications
  Letters}, vol.~18, no.~5, pp.~725--728, 2014.

\bibitem{Khanzadi2013}
M.~R. Khanzadi, R.~Krishnan, and T.~Eriksson, ``{Effect of synchronizing
  coordinated base stations on phase noise estimation},'' in {\em ICASSP, IEEE
  International Conference on Acoustics, Speech and Signal Processing -
  Proceedings}, pp.~4938--4942, IEEE, 2013.

\bibitem{Chang2012}
L.~Chang, J.~Zhang, X.~Li, B.~Liu, and K.~Sun, ``{Joint synchronization and
  channel estimation for the uplink coordinated multi-point systems},'' in {\em
  2012 7th International ICST Conference on Communications and Networking in
  China, CHINACOM 2012 - Proceedings}, pp.~384--389, IEEE, 2012.

\bibitem{Koivisto2013}
T.~Koivisto, T.~Kuosmanen, and T.~Roman, ``{Estimation of time and frequency
  offsets in LTE coordinated multi-point transmission},'' in {\em IEEE
  Vehicular Technology Conference}, pp.~3--7, IEEE, 2013.

\bibitem{OctoCloc26:online}
``Octoclock clock distribution module with gpsdo - ettus research | ettus
  research, a national instruments brand | the leader in software defined radio
  (sdr).'' \url{https://www.ettus.com/all-products/octoclock-g/}.
\newblock (Accessed on 05/25/2020).

\bibitem{USRP2Ett93:online}
``Usrp2 - ettus knowledge base.'' \url{https://kb.ettus.com/USRP2}.
\newblock (Accessed on 05/25/2020).

\bibitem{Quitin2012a}
F.~Quitin, U.~Madhow, M.~M.~U. Rahman, and R.~Mudumbai, ``{Demonstrating
  distributed transmit beamforming with software-defined radios},'' in {\em
  2012 IEEE International Symposium on a World of Wireless, Mobile and
  Multimedia Networks (WoWMoM)}, pp.~1--3, IEEE, jun 2012.

\bibitem{MunkyoSeo2008}
{Munkyo Seo}, M.~Rodwell, and U.~Madhow, ``{A feedback-based distributed phased
  array technique and its application to 60-GHz wireless sensor network},'' in
  {\em 2008 IEEE MTT-S International Microwave Symposium Digest}, pp.~683--686,
  IEEE, jun 2008.

\bibitem{Scherber2013}
D.~Scherber, P.~Bidigare, R.~ODonnell, M.~Rebholz, M.~Oyarzun, C.~Obranovich,
  W.~Kulp, D.~Chang, and D.~R.~B. III, ``{Coherent Distributed Techniques for
  Tactical Radio Networks: Enabling Long Range Communications with Reduced
  Size, Weight, Power and Cost},'' in {\em MILCOM 2013 - 2013 IEEE Military
  Communications Conference}, pp.~655--660, IEEE, nov 2013.

\bibitem{Bidigare2012}
P.~Bidigare, M.~Oyarzyn, D.~Raeman, D.~Chang, D.~Cousins, R.~O'Donnell,
  C.~Obranovich, and D.~R. Brown, ``{Implementation and demonstration of
  receiver-coordinated distributed transmit beamforming across an ad-hoc radio
  network},'' in {\em 2012 Conference Record of the Forty Sixth Asilomar
  Conference on Signals, Systems and Computers (ASILOMAR)}, pp.~222--226, IEEE,
  nov 2012.

\bibitem{Mghabghab2019}
S.~Mghabghab, H.~Ouassal, and J.~A. Nanzer, ``{Wireless frequency
  synchronization for coherent distributed antenna arrays},'' {\em 2019 IEEE
  International Symposium on Antennas and Propagation and USNC-URSI Radio
  Science Meeting, APSURSI 2019 - Proceedings}, pp.~1575--1576, 2019.

\bibitem{USRPX31043:online}
``Usrp x310 high performance software defined radio - ettus research | ettus
  research, a national instruments brand | the leader in software defined radio
  (sdr).'' \url{https://www.ettus.com/all-products/x310-kit/}.
\newblock (Accessed on 05/26/2020).

\bibitem{USRPN2105:online}
``Usrp n210 software defined radio (sdr) - ettus research | ettus research, a
  national instruments brand | the leader in software defined radio (sdr).''
  \url{https://www.ettus.com/all-products/un210-kit/}.
\newblock (Accessed on 05/26/2020).

\bibitem{Alemdar}
C.~Bocanegra, K.~Alemdar, S.~Garcia, C.~Singhal, and K.~R. Chowdhury,
  ``{NetBeam : Networked and Distributed 3-D Beamforming for Multi-user
  Heterogeneous Traffic},'' in {\em 2019 IEEE International Symposium on
  Dynamic Spectrum Access Networks (DySPAN)}, (Newark, NJ, USA, USA),
  pp.~1--10, IEEE, 2019.

\bibitem{Rahman2012}
M.~M. Rahman, H.~E. Baidoo-Williams, R.~Mudumbai, and S.~Dasgupta, ``{Fully
  wireless implementation of distributed beamforming on a software-defined
  radio platform},'' in {\em 2012 ACM/IEEE 11th International Conference on
  Information Processing in Sensor Networks (IPSN)}, pp.~305--315, IEEE, apr
  2012.

\bibitem{Pooler2018b}
R.~K. Pooler, J.~S. Sunderlin, R.~H. Tillman, and R.~L. Schmid, ``{A Precise RF
  Time Transfer Method for Coherent Distributed System Applications},'' in {\em
  2018 USNC-URSI Radio Science Meeting (Joint with AP-S Symposium)}, pp.~5--6,
  IEEE, jul 2018.

\bibitem{Schmid2017}
R.~L. Schmid, T.~M. Comberiate, J.~E. Hodkin, and J.~A. Nanzer, ``{A
  Distributed RF Transmitter Using One-Way Wireless Clock Transfer},'' {\em
  IEEE Microwave and Wireless Components Letters}, vol.~27, pp.~195--197, feb
  2017.

\bibitem{Peiffer2016}
B.~Peiffer, R.~Mudumbai, A.~Kruger, A.~Kumar, and S.~Dasgupta, ``{Experimental
  demonstration of a distributed antenna array pre-synchronized for
  retrodirective transmission},'' in {\em 2016 Annual Conference on Information
  Science and Systems (CISS)}, pp.~460--465, IEEE, mar 2016.

\bibitem{Yan2018}
H.~Yan, S.~Hanna, K.~Balke, R.~Gupta, and D.~Cabric, ``{Software Defined Radio
  Implementation of Carrier and Timing Synchronization for Distributed
  Arrays},'' nov 2018.

\bibitem{Frank1962}
R.~Frank, S.~Zadoff, and R.~Heimiller, ``{Phase shift pulse codes with good
  periodic correlation properties},'' {\em IRE Transactions on Information
  Theory}, vol.~8, no.~6, pp.~381--382, 1962.

\bibitem{Overdick2017}
M.~W.~S. Overdick, J.~E. Canfield, A.~G. Klein, and D.~R.~B. III, ``{A
  software-defined radio implementation of timestamp-free network
  synchronization},'' in {\em IEEE International Conference on Acoustics,
  Speech, and Signal Processing (ICASSP) 2017}, (New Orleans, LA, USA),
  pp.~1193--1197, IEEE, 2017.

\bibitem{Alvarez2015}
M.~A. Alvarez, W.~Thompson, and U.~Spagnolini, ``{Distributed time and
  frequency synchronization: USRP hardware implementation},'' in {\em 2015 IEEE
  International Conference on Communication Workshop (ICCW)}, pp.~2157--2162,
  IEEE, jun 2015.

\bibitem{Kim2009}
J.~Kim and S.~W. Lee, ``{Flight performance analysis of GRACE K-band ranging
  instrument with simulation data},'' {\em Acta Astronautica}, vol.~65,
  pp.~1571--1581, 12 2009.

\bibitem{Thomas1999}
J.~B. Thomas, ``{An Analysis of Gravity-Field Estimation Based on
  Intersatellite Dual-1-Way Biased Ranging},'' p.~196, 5 1999.

\bibitem{LISAPage}
``{LISA - Laser Interferometer Space Antenna - NASA Home Page}.''

\bibitem{Budianu2011OLFAR:Swarms}
A.~Budianu, R.~T. Rajan, S.~Engelen, A.~Meijerink, C.~J. Verhoeven, and M.~J.
  Bentum, ``{OLFAR: Adaptive topology for satellite swarms},'' in {\em 62nd
  International Astronautical Congress 2011, IAC 2011}, vol.~9, pp.~7086--7094,
  International Astronautical Federation (IAF), 10 2011.

\bibitem{Rajan2015JointNodes}
R.~T. Rajan, G.~Leus, and A.-J. van~der Veen, ``{Joint relative position and
  velocity estimation for an anchorless network of mobile nodes},'' {\em Signal
  Processing}, vol.~115, pp.~66--78, 10 2015.

\bibitem{Enzer2010}
D.~G. Enzer, R.~T. Wang, and W.~M. Klipstein, ``{GRAIL- A microwave ranging
  instrument to map out the lunar gravity field},'' in {\em 2010 IEEE
  International Frequency Control Symposium, FCS 2010}, pp.~572--577, IEEE,
  2010.

\bibitem{Oudrhiri2014}
K.~Oudrhiri, S.~Asmar, S.~Esterhuizen, C.~Goodhart, N.~Harvey, D.~Kahan,
  G.~Kruizinga, M.~Paik, D.~Shin, and L.~White, ``{An innovative direct
  measurement of the GRAIL absolute timing of Science Data},'' in {\em IEEE
  Aerospace Conference Proceedings}, pp.~1--9, IEEE, 2014.

\bibitem{Wu2019DL}
H.~Wu, Z.~Sun, and X.~Zhou, ``Deep learning-based frame and timing
  synchronization for end-to-end communications,'' {\em Journal of Physics:
  Conference Series}, vol.~1169, pp.~012--060, Feb 2019.

\bibitem{Wang2013learning}
Y.~Wang, C.~Zhang, Q.~Peng, and Z.~Wang, ``Learning to detect frame
  synchronization,'' in {\em Neural Information Processing} (M.~Lee, A.~Hirose,
  Z.-G. Hou, and R.~M. Kil, eds.), (Berlin, Heidelberg), pp.~570--578, Springer
  Berlin Heidelberg, 2013.

\bibitem{Zibar2015}
D.~{Zibar}, L.~H.~H. {de Carvalho}, M.~{Piels}, A.~{Doberstein}, J.~{Diniz},
  B.~{Nebendahl}, C.~{Franciscangelis}, J.~{Estaran}, H.~{Haisch}, N.~G.
  {Gonzalez}, J.~C. R.~F. {de Oliveira}, and I.~T. {Monroy}, ``Application of
  machine learning techniques for amplitude and phase noise characterization,''
  {\em Journal of Lightwave Technology}, vol.~33, no.~7, pp.~1333--1343, 2015.

\bibitem{Nguyen2021IoT}
V.~D. {Nguyen}, S.~K. {Sharma}, T.~X. {Vu}, S.~{Chatzinotas}, and
  B.~{Ottersten}, ``Efficient federated learning algorithm for resource
  allocation in wireless iot networks,'' {\em IEEE Internet of Things Journal},
  vol.~8, no.~5, pp.~3394--3409, 2021.

\bibitem{Zhang2018}
J.~{Zhang}, H.~{Tu}, Y.~{Ren}, J.~{Wan}, L.~{Zhou}, M.~{Li}, and J.~{Wang},
  ``An adaptive synchronous parallel strategy for distributed machine
  learning,'' {\em IEEE Access}, vol.~6, pp.~19222--19230, 2018.

\bibitem{Gerbessiotis1994}
A.~Gerbessiotis and L.~Valiant, ``Direct bulk-synchronous parallel
  algorithms,'' {\em Journal of Parallel and Distributed Computing}, vol.~22,
  no.~2, pp.~251--267, 1994.

\bibitem{DeepRL2016}
V.~Mnih, A.~P. Badia, M.~Mirza, A.~Graves, T.~Lillicrap, T.~Harley, D.~Silver,
  and K.~Kavukcuoglu, ``Asynchronous methods for deep reinforcement learning,''
  in {\em Proceedings of The 33rd International Conference on Machine Learning}
  (M.~F. Balcan and K.~Q. Weinberger, eds.), vol.~48 of {\em Proceedings of
  Machine Learning Research}, (New York, New York, USA), pp.~1928--1937, 20--22
  Jun 2016.

\bibitem{Qirong2013}
Q.~Ho, J.~Cipar, H.~Cui, J.~K. Kim, S.~Lee, P.~B. Gibbons, G.~A. Gibson, G.~R.
  Ganger, and E.~P. Xing, ``More effective distributed ml via a stale
  synchronous parallel parameter server,'' in {\em Proceedings of the 26th
  International Conference on Neural Information Processing Systems - Volume
  1}, NIPS'13, (Red Hook, NY, USA), p.~1223–1231, Curran Associates Inc.,
  2013.

\bibitem{Wang2017}
J.~{Wang}, W.~{Tu}, L.~C.~K. {Hui}, S.~M. {Yiu}, and E.~K. {Wang}, ``Detecting
  time synchronization attacks in cyber-physical systems with machine learning
  techniques,'' in {\em 2017 IEEE 37th International Conference on Distributed
  Computing Systems (ICDCS)}, pp.~2246--2251, 2017.

\bibitem{Bae2016}
S.~Bae, S.~Kim, and J.~Kim, ``Efficient frequency-hopping synchronization for
  satellite communications using dehop-rehop transponders,'' {\em IEEE
  Transactions on Aerospace and Electronic Systems}, vol.~52, no.~1,
  pp.~261--274, 2016.

\bibitem{Lee2019}
S.~Lee, S.~Kim, M.~Seo, and D.~Har, ``Synchronization of frequency hopping by
  lstm network for satellite communication system,'' {\em IEEE Communications
  Letters}, vol.~23, no.~11, pp.~2054--2058, 2019.

\bibitem{Tong2020}
J.~Tong, R.~Song, Y.~Liu, C.~Wang, Q.~Zhou, and W.~Wang, ``Enhanced
  synchronization of 5g integrated satellite systems in multipath channels,''
  in {\em 2020 International Conference on Computer Vision, Image and Deep
  Learning (CVIDL)}, pp.~617--621, 2020.

\bibitem{Guo2021}
H.~Guo, Q.~Yang, H.~Wang, Y.~Hua, T.~Song, R.~Ma, and H.~Guan, ``Spacedml:
  Enabling distributed machine learning in space information networks,'' {\em
  IEEE Network}, vol.~35, no.~4, pp.~82--87, 2021.

\bibitem{Sun2019}
L.~Sun, Y.~Gao, W.~Huang, P.~Li, Y.~Zhou, and J.~Yang, ``{Autonomous Time
  Synchronization Using BeiDou Inter-satellite Link Ranging},'' {\em ICSIDP
  2019 - IEEE International Conference on Signal, Information and Data
  Processing 2019}, 2019.

\bibitem{Giorgi2019}
G.~Giorgi, B.~Kroese, and G.~Michalak, ``{Future GNSS constellations with
  optical inter-satellite links. Preliminary space segment analyses},'' in {\em
  IEEE Aerospace Conference Proceedings}, vol.~2019-March, pp.~1--13, IEEE,
  2019.

\bibitem{Henkel2020}
P.~Henkel, ``{Precise point positioning for next-generation GNSS},'' {\em 2020
  European Navigation Conference, ENC 2020}, pp.~1--11, 2020.

\bibitem{Ruiz-De-Azua2018}
J.~A. Ruiz-De-Azua, A.~Camps, and A.~{Calveras Auge}, ``{Benefits of Using
  Mobile Ad-Hoc Network Protocols in Federated Satellite Systems for Polar
  Satellite Missions},'' {\em IEEE Access}, vol.~6, pp.~56356--56367, 2018.

\bibitem{Ruiz-de-Azua22}
J.~A.~R. de~Azua, A.~Calveras, and A.~Camps, ``From monolithic satellites to
  the internet of satellites paradigm: When space, air, and ground networks
  become interconnected,'' in {\em Computer-Mediated Communication} (I.~Dey,
  ed.), ch.~2, Rijeka: IntechOpen, 2022.

\bibitem{rs13194014}
L.~Fernandez, J.~A. Ruiz-de Azua, A.~Calveras, and A.~Camps, ``On-demand
  satellite payload execution strategy for natural disasters monitoring using
  lora: Observation requirements and optimum medium access layer mechanisms,''
  {\em Remote Sensing}, vol.~13, no.~19, 2021.

\bibitem{DeAzua2018}
J.~A. {De Azua}, A.~Calveras, and A.~Camps, ``{Internet of Satellites (IoSat):
  Analysis of Network Models and Routing Protocol Requirements},'' {\em IEEE
  Access}, vol.~6, pp.~20390--20411, 2018.

\bibitem{Ghasempour2017}
A.~Ghasempour and S.~K. Jayaweera, ``{Data synchronization for throughput
  maximization in distributed transmit beamforming},'' in {\em 2017 Cognitive
  Communications for Aerospace Applications Workshop (CCAA)}, pp.~1--4, IEEE,
  jun 2017.

\bibitem{Sriploy2018}
P.~Sriploy and T.~Chanpuek, ``{The enhancement of wireless sensor network in
  smart farming using distributed beamforming},'' in {\em 1st International
  ECTI Northern Section Conference on Electrical, Electronics, Computer and
  Telecommunications Engineering, ECTI-NCON 2018}, pp.~5--9, IEEE, 2018.

\bibitem{Diao2019}
J.~Diao, M.~Hedayati, and Y.~E. Wang, ``{Experimental Demonstration of
  Distributed Beamforming on Two Flying Mini-Drones},'' in {\em 2019 United
  States National Committee of URSI National Radio Science Meeting (USNC-URSI
  NRSM)}, (Boulder, CO, USA), pp.~97--98, 2019.

\bibitem{Mohanti2019}
S.~Mohanti, C.~Bocanegra, J.~Meyer, G.~Secinti, M.~Diddi, H.~Singh, and
  K.~Chowdhury, ``{AirBeam: Experimental demonstration of distributed
  beamforming by a swarm of UAVs},'' in {\em IEEE 16th International Conference
  on Mobile Ad Hoc and Smart Systems, MASS 2019}, (Monterey, CA, USA),
  pp.~162--170, 2019.

\bibitem{Lan2018}
A.~Lan, J.~Yan, L.~Wu, F.~Zhao, and J.~Wu, ``{A study on in-orbit calibration
  for a spaceborne distributed interferometer},'' in {\em International
  Geoscience and Remote Sensing Symposium (IGARSS)}, vol.~2018-July,
  pp.~1043--1046, 2018.

\bibitem{Dreifuerst2020}
R.~M. Dreifuerst, R.~W. Heath, M.~N. Kulkarni, and J.~Charlie, ``{Deep
  Learning-based Carrier Frequency Offset Estimation with One-Bit ADCs},'' in
  {\em IEEE Workshop on Signal Processing Advances in Wireless Communications,
  SPAWC}, vol.~2020-May, 2020.

\end{thebibliography}

\begin{IEEEbiography}[{\includegraphics[width=1in,height=1.25in,clip,keepaspectratio]{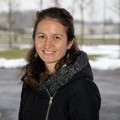}}]{Liz Martinez Marrero} (S'18) was born in Havana, Cuba, in 1989. She received the M.Sc. degree in telecommunications and Telematics from the Technological University of Havana (CUJAE), Cuba, in 2018. She is currently working toward the Ph.D. degree as a Doctoral Researcher at the Interdisciplinary Centre for Security, Reliability, and Trust (SnT) of the University of Luxembourg. Her research interests include digital signal processing for wireless communications, focusing on the physical layer, satellite communications, and carrier synchronization for distributed systems. 

During the 37\textsuperscript{th} International Communications Satellite Systems Conference (ICSSC2019) she received the Best Student Paper Award.
\end{IEEEbiography}

\begin{IEEEbiography}[{\includegraphics[width=1in,height=1.25in,clip,keepaspectratio]{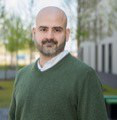}}]{Juan Carlos Merlano Duncan}
(S'09-M'12) received the Diploma degree in
electrical engineering from the Universidad
del Norte, Barranquilla, Colombia, in 2004,
the M.Sc. and Ph.D. Diploma (Cum Laude)
degrees from the Universitat Polit\`ecnica de
Catalunya (UPC), Barcelona, Spain, in 2009
and 2012, respectively. His research
interests are wireless communications, remote sensing,
distributed systems, frequency distribution and carrier
synchronization systems, software-defined radios, and
embedded systems.

At UPC, he was responsible for the design and implementation
of a radar system known as SABRINA, which was the first
ground-based bistatic radar receiver using space-borne
platforms, such as ERS-2, ENVISAT, and TerraSAR-X as
opportunity transmitters (C and X bands). He was also in charge
of the implementation of a ground-based array of transmitters,
which was able to monitor land subsidence with subwavelength
precision. These two implementations involved
FPGA design, embedded programming, and analog
RF/Microwave design. In 2013, he joined the Institute National
de la Recherche Scientifique, Montreal, Canada, as a Research
Assistant in the design and implementation of cognitive radio
networks employing software development and FPGA
programming. He joined the University of Luxembourg since
2016, where he currently works as a Research Scientist in the
COMMLAB laboratory working on SDR implementation of
satellite and terrestrial communication systems and passive
remote sensing systems.
\end{IEEEbiography}

\begin{IEEEbiography}[{\includegraphics[width=1in,height=1.25in,clip,keepaspectratio]{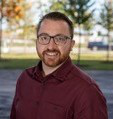}}]{Jorge Querol}
(S'13-M'18) was born in
Forcall, Castell\'o, Spain, in 1987. He
received the B.Sc. (+5) degree in
telecommunication engineering, the
M.Sc. degree in electronics engineering,
the M.Sc. degree in photonics, and the
Ph.D. degree (Cum Laude) in signal
processing and communications from the
Universitat Polit\`ecnica de Catalunya - BarcelonaTech (UPC),
Barcelona, Spain, in 2011, 2012, 2013 and 2018 respectively.
His research interests include Software Defined Radios (SDR),
real-time signal processing, satellite communications, 5G nonterrestrial
networks, satellite navigation, and remote sensing.
His Ph.D. thesis was devoted to the development of novel antijamming
and counter-interference systems for Global
Navigation Satellite Systems (GNSS), GNSS-Reflectometry,
and microwave radiometry. One of his outstanding
achievements was the development of a real-time standalone
pre-correlation mitigation system for GNSS, named FENIX, in
a customized Software Defined Radio (SDR) platform. FENIX
was patented, licensed and commercialized by MITIC
Solutions, a UPC spin-off company.

Since 2018, he is Research Associate at the SIGCOM research
group of the Interdisciplinary Centre for Security, Reliability,
and Trust (SnT) of the University of Luxembourg,
Luxembourg. He is involved in several ESA and
Luxembourgish national research projects dealing with signal
processing and satellite communications.

He received the best academic record award of the year in
Electronics Engineering at UPC in 2012, the first prize of the
European Satellite Navigation Competition (ESNC) Barcelona
Challenge from the European GNSS Agency (GSA) in 2015,
the best innovative project of the Market Assessment Program
(MAP) of EADA business school in 2016, the award Isabel P.
Trabal from Fundacio Caixa d'Enginyers for its quality
research during his Ph.D. in 2017, and the best Ph.D. thesis
award in remote sensing in Spain from the IEEE Geoscience
and Remote Sensing (GRSS) Spanish Chapter in 2019.
\end{IEEEbiography}

\begin{IEEEbiography}[{\includegraphics[width=1in,height=1.25in,clip,keepaspectratio]{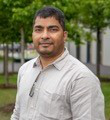}}]{Sumit Kumar}
(S'14-M'19) received his
Bachelor of Technology and Master of
Science in Electronics \& Communication
Engineering from Gurukula Kangri
University, Haridwar, India (2008) and
the International Institute of Information
Technology, Hyderabad, India (2014),
respectively, and his PhD degree from
Eurecom (France) in 2019. His research interests are in wireless
communication, interference management and software defined
radio.
\end{IEEEbiography}

\begin{IEEEbiography}[{\includegraphics[width=1in,height=1.25in,clip,keepaspectratio]{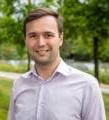}}]{Jevgenij Krivochiza} (S'13-M'20)
received the B.Sc. and M.Sc. degrees in electronic engineering in telecommunications physics and electronics from the Faculty of Physics, Vilnius University, in 2011 and 2013, respectively. He received the Ph.D. degree in electrical engineering from the Interdisciplinary Centre for Security, Reliability, and Trust (SnT), University of Luxembourg, in 2020. Currently, he is a Research Associate at SNT, University of Luxembourg. His main research topics are coming from development for FPGA silicon, software-defined radios, digital signal processing, precoding, interference mitigation, DVB-S2X, DVB-S2, and LTE systems. He works on \acrshort{DSP} algorithms for SDR platforms for advanced precoding and beamforming techniques in next-generation satellite communications.
\end{IEEEbiography}

\begin{IEEEbiography}[{\includegraphics[width=1in,height=1.25in,clip,keepaspectratio]{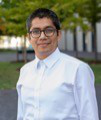}}]{Shree Krishna Sharma} (S'12-M'15-SM'18) obtained his PhD degree in wireless communications from University of Luxembourg in 2014. He held various research and academic positions at the SnT, University of Luxembourg; Western University, Canada; and Ryerson University, Canada. He has published more than 100 technical papers in scholarly journals, international conferences, and book chapters, and has over 4200 google scholar citations with an h-index of 30. He is the recipient of several prestigious awards including ``FNR Award for Outstanding PhD Thesis 2015'' from FNR, Luxembourg, ``Best Paper Award'' in CROWNCOM 2015 conference, ``2018 EURASIP JWCN Best Paper Award'' and the co-recipient of ``FNR Award for Outstanding Scientific Publication 2019''. He is a Senior Member of IEEE and a lead editor of two IET books on “Satellite Communications in the 5G Era” and “Communications Technologies for Networked Smart Cities”.
\end{IEEEbiography}

\begin{IEEEbiography}[{\includegraphics[width=1in,height=1.25in,clip,keepaspectratio]{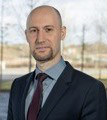}}]{Symeon Chatzinotas}
(S'06-M'09-SM'13) received the M.Eng. degree in
telecommunications from the Aristotle
University of Thessaloniki, Thessaloniki,
Greece, in 2003, and the M.Sc. and Ph.D.
degrees in electronic engineering from the
University of Surrey, Surrey, U.K., in 2006
and 2009, respectively. He is currently
Full-Professor, and the Deputy Head of the SIGCOM Research
Group, Interdisciplinary Centre for Security, Reliability, and
Trust, University of Luxembourg, Luxembourg, and a Visiting
Professor with the University of Parma, Italy. His research
interests include multiuser information theory, cooperative/
cognitive communications, and wireless network
optimization.

He has been involved in numerous research and development
projects with the Institute of Informatics Telecommunications,
National Center for Scientific Research Demokritos, Institute
of Telematics and Informatics, Center of Research and
Technology Hellas, and Mobile Communications Research
Group, Center of Communication Systems Research,
University of Surrey.

Dr. Chatzinotas has coauthored more than 400 technical papers in refereed international journals, conferences and scientific books. He was the co-recipient of the 2014 IEEE Distinguished Contributions to Satellite Communications Award, the CROWNCOM 2015 Best Paper Award, and the 2018 EURASIP JWCN Best Paper Award. He is currently on the Editorial Board of the IEEE Open Journal of Vehicular Technology and the International Journal of Satellite Communications and Networking.
\end{IEEEbiography}

\begin{IEEEbiography}[{\includegraphics[width=1in,height=1.25in,clip,keepaspectratio]{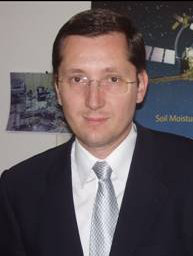}}]{Adriano Camps}
(S'91-A'97-M'00-SM'03-F'11) was  born  in Barcelona, Spain, in 1969. He received the degree in Telecommunications Engineering and a Ph.D. degree in  Telecommunications  Engineering  from  the  Universitat Polit\'ecnica de Catalunya (UPC), Barcelona, Spain, in  1992  and  1996,  respectively.  From  1991  to 1992,  he  was  at  the  ENS  des  Télécommunications de Bretagne, Brest, France,  with  an  Erasmus  Fellowship. Since 1993, he  has been  with the  Electromagnetics and  Photonics  Engineering  Group,  Department  of Signal Theory and Communications, UPC, where he was first Assistant Professor, Associate Professor in 1997, and Full Professor since  2007.  In  1999,  he  was  on  sabbatical  leave  at  the  Microwave  Remote Sensing  Laboratory,  of  the  University  of  Massachusetts,  Amherst.  Since 1993,  he  has  been  deeply  involved  in  the  European  Space  Agency  SMOS Earth Explorer Mission, from the instrument and algorithmic points of view, performing  field  experiments,  and  since  2001  studying  the  use  of  GNSS-R  techniques  to  perform  the  sea  state  correction  needed  to  retrieve  salinity from  L-band  radiometric  observations.  His  research  interests  are  focused  in microwave  remote  sensing,  with  special  emphasis  in  microwave  radiometry by aperture synthesis techniques, remote sensing using signals of opportunity (GNSS-R), radio-frequency detection and mitigation techniques for microwave radiometry and GNSS, and CubeSats as platforms to test novel remote sensing concepts. He has published over 216 papers in peer-reviewed journals, 6 book chapters,  1  book,  and  more  than  450  international  conference  presentations, holds 12 patents, and has advised 25 Ph.D. Thesis students (+ 8 on-going), and more than 140 final projects and M.Eng. Thesis. According to Publish or Perish (Google Scholar) his publications have received more than 7161/10706 citations, and his h-index is 39/51 according to Scopus/Google Scholar. He was the IEEE GRSS President in 2017-2018, Technical Program Committee chair of IGARSS 2017, general co-chair of IGARSS 2020, and he has been involved in the organization of several other conferences. He has also received several  awards  including  the  EURopean  Young  Investigator  (EURYI)  award 2004, the Duran Farell award for technology transfer in 2000 and 2010, the ICREA Acad\`emia research award 2008 and 2014, etc.
\end{IEEEbiography}

\begin{IEEEbiography}[{\includegraphics[width=1in,height=1.25in,clip,keepaspectratio]{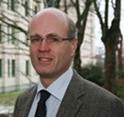}}]{Bj\"orn Ottersten}
(S'87-M'89-SM'99-F'04) received the M.S. degree in electrical engineering and applied physics from Linköping University, Linköping, Sweden, in 1986, and the Ph.D. degree in electrical engineering from Stanford University, Stanford, CA, USA, in 1990. He has held research positions with the Department of Electrical Engineering, Linköping University, the Information Systems Laboratory, Stanford University, the Katholieke Universiteit Leuven, Leuven, Belgium, and the University of Luxembourg, Luxembourg. From 1996 to 1997, he was the Director of Research with ArrayComm, Inc., a start-up in San Jose, CA, USA, based on his patented technology. In 1991, he was appointed Professor of signal processing with the Royal Institute of Technology (KTH), Stockholm, Sweden. Dr. Ottersten has been Head of the Department for Signals, Sensors, and Systems, KTH, and Dean of the School of Electrical Engineering, KTH. He is currently the Director for the Interdisciplinary Centre for Security, Reliability and Trust, University of Luxembourg. He is a recipient of the IEEE Signal Processing Society Technical Achievement Award, the EURASIP Group Technical Achievement Award, and the European Research Council advanced research grant twice. He has co-authored journal papers that received the IEEE Signal Processing Society Best Paper Award in 1993, 2001, 2006, 2013, and 2019, and 8 IEEE conference papers best paper awards. He has been a board member of IEEE Signal Processing Society, the Swedish Research Council and currently serves of the boards of EURASIP and the Swedish Foundation for Strategic Research. Dr. Ottersten has served as Editor in Chief of EURASIP Signal Processing, and acted on the editorial boards of IEEE Transactions on Signal Processing, IEEE Signal Processing Magazine, IEEE Open Journal for Signal Processing, EURASIP Journal of Advances in Signal Processing and Foundations and Trends in Signal Processing. He is a fellow of EURASIP. 
\end{IEEEbiography}

\EOD

\end{document}